\documentclass[sigconf, nonacm]{acmart}

\usepackage{xspace}
\xspaceaddexceptions{]}
\usepackage{multirow} 
\usepackage{array}
\usepackage{listings}
\usepackage{marginnote}
\usepackage{balance}
\usepackage{enumitem}
\usepackage{fontawesome5}
\usepackage{color, colortbl}
\usepackage{pifont}
\usepackage[draft,commandnameprefix=ifneeded]{changes}
\usepackage[table]{xcolor}

\usepackage{amssymb}
\usepackage{pifont} 
\usepackage[x11names]{xcolor}
\usepackage[dvipsnames]{xcolor}
\usepackage{pgfplotstable}
\usepackage{pgfplots}
\usepgfplotslibrary{groupplots}
\usepgfplotslibrary{statistics}
\usepackage{booktabs}
\usepackage{siunitx} 
\usepackage{multirow}
\usepackage{tabularx}
\usepackage[most]{tcolorbox}
\DeclareMathOperator*{\argmax}{arg\,max}
\definecolor{darkgreen}{RGB}{0,100,0}
\newsavebox{\paqllabelbox}
\newlength{\paqllabelindent}
\newlength{\paqlskip}
\newcommand{\paqllabel}[1]{\ifnum#1=1 \usebox{\paqllabelbox}\fi}
\makeatletter
\def\paql@NewLine{\ifx\lst@OutputBox\@gobble\else
    \par\nobreak\noindent\hbox{}\fi
  \global\advance\lst@newlines\m@ne\lst@newlinetrue}
\lstnewenvironment{paqllisting}[1]
  {\sbox{\paqllabelbox}{\normalfont\small #1}%
   \setlength{\paqllabelindent}{\wd\paqllabelbox}%
   \addtolength{\paqllabelindent}{2mm}%
   \let\lst@NewLine\paql@NewLine
   \lstset{basicstyle=\normalfont\ttfamily\small,
     aboveskip=\paqlskip, belowskip=\paqlskip,
     numbers=left, firstnumber=1, numberstyle=\paqllabel, numbersep=2mm,
     xleftmargin=\paqllabelindent}}
  {}
\makeatother
\usepackage{caption}
\usepackage{makecell}
\makeatletter
\let\c@table\c@figure
\makeatother
\renewcommand{\thetable}{\thefigure}

\definecolor{LightGreen}{RGB}{200,255,200}
\definecolor{LightRed}{RGB}{255,200,200}
\definecolor{LightYellow}{RGB}{255,255,191} 
\definecolor{LightBlue}{RGB}{191,223,255}

\newcommand\vldbavailabilityurl{https://github.com/kuangfei-long/ex2bundle}
\newcommand\vldbpagestyle{plain} 

\newtheorem{example}{Example}
\newtheorem{problem}{Problem}[section]

\newcommand{\dataset}{\textsc{SubSumE}\xspace}
\newcommand{\cnndm}{CNN/\allowbreak DailyMail\xspace}
\newcommand{\system}{\textsc{Ex2Bundle}\xspace}
\newcommand{\sudocu}{\textsc{SuDocu}\xspace}

\newcommand{\SBERT}{\textsc{SBERT}\xspace}

\newcommand{\presumm}{\textsc{BERTSumExt}\xspace}
\newcommand{\memsumm}{\textsc{MemSum}\xspace}

\newcommand{\squid}{\textsc{SQuID}\xspace}

\newcommand{\paql}{PaQL\xspace}
\newcommand{\sketchref}{\textsc{SketchRefine}\xspace}

\newcommand{\qbe}{QbE\xspace}
\newcommand{\bqbe}{BQbE\xspace}

\newcommand{\topic}[1]{\textsc{#1}} 

\newcommand{\numtriplets}{2{,}200\xspace}
\newcommand{\numpeople}{103\xspace}
\newcommand{\numintents}{ten\xspace}

\newcommand{\paratitle}[1]{\smallskip\noindent\textbf{#1.}~}
\newcommand{\code}[1]{\texttt{#1}}

\definecolor{teal}{rgb}{0.1,0.6,0.7}

\RenewDocumentCommand{\marginnote}{o m o}{}

\definechangesauthor[name={MB}, color=cyan]{MB}

\usepackage[linesnumbered, lined, ruled, noend]{algorithm2e}
\SetKwFor{MyWhile}{while}{do}{}
\SetNlSty{bfseries}{\color{black}}{}

\begin{document}
\title{Example-Driven Intent Synthesis for Constrained Data Bundle Retrieval: Focused Text Snippet Extraction and Beyond}

\author{Whanhee Cho}
\affiliation{%
  \institution{University of Utah}
  \city{}
  \state{}
  \country{}
} 
\email{whanhee.cho@utah.edu} 

\author{Kuangfei Long}
\orcid{}
\affiliation{%
  \institution{Boston University}
  \city{}
  \state{}
  \country{}
}
\email{longkuangfei@outlook.com}

\author{Mahmood Jasim}
\orcid{}
\affiliation{%
  \institution{Louisiana State University}
  \city{}
  \state{}
  \country{}
}
\email{mjasim@lsu.edu}

\author{Matteo Brucato}
\affiliation{%
  \institution{OSM Data}
  \city{}
  \state{}
  \country{}
}
\email{matteo@osm-data.com}

\author{Alexandra Meliou}
\affiliation{%
  \institution{UMass Amherst}
  \city{}
  \state{}
  \country{}
}
\email{ameli@cs.umass.edu}

\author{Peter J.~Haas}
\affiliation{%
  \institution{UMass Amherst}
  \city{}
  \state{}
  \country{}
}
\email{phaas@cs.umass.edu}

\author{Anna Fariha}
\affiliation{%
  \institution{University of Utah}
  \city{}
  \state{}
  \country{}
}
\email{afariha@cs.utah.edu}

\begin{abstract}
\looseness-1 Selecting a \emph{bundle} of items that collectively satisfies
constraints is a fundamental task across databases, recommender systems, and
text summarization. Unlike traditional top-$k$ retrieval, bundle retrieval is
inherently combinatorial and, in general, NP-hard. Although package queries can
efficiently retrieve bundles given a well-formed query, two key user-centric
challenges remain: (1)~expressing and tuning multi-dimensional bundle intent
through a user-friendly interface, and (2)~ensuring feasibility when the query
yields empty results. We introduce \system, an \emph{\underline{Ex}ample-driven
\underline{Bundle} retrieval} framework that enables users to specify their
intent through \textit{example bundles} and automatically synthesizes package
queries that capture the intent implicit in those examples via aggregate
constraints. \system also addresses a challenge unique to bundle retrieval:
when inferred constraints are infeasible over the target data, our
\textit{data-aware constraint relaxation} adjusts a minimal number of
constraints while preserving alignment with user intent. We instantiate a
specific application of \textit{focused text snippet extraction} to demonstrate
the efficacy of \system. Extensive experiments over two real-world datasets and
a user study demonstrate that \system improves usability and consistently
returns intent-aligned bundles even under distributional shifts of the target
database.
\end{abstract}

\maketitle

\pagestyle{\vldbpagestyle}

\ifdefempty{\vldbavailabilityurl}{}{
\vspace{-2mm}
\begingroup\small\noindent\raggedright\textbf{Artifact Link:} \url{\vldbavailabilityurl}.
\endgroup
}


\vspace{-1mm}
\section{Introduction}\label{sec:intro}

\looseness-1 Data \textit{bundle} retrieval is a fundamental task across
multiple domains: package queries retrieve packages (sets of tuples) from
relational databases under aggregate constraints~\cite{BrucatoBAM16};
recommender systems suggest playlists (sets of songs) or combo offers (sets of
products) tailored to user preferences~\cite{BaiWWW19Personalized}; and
extractive text summarization systems retrieve subsets of sentences from
documents that form coherent summaries based on user
queries~\cite{automatic_summarization, XiaoEMNLP19Extractive, sudocu}. Unlike
traditional retrieval that returns individual or top-k items, bundle retrieval
involves selecting a \textit{set} (or package~\cite{BrucatoBAM16}) whose
elements must collectively optimize global objectives while satisfying
set-level constraints that align with user preferences. Consequently, the
inclusion or exclusion of any single item affects the feasibility and overall
quality of the bundle, making the problem inherently combinatorial and
NP-hard~\cite{BrucatoBAM16}. While systems exist for the efficient retrieval of
data bundles from this combinatorial search space~\cite{BrucatoBAM16}, two
user-centric challenges remain as primary obstacles: (1)~the \emph{interface}
challenge---how can humans easily express their intent for the desired data
bundles---and (2)~the \emph{feasibility} challenge---how to ensure non-empty
answers for the user queries.

\smallskip\noindent\textbf{Challenge 1: Usable interface.} \looseness-1 The
first challenge is communicating the user intent for the desired data bundle.
Like SQL, the Package Query Language (\paql)~\cite{BrucatoBAM16} allows
specification of a bundle query through linear constraints and objective
functions over data attributes, but the users must know the exact query
parameters: attribute names and their numerical bounds. This is particularly
difficult for non-experts who must (i)~learn the \paql~\cite{BrucatoBAM16}
syntax, (ii)~know the database schema, (iii)~determine the attributes of
interest, (iv)~have a good sense of the value distributions, and (v)~translate
their intent precisely to parameterized constraints and global objectives of a
\paql query. We show in Example~\ref{ex:one} that even for experts, step~(v) is
particularly difficult for \paql because constraints are over \emph{bundle
aggregates} (SUM, AVG, COUNT): users must reason at the bundle
level---predicting what a SUM should be over a bundle of unspecified
size---rather than at the tuple level as in SQL. For certain applications, such
as extractive document summarization~\cite{XiaoEMNLP19Extractive}, an
alternative is to specify the intent in natural language. However, as prior
work shows, natural language is too generic to express specific user
intents~\cite{sudocu} and subsequent conversational tuning of intent is
imprecise and frustrating for humans~\cite{DBLP:conf/sigir/ZamaniMCLDBCD20}.

\begin{example}[Synthesizing a package query]
\label{ex:one}
Morpheus, a department Chair, seeks to fill two tenure-track assistant
professor positions to strengthen AI and Databases. As he plans for the two new
hires to jointly teach an ``AI in Databases'' course, he wants them to have
reasonable collective teaching experience, but not too much, as that may
indicate teaching-focused or senior candidates unlikely to accept the position.
He also wants to maximize the total recommendation score of the new hires. This
is a bundle retrieval problem: Morpheus must pick two of $200$ candidates that
best satisfy his criteria, as shown via the following under-specified package
query:\footnote{While this toy example involves 2 hires over 4 criteria, faculty
searches typically involve 5--10 candidates and 15+ criteria, yielding a
combinatorial search space. E.g., choosing $5$ from $200$ candidates yields
$\binom{200}{5} \approx 2.5$ billion possible bundles.}

\newcommand{\vaguevalue}[1]{\textcolor{magenta}{#1}}
\begin{paqllisting}{\texttt{\textbf{Q1}:}}
SELECT PACKAGE(*) FROM CANDIDATES
SUCH THAT COUNT(*) = 2
  AND SUM(ai_score) is |\vaguevalue{high}|
  AND SUM(db_score) is |\vaguevalue{high}|
  AND SUM(teaching_score) is |\vaguevalue{reasonable}|
MAXIMIZE SUM(reco_score);
\end{paqllisting}

{\normalfont
\begin{table}[t]
\centering
\small
\renewcommand{\arraystretch}{0.8}
\resizebox{0.9\columnwidth}{!}{
\begin{tabular}{@{}l@{}ccc@{}c@{}}
\toprule
& \textbf{AI} & \textbf{Databases} & \textbf{Teaching} & \textbf{Recommendation} \\
& \texttt{\small ai\_score} & \texttt{\small db\_score} & \texttt{\small teaching\_score} & \texttt{\small reco\_score} \\
\midrule
Smith   & 0.8           & 0.4           & 0.1    & 0.4  \\
Jones   & 0.4           & 0.7           & 0.2    & 0.9  \\
Neo     & 0.4           & 0.5           & 0.4    & 0.8  \\
Brown   & 0.5           & 0.4           & 0.4    & 0.9  \\
\bottomrule
\end{tabular}}
\caption{\small Partial list of candidates for Example~\ref{ex:one}.}
\label{tab:one}
\vspace{-7mm}
\end{table}}

\looseness-1 However, Morpheus has only the raw application materials (resumes,
statements, letters, etc.) of these candidates. Before querying the candidate
database, he must map each candidate's materials to a 4-dimensional vector
space: \texttt{\small ai\_score}, \texttt{\small db\_score}, \texttt{\small
teaching\_score}, and \texttt{\small reco\_score}. He uses an off-the-shelf
embedding technique to obtain numeric scores for each dimension for every
candidate (Table~\ref{tab:one}). But he now faces another problem: he does not
know what values should replace the underspecified \vaguevalue{high} and
\vaguevalue{reasonable} in \texttt{\small {Q1}}. Is $0.7$ a \vaguevalue{high}
AI expertise score? What range is \vaguevalue{reasonable} for teaching score?
He can examine the embeddings' value distributions, but still cannot determine
which values accurately capture his intended criteria.

\end{example}

Example~\ref{ex:one} highlights two key struggles in formulating package queries
to express bundle query intent: (1)~the lack of means to accurately map a
database to a vector database with appropriate dimensions of interest, which
primarily affects non-experts; and (2)~the lack of knowledge to correctly
parameterize the \paql query, which affects experts and non-experts alike. In
Example~\ref{ex:one}, Morpheus was interested in only 4 criteria; however, his
struggle would be even worse if there were a larger number of criteria.

\smallskip\noindent\textbf{Our solution to Challenge 1: bundle-query by
example.} \looseness-1 To overcome the usability challenge in the interface of
bundle query intent specification, we build on the \emph{Query by Example}
(\qbe) paradigm~\cite{DBLP:conf/vldb/Zloof75}, which has seen significant
success in traditional SQL querying~\cite{squid} and other
domains~\cite{DBLP:conf/aplas/GulwaniJ17, DBLP:journals/pvldb/RezigBFPVGS21,
DBLP:conf/popl/Gulwani11, sudocu} by lowering the barrier to task
specification. We extend \qbe to bundle queries (\bqbe): the user provides a
few exemplar data bundles to convey their query intents implicitly, thereby
bypassing the construction of a precisely parameterized PaQL query. \bqbe is
particularly useful when users cannot articulate precise criteria but can
instead provide representative examples of what they seek.

\begin{example}[Parameterizing a package query]
\label{ex:two}
Continuing from Example~\ref{ex:one}, Morpheus recalls that strong new hires
resemble recent hires at top universities, such as \{Trinity, Cypher\}
(University X) and \{Link, Niobe, Seraph\} (University Y). These exemplars
excel in Databases and AI with reasonable collective teaching experience.
Morpheus computes their scores across 4 dimensions using the same embedding of
Example~\ref{ex:one} (Table~\ref{tab:two}). Guided by the aggregated scores, he
parameterizes the underspecified query \texttt{\small {Q1}} to obtain:

\begin{paqllisting}{\texttt{\textbf{Q2}:}}
SELECT PACKAGE(*) FROM CANDIDATES
SUCH THAT COUNT(*) = 2
  AND SUM(ai_score) BETWEEN 0.8 AND 1.2
  AND SUM(db_score) BETWEEN 1.0 AND 1.3
  AND SUM(teaching_score) BETWEEN 0.7 AND 1.3
MAXIMIZE SUM(reco_score);
\end{paqllisting}
\end{example}

Example~\ref{ex:two} shows a scenario where a user cannot specify precise
criteria via \paql parameters but can provide valid \textit{examples} that
implicitly convey them, from which user preferences can be learned and
translated into query parameters. Yet another key challenge remains, as
Example~\ref{ex:three} shows.

{\normalfont
\begin{table}[t]
\centering
\small
\renewcommand{\arraystretch}{0.9}
\resizebox{0.9\columnwidth}{!}{
\begin{tabular}{c@{\hspace{8pt}}l@{}cccc}
\toprule
& & \texttt{\small ai\_score} & \texttt{\small db\_score} & \texttt{\small teaching\_score} & \texttt{\small reco\_score} \\
\midrule
& Trinity  & 0.6 & 0.5 & 0.3 & 0.7 \\
& Cypher   & 0.2 & 0.8 & 0.4 & 0.9 \\
\rowcolor{gray!20}
\cellcolor{white}\multirow{-3}{*}{\rotatebox[origin=c]{90}{\scriptsize\textbf{\shortstack{Univ.~X}}}}
& \textbf{SUM} & \textbf{0.8} & \textbf{1.3} & \textbf{0.7} & \textbf{1.6} \\
\midrule
& Link    & 0.9 & 0.3 & 0.4 & 0.6 \\
& Niobe   & 0.2 & 0.3 & 0.6 & 0.8 \\
& Seraph  & 0.1 & 0.4 & 0.3 & 0.7 \\
\rowcolor{gray!20}
\cellcolor{white}\multirow{-4}{*}{\rotatebox[origin=c]{90}{\scriptsize\textbf{\shortstack{Univ.~Y}}}}
& \textbf{SUM} & \textbf{1.2} & \textbf{1.0} & \textbf{1.3} & \textbf{2.1} \\
\midrule
\multicolumn{2}{l}{\textbf{Implicit intent}} & \textbf{[0.8, 1.2]} & \textbf{[1.0, 1.3]} & \textbf{[0.7, 1.3]} & \textbf{Maximize} \\
\bottomrule
\end{tabular}}
\caption{\small Example bundles help discover query parameters.}
\vspace{-7mm}
\label{tab:two}
\end{table}
}

{\normalfont
\begin{table}[t]
\centering
\small
\resizebox{0.94\columnwidth}{!}{
\begin{tabular}{@{}llllcc}
\toprule
bundle & \texttt{\small ai} & \texttt{\small db} & \texttt{\small teaching} & \texttt{\small reco} & \textbf{valid?}\\
\midrule
b$_1$ \phantom{s} \{Smith, Jones\} & 1.2                                   & 1.1                                     & 0.3 \textcolor{red}{($\downarrow$ 0.4)} & 1.3 & \textbf{no}  \\
b$_2$ \phantom{s} \{Smith, Neo\}   & 1.2                                   & 0.9 \textcolor{red}{($\downarrow$ 0.1)} & 0.5 \textcolor{red}{($\downarrow$ 0.2)} & 1.2 & \textbf{no}  \\
b$_3$ \phantom{s} \{Smith, Brown\} & 1.3 \textcolor{red}{($\uparrow$ 0.1)} & 0.8 \textcolor{red}{($\downarrow$ 0.2)} & 0.5 \textcolor{red}{($\downarrow$ 0.2)} & 1.3 & \textbf{no}  \\
b$_4$ \phantom{s} \{Jones, Neo\}   & 0.8                                   & 1.2                                     & 0.6 \textcolor{red}{($\downarrow$ 0.1)} & 1.7 & \textcolor{black}{\textbf{almost}}  \\
b$_5$ \phantom{s} \{Jones, Brown\} & 0.9                                   & 1.1                                     & 0.6 \textcolor{red}{($\downarrow$ 0.1)} & 1.8 & \textcolor{black}{\textbf{almost}}  \\
b$_6$ \phantom{s} \{Neo, Brown\}   & 0.9                                   & 0.9 \textcolor{red}{($\downarrow$ 0.1)} & 0.8                                     & 1.7 & \textcolor{black}{\textbf{almost}}  \\
\bottomrule
\end{tabular}} 
\caption{\small No candidate bundle fully satisfies the constraints of
\texttt{\small Q2}. For instance, $b_6$ violates the constraint
``\code{SUM(db\_score) BETWEEN 1.0 AND 1.3}'', as its \code{db\_score} (0.9)
is 0.1 below the lower bound (1.0).}
\vspace{-8mm}
\label{tab:three}
\end{table}
}


\begin{example}[Ensuring query feasibility to obtain results]
\label{ex:three}
Morpheus issues \texttt{\small Q2} to a package query execution
engine~\cite{BrucatoBAM16}, but it returns no result. A close inspection reveals
that indeed none of the 2-candidate bundles fully satisfies \texttt{\small Q2}
(Table~\ref{tab:three}). However, the last three bundles ``almost'' satisfy the
constraints. If the \texttt{\small db\_score} constraint had a slightly relaxed
lower bound (0.9 instead of 1.0), then b$_6$ would be a valid bundle. Similarly,
slightly relaxing the lower bound of the \texttt{\small teaching\_score}
constraint (0.6 instead of 0.7) would result in two additional valid bundles:
b$_4$ and b$_5$. Then following the goal of maximizing the \texttt{\small
reco\_score}, the bundle $b_5$ would be the final result. \end{example}

\smallskip\noindent\textbf{Challenge 2: ensuring feasibility.} \looseness-1
While \bqbe addresses the first challenge, it brings forth a second one that is
particular to bundle retrieval and absent from traditional single-tuple \qbe:
constraints inferred from example bundles may yield queries that are
\textit{infeasible} over the target data, returning empty
answers~\cite{VLDB2006LuoEfficient, VLDB2011NandiGuided}, because aggregate
constraints can be jointly unsatisfiable even when individually plausible,
especially under distribution shift between the example and target domains. In
Example~\ref{ex:three}, even with a fully formed \paql query, with the desired
constraints and specific parameters, it turned out to be \textit{infeasible}
over the target data (Morpheus' University). This happened because the data
distributions differed between the example domains (Universities X and Y) and
the target domain (Morpheus' University), a common scenario in
practice~\cite{SIGMOD15IdreosPC15Overview, CIDR13SellamK13Meet}. Existing
package query execution systems~\cite{BrucatoBAM16} provide no further insight
into how the query parameters interact with the target data, leaving users
uncertain about how to tune the parameters to make the query feasible such that
valid results are retrieved.

\smallskip\noindent\textbf{Our solution to Challenge 2: data-aware relaxation
of query parameters.} \looseness-1 To ensure feasibility even when the original
query is infeasible, we introduce \textit{data-aware constraint relaxation
techniques} that adapt the synthesized \paql query to the target data by
adjusting constraint bounds. By analyzing the data distribution of the target
data, we identify which constraints act as bottlenecks that prevent retrieval
of ``almost-valid'' bundles. This analysis guides us to relax the appropriate
constraints, while limiting deviation from the user's original intent, but
without altering too many constraints.

\subsubsection*{\textbf{Why existing approaches fall short.}}
Three classes of relevant approaches exist, but all of them fail to support \bqbe.

\smallskip\noindent\emph{\paql and \qbe systems.}
\looseness-1 \paql engines \cite{BrucatoBAM16} require fully parameterized
queries and return empty results when infeasible; \qbe systems~\cite{DBLP:conf/vldb/Zloof75,
squid} reduce the parameter burden for single-tuple retrieval but do not
extend to bundle queries with aggregate constraints.

\smallskip\noindent\emph{Top-$k$ retrieval.} For \bqbe, an alternative is to
treat the entire example bundle as a single query vector and retrieve top-$k$
similar tuples based on vector similarity~\cite{XuanEMLNP25MCIR,
YunxiaoEMNLP25Answering} to assemble the result bundle. However, this approach
can miss globally optimal bundles and violate intended constraints, since
top-$k$ retrieval evaluates items independently using an implicit scoring
function---without guarantees of global optimality---whereas bundle retrieval
requires reasoning over combinations of tuples under globally
enforced~\cite{ZhaoICLR25DO}, multi-dimensional aggregate constraints. RAG
systems inherit the same limitation, as they perform top-$k$ nearest-neighbor
search over individual tuples (but not their combinations) before passing
results to a generative model. Another alternative is to match each tuple in
the example bundle independently and combine their top matches to form the
result bundle; yet this approach breaks down when the example and target bundle
sizes differ. In summary, bundle retrieval is fundamentally an NP-hard
combinatorial optimization problem, and similarity-based greedy top-$k$ search
fails to address it.

\smallskip\noindent\emph{Conversational LLMs.} \looseness-1 An LLM prompted
with examples (few-shot learning) could produce the bundle directly, but this
requires reasoning over the entire target data and solving a combinatorial
optimization problem, which LLMs handle unreliably~\cite{constraintbench}. An
LLM could instead synthesize \paql queries; however, the bounds would reflect
an opaque mapping between the user's examples and constraint values. LLMs also
have no native mechanism to detect infeasibility or relax bounds in an
intent-preserving way, and per-query latency limits interactive use. In a pure
NL interface, without examples to anchor the bounds, the user must rely on
conversational refinement, which is imprecise and
tedious~\cite{DBLP:conf/sigir/ZamaniMCLDBCD20}.

\subsubsection*{\textbf{Desiderata.}} 
Examples~\ref{ex:one}--\ref{ex:three} and the limitations of alternative
approaches motivate the desiderata of an ideal \bqbe system:

\begin{itemize}[leftmargin=5.7mm]
	
\item[\textbf{D1}] Ensure a user-friendly interface for specifying intents,
where users can provide examples to implicitly convey their bundle query
intents, without requiring any knowledge of the database schema and complex
query syntax and parameters.

\item[\textbf{D2}] Model the user intent implicit in the examples into a
representation with explicit parameters, such as a \paql query.

\item[\textbf{D3}] Ensure feasibility of the synthesized \paql query over the
target database, yielding a non-empty result even when no bundle perfectly
matches the user's intent.

\item[\textbf{D4}] Provide an intuitive interface for iterative intent
refinement, allowing users to adjust query constraints.

\end{itemize}




\subsubsection*{\textbf{\system}} 
We introduce \system, an example-driven data bundle retrieval system that
(i)~enables users to specify bundle query intent through example bundles, addressing D1;
(ii)~synthesizes a package query from these examples to transparently capture user intent, addressing D2;
(iii)~applies data-aware constraint relaxation to ensure feasibility of the synthesized query w.r.t.\ the target data domain, addressing D3; and
(iv)~provides an intuitive interface for interactive intent refinement, addressing D4. 

\subsubsection*{\textbf{Use cases.}} 
\label{usecase}
We now present three real-world applications where \bqbe lowers the barrier for
data bundle retrieval (more details are in Appendix~\ref{appx:appendix_applications}):

\vspace{0.5mm}\noindent $\bullet$ \emph{Supplier selection for business}.
Businesses entering new markets struggle to identify a set of suppliers via
explicit constraints over acceptable price ranges, inventory availability, or
financial stability thresholds. \system can infer these constraints implicitly
from examples of supplier bundles drawn from existing markets where the
business is successful, adapt them to the target market, and retrieve an
optimal set of suppliers that satisfies the constraints.

\vspace{0.5mm}\noindent $\bullet$ \emph{Focused snippet extraction}. In the
context of text summarization, \bqbe arises as \textit{focused text snippet
extraction}: given a ``focus'' in the form of several source document--summary
pairs, where each summary is expressed as a set of extracted sentences
(snippet), the goal is to select a snippet from a target document that is
consistent with the examples. \system builds on our prior work
\sudocu~\cite{sudocu} to support personalized extractive
summarization~\cite{subsume}.

\vspace{0.5mm}\noindent $\bullet$ \emph{Playlist recommendation.} Existing
playlist recommenders---e.g., Spotify's Discover Weekly~\cite{spotify_music},
YouTube's Daily Discover~\cite{youtube_music}, and Netflix's Top
Picks~\cite{netflix}---rely on item-level or single-playlist similarity, which
often fail to capture diverse tastes and cause user
frustration~\cite{spotify-frustration}. \system allows users to provide example
playlists reflecting their desired spread across genres, artists, tempo, etc.,
and generates playlists that satisfy their preferences.

\subsubsection*{\textbf{Contributions}} \label{contrib}
We make the following contributions:

\begin{itemize}[leftmargin=*]


     \item We formalize the problem of \textit{example-driven bundle retrieval}
     and show how it translates to the package query framework
     (\S\ref{sec:problem}).
     
     \item We present \system, an end-to-end framework for \emph{example-driven
     bundle retrieval} that synthesizes \paql query constraints from user
     examples. We contribute a \textit{data-aware constraint-bound relaxation
     technique} that ensures query feasibility. We further introduce design
     principles for an interactive slider-based interface for intent
     refinement, along with effective techniques to translate between sliders
     and constraint bounds (\S\ref{sec:system}).

     \item We evaluate \system on two use cases---package queries over the
     TPC-H dataset~\cite{tpc-h} and focused text snippet extraction over
     real-world datasets~\cite{subsume,hermann2015teaching}---showing that
     \system achieves 100\% constraint satisfaction while maintaining
     competitive quality, and outperforms retrieval and extractive baselines.
     Notably, \system remains competitive with LLM-based approaches without
     incurring any additional token or inference cost, while scaling to
     realistic workloads (\S\ref{sec:experiments}).
     
     \item Our user study shows that \system achieves high user satisfaction,
     due to improved ease of use and customizable sliders, for the task of
     focused text snippet extraction (\S\ref{sec:userstudy}).

\end{itemize}
\newcommand{\score}{quality score\xspace}
\newcommand{\scorefunction}{\sigma\xspace}
\newcommand{\schema}{\mathbf{f}\xspace}
\newcommand{\featureSchema}{f\xspace}
\newcommand{\sourcedata}{T^s\xspace}
\newcommand{\sourcedataset}{\mathbf{T}^s\xspace}
\newcommand{\targetdata}{T^q\xspace}
\newcommand{\tuple}{t\xspace}
\newcommand{\lowerbound}{\mathtt{lb}\xspace}
\newcommand{\upperbound}{\mathtt{ub}\xspace}
\newcommand{\satisfies}{\vdash\xspace}
\newcommand{\featureprofile}{\mathbf{F}\xspace}
\newcommand{\featureBundle}{F\xspace}
\newcommand{\aggregate}{\operatorname*{\scalebox{1.1}{$\mathcal{A}$}}}
\newcommand{\synthesisfunction}{G\xspace}

\section{Problem Formulation}
\label{sec:problem}

In this section, we formalize the problem of example-driven bundle retrieval.
We begin by introducing how we model the \textit{source} and \textit{target
data}---from which bundles are retrieved---and \textit{example bundles}---which
communicate the user's intent. We then show how bundle retrieval can be
naturally modeled within the package query framework and define the problem of
\textit{example-driven package query synthesis} for bundle retrieval, our focus
in this paper. Table~\ref{fig:notation} provides a summary of notations used
throughout the paper.

\begin{table}[t]
    \centering
    \resizebox{0.94
    \linewidth}{!}{%
    \begin{tabular}{ll}
    \toprule
    \textbf{Symbol} & \textbf{Description} \\
    \midrule
    $\schema = \{\featureSchema_1, \dots, \featureSchema_K\}$                   & Common schema over $K$ attributes/features \\
    $\sourcedata_i \models \schema$                                             & A source data over the schema $\schema$ \\    
    $\tuple \models \schema$                                                    & A single tuple over the schema $\schema$ \\
    $\schema(\tuple) = [\featureSchema_1(\tuple), \dots, \featureSchema_K(\tuple)] \in \mathbb{R}^K$      & Feature vector of tuple $\tuple$ \\
    $E_i \subseteq \sourcedata_i$                                               & Example bundle for the source data $\sourcedata_i$ \\    
    $\sourcedataset = \{\sourcedata_1, \dots, \sourcedata_N\}$                  & Set of $N$ source data \\
    $\mathbf{E} = \{E_1, \dots, E_N\}$                                          & Set of $N$ user-provided example bundles \\
    \midrule                
    $\targetdata \models \schema$                                               & A target data over $\schema$ to retrieve a bundle from \\
    $B \models \schema$                                                         & A bundle over the schema $\schema$ \\
    $\Theta = \{\langle \lowerbound_j, \upperbound_j \rangle\}_{j=1}^K$         & Constraint bounds for $K$ attributes \\  
    $\featureprofile(B) = [\featureBundle_1(B), \dots, \featureBundle_K(B)]$    & Feature profile of bundle $B$ \\  
    $\featureprofile(B) \satisfies \Theta$                                      & $\featureprofile(B)$ satisfies the constraints given by $\Theta$\\
    $\scorefunction(\tuple) \mapsto \mathbb{R}$                                 & Domain-specific tuple-level scoring function \\    
    
    \bottomrule
    \end{tabular}
    }
    \caption{\small Table of notations. Bold letters denote sets or vectors.}
    \vspace{-7mm}
    \label{fig:notation}
\end{table}

\smallskip\noindent{\textbf{Source and target data.}} In example-driven bundle
retrieval, each example bundle is taken from a corresponding \emph{source
data}. In Example~\ref{ex:two}, \{Trinity, Cypher\} is an example bundle from
the source data of University X and \{Link, Niobe, Seraph\} from the source
data of University Y. One key requirement here is that the source data for all
the example bundles must have the same schema, to allow for effective intent
discovery across the shared attributes.

\looseness-1 We define a schema $\schema$ of a single table\footnote{
While our formalization focuses on a single-table schema, it is not an inherent
limitation: we support relational databases by joining tables into a
denormalized table, as shown in our experiments over the TPC-H dataset
(\S\ref{sec:experiments}).}
over $K$ numerical\footnote{
For unstructured data such as text, we derive numerical features through
domain-specific methods such as topic modeling or learned vector embeddings
(\S\ref{sec:system}) to produce a structured representation under a fixed
numerical schema.}
attributes/features $\{\featureSchema_1, \dots, \featureSchema_K\}$. Each
source data $\sourcedata_i$ must conform to $\schema$, denoted as
$\sourcedata_i \models \schema$. In Table~\ref{tab:two}, the source data for
all the Universities conform to the schema \{\texttt{\small ai\_score},
\texttt{\small db\_score}, \texttt{\small teaching\_score}\}. A tuple $\tuple$ over $\schema$, denoted as $\tuple \models
\schema$, is characterized by a feature vector $\schema(\tuple) {=}
[\featureSchema_1(\tuple), \dots, \featureSchema_K(\tuple)] \in \mathbb{R}^K$.
In Table~\ref{tab:two}, Trinity's feature vector is $[0.6, 0.5, 0.3]$. We
use $\targetdata \models \schema$ to denote the \textit{target data} from which
the user wants to extract their desired bundle. In Example~\ref{ex:two},
Table~\ref{tab:one} represents the target data.

\smallskip \noindent \textbf{Example bundles.}
In example-driven bundle retrieval, users convey their intent via a set of \emph{example bundles} $\mathbf{E} = \{E_1, \dots, E_N\}$, where each $E_i \subseteq \sourcedata_i$ consists of a subset of tuples from the corresponding source data $\sourcedata_i$. In Example~\ref{ex:two}, $\mathbf{E} = $ \{\{Trinity, Cypher\}, \{Link, Niobe, Seraph\}\} and the source data set $\sourcedataset = 
\{\sourcedata_{\text{UofX}}, \sourcedata_{\text{UofY}}\}$.

\subsection{Example-driven bundle retrieval} We are now ready to (informally) define our problem:

\vspace{-1mm}
\begin{problem}[Example-Driven Bundle Retrieval]\label{prb:underspecified}
Given a set of source data $\sourcedataset$ over the same schema $\schema$, corresponding example bundles $\mathbf{E}$, where $E_i \in \mathbf{E}$ is an example bundle for the source data $\sourcedata_i \in \sourcedataset$, and a single target data $\targetdata \models \schema$, identify a bundle $B^* \subseteq \targetdata$ that (i)~satisfies the ``user intent'' implicit in $\mathbf{E}$ w.r.t $\sourcedataset$ and (ii)~is the ``best'' among such bundles.
\end{problem}
\vspace{-1mm}

However, this problem is underspecified: it is unclear how to model the user
intent and define ``best'' or the notion of optimality.

\subsubsection{Modeling user intent} \looseness-1 \label{sec:intent_modeling}
Capturing user intent from example bundles may require complex models involving
non-linear constraints. However, many practical use cases share the following
common structure: (i)~the desired bundle is a \emph{set} of items, (ii)~quality
of the bundle depends on some \emph{aggregate} properties (e.g., total score)
of the set, and (iii)~the user preference is expressible as \emph{bounds} on
those aggregates. These align perfectly with \emph{package
query}~\cite{BrucatoBAM16}---which retrieves subsets of tuples from relational
tables that collectively satisfy (linear) aggregate constraints while
optimizing a (linear) global objective---as illustrated by~\texttt{\small Q2}
in Example~\ref{ex:two}. Thus, we model user intent as \textit{linear
constraints} on the bundle's ``feature profile'' (described next), with an
optional cardinality constraint on the bundle size. The benefit of this
modeling is twofold: it provides \textit{interpretability} ---as linear
constraints can be easily understood and inspected by
humans~\cite{DBLP:journals/datamine/ScholbeckCMBH24, DBLP:conf/iclr/GurneeT24,
DBLP:conf/fat/SemenovaRP22, molnar2020interpretable,
DBLP:journals/tiis/MohseniZR21, DBLP:conf/chi/Poursabzi-Sangdeh21}---and
\textit{tractability}---it maps to Integer Linear Programming (ILP) with
established solvers and foundations~\cite{BrucatoBAM16}.

\smallskip\noindent\textbf{Feature profile of a bundle.}
To capture the feature-wise properties of a bundle $B$, we define its
\emph{feature profile} $\featureprofile$ as a vector obtained by aggregating, for each feature, the values of that feature across all tuples in $B$. Formally, 
\vspace{-1mm}
\begin{equation*}
    \featureprofile(B) = [\featureBundle_1(B), \dots, \featureBundle_K(B)],\\
    \;\;\text{where  } \featureBundle_j(B) = \textstyle\aggregate_{t \in B} \featureBundle_j(\tuple)
\end{equation*}
Here, $\aggregate$ is an aggregate such as \texttt{\small SUM}. In
Table~\ref{tab:two}, using \texttt{\small SUM} as the aggregate, the feature
profile of the bundle \{Trinity, Cypher\} is $[0.8,\, 1.3,\, 0.7]$---the
column-wise sums for University X hires.

\smallskip\noindent\textbf{Formulating constraints.} Following prior work in \qbe~\cite{squid}, given a set of user-provided example bundles, we posit that the user's intended result bundle would exhibit a feature profile similar to that of the example bundles. Accordingly, we model the user's intent as a \emph{set of bounded constraints over the feature profile} of the desired bundle, where the derivation of the bounds should be guided by the feature profiles of the example bundles in $\mathbf{E}$, the source data set $\sourcedataset$, and the target data $\targetdata$. Formally:
\begin{align*}
    \{\lowerbound_1 &\le \featureprofile_1(B) \le \upperbound_1, \dots,
    \lowerbound_K \le \featureprofile_K(B) \le \upperbound_K\}\\
    \text{where } \lowerbound_j  &\text{ and } \upperbound_j \text{ depend on $\mathbf{E}$, $\sourcedataset$, and $\targetdata$ for $1\le j\le K$}
\end{align*}
We use $\Theta = \{\langle \lowerbound_1, \upperbound_1 \rangle, \dots, \langle \lowerbound_K, \upperbound_K \rangle\}$ to denote constraint bounds for all $K$ attributes over the schema $\schema$. The notation $\featureprofile(B) \satisfies \Theta$ denotes that the feature profile of the bundle $B$ satisfies the constraint bounds specified by $\Theta$, i.e., $\forall\, 1 \le j \le K,\; \lowerbound_j \le \featureBundle_j(B) \le \upperbound_j$.

\subsubsection{Defining bundle optimality.}
\label{sec:define-bundle-optimality}\looseness-1 The second issue in
Problem~\ref{prb:underspecified} is that, when multiple candidate bundles
satisfy $\Theta$, a principled way is required to quantify their quality and
ensure optimality. To this end, we assume knowledge of an application-specific
tuple-level function $\scorefunction(\tuple) \mapsto \mathbb{R}$, typically
provided by a domain expert who configures the system for end users. We define
the quality of a bundle $B$ by aggregating $\scorefunction$ over all $\tuple
\in B$, which naturally fits the linear objective function optimized by the
\paql framework. For example, this could correspond to maximizing the total
recommendation score in Example~\ref{ex:one}, or minimizing the word count in
text summarization. Alternatively, $\scorefunction$ can be learned from user
feedback on bundle quality, e.g., for playlist recommendation in music
streaming.

\subsection{Example-driven package query synthesis}\label{sec:paqlsynthesis}

With our models of user intent and bundle optimality, which naturally align with the package query framework, we reformulate Problem~\ref{prb:underspecified} of example-driven bundle retrieval as that of synthesizing a package query---specifically, its parameters---from example bundles. The synthesized package query serves as a \textit{mechanism} for efficiently retrieving the optimal result bundle.

Given a target data $\targetdata$, constraint bounds $\Theta = \{\langle
\lowerbound_j, \upperbound_j\rangle\}_{j=1}^K$, and (optional) cardinality
constraint bounds $\mathcal{C} = \langle \lowerbound_c, \upperbound_c\rangle$,
we fix the following parameterized package query:

\begin{paqllisting}{$\mathrm{PQ}(\targetdata, \Theta, \mathcal{C})$:}
SELECT PACKAGE(*) AS B FROM |$\targetdata$|
SUCH THAT COUNT(B) BETWEEN |$\lowerbound_c$| AND |$\upperbound_c$|
  AND |$\featureBundle_j(B)$| BETWEEN |$\lowerbound_j$| AND |$\upperbound_j$| |$\forall 1 \le j \le K$|
MAXIMIZE |$\displaystyle \aggregate_{t \in B} \sigma(\tuple)$|
\end{paqllisting}

\looseness-1 Here, $\aggregate$ aggregates the tuples $\tuple \in B$ using
$\scorefunction$ to compute the quality of $B$, which the package query aims to
maximize as its objective. Without loss of generality, a minimization objective
can be expressed as a maximization objective by simply negating the objective.
The choice of aggregation functions (including those used to compute the
feature profile $\featureprofile(B)$) and the scoring function $\scorefunction$
is typically guided by the application domain. These are system parameters
determined during setup and are not optimized over. We discuss how to choose
the aggregation and scoring functions in \S\ref{sec:quality}. Also,
without loss of generality, the cardinality constraint $\lowerbound_c \le
\texttt{COUNT}(B) \le \upperbound_c$ can be subsumed into the feature-profile
constraints by augmenting the feature profile with an additional dimension
representing bundle cardinality. Hence, we do not explicitly include the
cardinality constraint in the remainder of the paper.

In Example~\ref{ex:two}, \texttt{\small Q2} is a package query over the schema $\schema {=} \{$\texttt{\small ai\_score}, \texttt{\small db\_score}, \texttt{\small teaching\_score}$\}$, so $K = 3$. The aggregation used to compute the feature profile is \texttt{\small SUM}, and the objective is to maximize \texttt{\small SUM(reco\_score)} (i.e., $\scorefunction =$ \texttt{\small reco\_score}). 

While Morpheus manually derived the constraint bounds in $\Theta$, the resulting query was infeasible over the target data $\targetdata$ representing his university. Thus, the main requirement is to determine the parameter $\Theta$, and use it to synthesize a package query that will retrieve the optimal bundle $B \subseteq \targetdata$, as specified in Problem~\ref{prb:underspecified}.

\begin{problem}[Example-driven package query synthesis]\label{prob2}
Giv\-en a schema $\schema$, 
a set of source data $\sourcedataset = \{\sourcedata_1, \dots, \sourcedata_N\}$ where each $\sourcedata_i \models \schema$, 
corresponding example bundles $\mathbf{E}$, 
a target data $\targetdata \models \schema$, 
and a tuple-level scoring function $\scorefunction: \tuple {\mapsto} \mathbb{R}$, together with the corresponding aggregator $\aggregate$,
determine a synthesis function $\synthesisfunction: (\mathbf{E}, \sourcedataset, \targetdata) \mapsto \langle \mathbb{R}, \mathbb{R}\rangle^{K}$ to derive the constraint bounds $\Theta$ such that:
\begin{enumerate}[leftmargin=*,noitemsep]

    \item \textbf{Feasibility}: the package query $PQ(\targetdata, \Theta)$
    returns a non-empty result over $\targetdata$, ensuring $\exists\, B
    \subseteq \targetdata \;\text{s.t.}\; \featureprofile(B) \satisfies \Theta$.

    \item \textbf{Optimality}: the retrieved bundle $B^* = PQ(\targetdata,
    \Theta)(\targetdata)$ maximizes the objective, i.e., $B^* {=}
    \displaystyle\argmax_{B \subseteq \targetdata \;\text{s.t.}\;
    \featureprofile(B) \satisfies \Theta}\;\; \displaystyle\aggregate_{\tuple
    {\in} B} \scorefunction(\tuple)$.
    
\end{enumerate}
\end{problem}

The core challenge here is to determine the synthesis function
$\synthesisfunction: (\mathbf{E}, \sourcedataset, \targetdata) \mapsto
\langle\mathbb{R}, \mathbb{R}\rangle^{K}$ to derive the constraint bounds
$\Theta$ by capturing the intent implicit in $\mathbf{E}$ while generalizing to
the target data $\targetdata$. Unlike general \qbe, we do not focus on
synthesizing package queries with arbitrary structures, which would require
inferring joins and aggregates. Instead, we focus on inferring the parameters
of the constraint bounds $\Theta$. This is challenging because the distribution
of $\targetdata$ may differ from $\sourcedataset$: overly tight constraints
(Example~\ref{ex:two}) can yield infeasible queries on the target data, while
overly relaxed constraints allow bundles that deviate from user intent.
Section~\ref{sec:system} describes the techniques to ensure query feasibility.



\begin{figure}[t]
\centering
\includegraphics[width=0.97\linewidth]{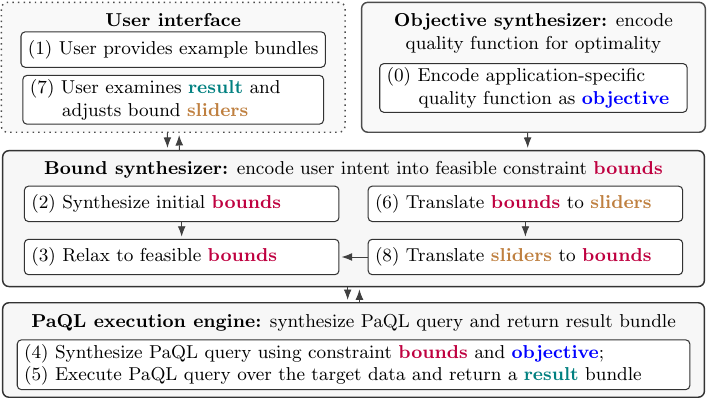}
\vspace{-3mm}
\caption{\small \looseness-1 \system workflow: before end-user interaction, 
a domain expert provides a quality function to be used as the objective and 
specifies an aggregate function to be used in the constraints. 
(0)~\system encodes the quality function as objective, 
(1)~the user provides example bundles, from which \system formulates constraints using the expert-specified aggregate,
(2)~synthesizes initial constraint bounds and 
(3)~relaxes them to ensure feasibility. A \paql query is 
(4)~formed and 
(5)~executed to retrieve a result bundle. For intent refinement, 
(6)~an interactive slider interface is generated, 
(7)~the user inspects results and adjusts sliders, 
(8)~\system maps adjustments back to bounds, and repeats (3)--(5).}
\vspace{-4mm}
\label{fig:system-framework}
\end{figure}

\newcommand{\slider}{\gamma}

\section{The \system Framework}
\label{sec:system}

\system consists of four components (Fig.~\ref{fig:system-framework}):
\textit{bound slider}~(\S\ref{sec:interface}), an interface that allows users
to refine their intent by adjusting constraint bounds;
\textit{bound synthesizer}~(\S\ref{sec:intent}), which encodes user intent into
\emph{feasible} constraint \textcolor{black}{{bounds}};
\textit{objective synthesizer}~(\S\ref{sec:quality}), which encodes an
application-specific quality function as the \textcolor{black}{{objective}} of
the \paql query;
and \textit{\paql execution engine}~(\S\ref{sec:execution}), which synthesizes
and executes package queries to retrieve \textcolor{black}{{result}} bundles.

\smallskip\noindent\emph{Running example for focused text snippet extraction.}
\looseness-1
Throughout the rest of the paper, we use a running example centered on the task
of \textit{Focused Text Snippet Extraction} (FTSE), a primary application
through which we evaluate \system (\S\ref{sec:experiments}). Because \system
requires numeric features and text is non-numeric, we apply
\textit{contextualized topic modeling} (CTM)~\cite{ctm_1} with sentence
embeddings from \SBERT~\cite{reimers-2019-sentence-bert} to map sentences into a numeric vector
space of topic-based features. Under the resulting topic model defined over
schema $\schema$, $f_j(t) \in [0, 1]$ denotes the relevance of a sentence $t$
to topic $f_j$, and $\schema(t)$ denotes the vector representation of $t$ in
the topic space $\schema$. For ease of understanding, we use the following,
more specific and domain-relevant terms for FTSE, instead of generic terms:

\noindent{
\begin{center}
\vspace{1mm}
\resizebox{0.99\columnwidth}{!}{
    \begin{tabular}{ll@{\hspace{0.5cm}}|@{\hspace{0.5cm}}ll}
    \toprule
    \textbf{General term} &  \textbf{FTSE term} & \textbf{General term} &  \textbf{FTSE term}\\
    \midrule
    source/target data  & source/target document     & tuple   & sentence\\
    bundle              & snippet or summary         & schema & topic model\\
    attribute/feature             & topic                      & feature profile & topic profile\\    
    \bottomrule
    \end{tabular}}
\vspace{-1mm}
\end{center}
    
}


\begin{example}
Consider a topic model $\schema$ over \texttt{\small Demographics},
\texttt{\small Geography}, and \texttt{\small Economy} applied to documents
describing U.S.\ states. Let $\sourcedata_{\text{UT}}$ denote the Utah document
and $E_{\text{UT}} = \{t_{\text{UT}}^{1}\,, t_{\text{UT}}^{2}\,,
t_{\text{UT}}^{3}\} \subseteq \sourcedata_{\text{UT}}$ be three sentences
(Table~\ref{tab:topic_distribution}). Under $\schema$, $t_{\text{UT}}^{1}$ maps
to $\schema(t_{\text{UT}}^{1}) = [0.90, 0.00, 0.00]$, indicating a demographics
focus; $t_{\text{UT}}^{2}$ to $[0.90, 0.90, 0.00]$, covering demographics and
geography; and $t_{\text{UT}}^{3}$ to $[0.90, 0.30, 0.10]$, primarily
demographics with weaker geography and economy associations.
\end{example}
\vspace{-3mm}

\subsection{Bound slider interface for intent refinement}
\label{sec:interface}
A key requirement (D4 in \S\ref{sec:intro}) for \system is an intuitive
interface for iterative intent refinement. A simple approach is to let users
directly edit the synthesized bounds (\S\ref{sec:intent}). However, non-experts
often do not know the precise mapping between their intent and constraint
bounds (Example~\ref{ex:one}). Based on studies~\cite{10.1145/3706598.3714292,
DBLP:conf/eccv/GandikotaMZTB24, DBLP:conf/uist/SetlurBTGC16,
DBLP:conf/chi/WangHS0024, 10.1145/3586183.3606777,
DBLP:journals/tvcg/StrobeltWSHBPR23} establishing that humans are better at
relative judgments (more vs.\ less) than absolute ones (assigning
ratings)~\cite{thurstone1974comparative}, we introduce a single-parameter
slider that abstracts the raw bounds.

Fig.~\ref{fig:slider-example} shows the slider interface, where each position
corresponds to a pair of upper and lower bounds, and the neutral point
corresponds to the feasible bounds synthesized from user examples (details in
\S\ref{sec:intent}). The slider instance in Fig.~\ref{fig:slider-example}
depicts the topic \topic{Demographics}, where the neutral point $0$ corresponds
to the feasible bounds $\langle 1.50, 2.70\rangle$. The slider further encodes
other representations on a scale from $-100$ to $+100$ around the neutral point
to allow fine-grained adjustments. Upon reviewing the result bundle, users
simply indicate whether they want more (less) relevance to the topic by moving
the slider right (left).

\newcommand{\maxval}{\mathtt{mx}}
\newcommand{\width}{\mathtt{w}}

\smallskip\noindent\textbf{Mapping between slider positions and bounds.}
\looseness-1 We use $\gamma_j$ to denote the position of the slider for topic
$f_j$. We map the initial feasible bounds $\langle \lowerbound_j, \upperbound_j
\rangle$---synthesized from user examples---to the midpoint of the slider at
$\gamma_j {=} 0$. The extremes of the left-most and right-most slider positions
map to the minimum and maximum attainable bundle scores for topic $f_j$ in
$\targetdata$: $0$ and $\maxval {=} \featureBundle_j(\targetdata)$. However,
while each side of the midpoint has 100 positions, the initial feasible bounds
are generally not equidistant from the two extremes. Thus, for slider positions
left of the midpoint, we determine the \textit{lower} bound by interpolating
based on the position's relative distance between the left-most endpoint
($\gamma_j {=} {-}100$) and the midpoint ($\gamma_j {=} 0$). Symmetrically, for
positions right of the midpoint, we determine the \textit{upper} bound by
interpolating based on the position's relative distance between the right-most
endpoint ($\gamma_j {=} {+}100$) and the midpoint ($\gamma_j {=} 0$). Assuming
a constant width $\width$, the mappings are as follows:
\begin{center}
	\vspace{-1mm}
\resizebox{0.85\columnwidth}{!}{
\begin{tabular}{lllr}
\toprule
\textbf{Slider Position} & \textbf{Lower bound} & \textbf{Upper bound} & $\boldsymbol{\gamma_j}$ \\
\midrule
(left-most)           & $0$ & $\width$ & $-100$ \\
(left of the center)  & $\lowerbound_j{\cdot}\frac{(100{-}x)}{100}$ & $\lowerbound_j{\cdot}\frac{(100{-}x)}{100} + \width$ & $-x$ \\
(center)              & $\lowerbound_j$ & $\upperbound_j$ & $0$ \\
(right of the center) & $\upperbound_j {+} \frac{(\maxval {-} \upperbound_j){\cdot}x}{100} - \width$ & $\upperbound_j {+} \frac{(\maxval {-} \upperbound_j){\cdot}x}{100}$ & $+x$ \\
(right-most)          & $\maxval - \width$ & $\maxval$ & $+100$ \\
\bottomrule
\end{tabular}}
\vspace{0mm}
\end{center}

\setlength{\textfloatsep}{10pt}

\noindent A question still remains: how to determine the value of $\width$?
Should it be constant for all slider positions? We use the width of the neutral
position $(\upperbound_j{-}\lowerbound_j)$ as a guide to derive the bound width
of slider position at $\gamma_j{=}x$ with the formula: $\width(x) =
e^{-\frac{\alpha{\cdot}|x|}{100}}{\cdot}(\upperbound_j {-} \lowerbound_j)$.


\newcommand{\topicscoretotal}[4][]{%
    \hspace{-7mm}
    \begin{tikzpicture}[baseline={([yshift=-0.8ex]current bounding box.center)}]
        \begin{axis}[
            ybar,
            bar width=10pt,
            width=4.0cm,
            height=2cm,
            ymin=0, ymax=3,
            xmin=0.4, xmax=3.6,
            xtick={1, 2, 3},
            xticklabels={Demographics, Geography, Economy},
            xticklabel style={rotate=270, anchor=west},
            ytick={0.0,0.5,1.0, 1.5, 2, 2.5},            
            yticklabels={0.0,,,,2.0,},            
            every node near coord/.append style={font=\tiny},
            nodes near coords={
                \pgfmathprintnumber[fixed, precision=2, zerofill]{\pgfplotspointmeta}
            },
            nodes near coords align={vertical},
            every node near coord/.append style={font=\tiny},
            axis x line=bottom,
            axis y line=left,
            axis line style={-}, 
            ylabel=\empty,
            xticklabel style={font=\scriptsize},
            yticklabel style={font=\scriptsize},
            #1 
        ]
            \addplot[fill=gray!40] coordinates {(1,#2) (2,#3) (3,#4)};
        \end{axis}
    \end{tikzpicture}%
}

\newcommand{\topicscore}[4][]{%
    \hspace{-7mm}
    \begin{tikzpicture}[baseline={([yshift=-0.8ex]current bounding box.center)}]
        \begin{axis}[
            ybar,
            bar width=10pt,
            width=4.0cm,
            height=2cm,
            ymin=0, ymax=1.0,
            xmin=0.4, xmax=3.6,
            xtick={1, 2, 3},
            xticklabels={Demographics, Geography, Economy},
            xticklabel style={rotate=270, anchor=west},
            ytick={0.0,0.5,1.0},            
            yticklabels={0.0,,1.0},            
            every node near coord/.append style={font=\tiny},
            nodes near coords={
                \pgfmathprintnumber[fixed, precision=2, zerofill]{\pgfplotspointmeta}
            },
            nodes near coords align={vertical},
            every node near coord/.append style={font=\tiny},
            axis x line=bottom,
            axis y line=left,
            axis line style={-}, 
            ylabel=\empty,
            xticklabel style={font=\scriptsize},
            yticklabel style={font=\scriptsize},
            #1 
        ]
            \addplot[fill=gray!40] coordinates {(1,#2) (2,#3) (3,#4)};
        \end{axis}
    \end{tikzpicture}%
}

\begin{table}[t]
    \centering
    \renewcommand{\tabularxcolumn}[1]{m{#1}} 
    \resizebox{0.49\textwidth}{!}{
    \footnotesize
    \begin{tabularx}{0.49\textwidth}{@{}X@{}c@{}}
        \toprule
        \textbf{Sentence/snippet} & \textbf{Topic score/profile} \\
        \midrule
        $(t_{\text{UT}}^{1})$ The largest ancestry groups in the state are: 26.0\% English, 11.9\% German, [...] & 
        \topicscore[xticklabels={,,}]{0.90}{0.00}{0.00} \\
        
        $(t_{\text{UT}}^{2})$ In comparison to all the U.S.\ states and territories, Utah, with a population of just over three million, is the 13th-largest by area, the 30th-most populous, and the 11th-least densely populated. & 
        \topicscore[xticklabels={,,}]{0.90}{0.90}{0.00} \\
        
        $(t_{\text{UT}}^{3})$ St.\ George was the fastest-growing metropolitan area in the United States from 2000 to 2005. & 
        \topicscore[xticklabels={,,}]{0.90}{0.30}{0.10} \\
        \midrule

        $(E_{\text{UT}})$ The largest ancestry groups in the state are: 26.0\% English, 11.9\% German, [...]. In comparison to all the U.S.\ states and territories, Utah, with a population of just over three million, is the 13th-largest by area, the 30th-most populous, and the 11th-least densely populated. St.\ George was the fastest-growing metropolitan area in the United States from 2000 to 2005. & \topicscoretotal{2.70}{1.20}{0.10}  \\
        \bottomrule
    \end{tabularx}
    } 
    \caption{\small Illustrative topic scores for sentences $t_{\text{UT}}^{1}$,
    $t_{\text{UT}}^{2}$, and $t_{\text{UT}}^{3}$; and topic profile for the example snippet $E_{\text{UT}} =
    \{t_{\text{UT}}^{1}, t_{\text{UT}}^{2}, t_{\text{UT}}^{3}\}$.}
    \vspace{-6mm}
    \label{tab:topic_distribution}
\end{table}

\looseness-1 Here, $\alpha$ is a tunable parameter between $0$ and $\infty$,
which determines how aggressively bound widths are narrowed as the slider moves
farther away from the neutral position. When $\alpha {=} 0$, all slider
positions use the same width as the neutral position, i.e., $(\upperbound_j {-}
\lowerbound_j)$. As $\alpha$ increases, the width shrinks more rapidly as the
slider moves away from the neutral position, resulting in increasingly tighter
bounds for extreme values of $\gamma_j$. Regardless, \system automatically
``snaps'' the slider to a feasible position in case the user's adjustment
leaves it at an infeasible state. Note that the bounds represented by different
slider positions can be overlapping (Fig.~\ref{fig:slider-example}).

\smallskip Overall, the slider interface offers: (1)~\emph{simplicity} by
reducing the number of controls, consistent with HCI findings that simpler
control spaces are cognitively less demanding~\cite{DBLP:conf/nime/HuntWP02};
(2)~\emph{semantic clarity} through an intuitive scale; (3)~\emph{feasibility
preservation} by automatically snapping user adjustments to a feasible state;
and (4)~\emph{reversibility} by allowing users to revert back to a previous
intent. We evaluate the effectiveness of the slider interface in
\S\ref{sec:userstudy}.


\begin{figure}[t]
\centering
\resizebox{0.9\columnwidth}{!}{
\begin{tikzpicture}[scale=1, font=\Large]

    \definecolor{slidercolor}{RGB}{70,130,180}
    \definecolor{veryred}{RGB}{180,30,30}         
    \definecolor{lightred}{RGB}{220,100,100}      
    \definecolor{neutralcolor}{RGB}{128,128,128}  
    \definecolor{lightgreen}{RGB}{100,180,100}    
    \definecolor{verygreen}{RGB}{30,130,30}       

    \draw[line width=3pt, slidercolor] (0,0) -- (10,0);

    \draw[line width=2.5pt] (0,-0.3) -- (0,0.3);
    \node[above=0pt, font=\Large] at (0,0.3) {$\gamma=-100$};
    \node[below=10pt, align=center, font=\Large] at (0,0) {Very little};
    \draw[gray!20, line width=0.5cm] (-0.9, -1.2) -- (0.9, -1.2);
    \draw[veryred, line width=0.5cm] (-0.9, -1.2) -- (-0.666, -1.2);
    \node[below=8pt, font=\Large] at (0, -1.2) {$\langle 0.00, 1.09\rangle$};

    \draw[line width=2.5pt] (2.5,-0.3) -- (2.5,0.3);
    \node[above=0pt, font=\Large] at (2.5,0.3) {$\gamma=-50$};
    \node[below=10pt, align=center, font=\Large] at (2.5,0) {Less};
    \draw[gray!20, line width=0.5cm] (1.6, -1.2) -- (3.4, -1.2);
    \draw[lightred, line width=0.5cm] (1.898, -1.2) -- (2.662, -1.2);
    \node[below=8pt, font=\Large] at (2.5, -1.2) {$\langle 0.75, 1.89\rangle$};

    \draw[line width=2.5pt] (5,-0.3) -- (5,0.3);
    \node[above=0pt, font=\Large] at (5,0.3) {$\gamma=0$};
    \node[below=10pt, align=center, font=\Large] at (5,0) {Neutral};
    \draw[gray!20, line width=0.5cm] (4.1, -1.2) -- (5.9, -1.2);
    \draw[neutralcolor, line width=0.5cm] (5-0.9+0.5*1.8, -1.2) -- (5-0.9+0.8*1.8, -1.2);
    \node[below=8pt, font=\Large] at (5, -1.2) {$\langle 1.50, 2.70\rangle$};

    \draw[line width=2.5pt] (7.5,-0.3) -- (7.5,0.3);
    \node[above=0pt, font=\Large] at (7.5,0.3) {$\gamma=+50$};
    \node[below=10pt, align=center, font=\Large] at (7.5,0) {More};
    \draw[gray!20, line width=0.5cm] (6.6, -1.2) -- (8.4, -1.2);
    \draw[lightgreen, line width=0.5cm] (7.518, -1.2) -- (8.238, -1.2);
    \node[below=8pt, font=\Large] at (7.5, -1.2) {$\langle 3.56, 4.70\rangle$};

    \draw[line width=2.5pt] (10,-0.3) -- (10,0.3);
    \node[above=0pt, font=\Large] at (10,0.3) {$\gamma=+100$};
    \node[below=10pt, align=center, font=\Large] at (10,0) {A lot};
    \draw[gray!20, line width=0.5cm] (9.1, -1.2) -- (10.9, -1.2);
    \draw[verygreen, line width=0.5cm] (10.054, -1.2) -- (10.9, -1.2);
    \node[below=8pt, font=\Large] at (10, -1.2) {$\langle 5.61, 6.70\rangle$};
\end{tikzpicture}
}
\vspace{-4mm}
\caption{Slider for \topic{Demographics}. Users can adjust topic emphasis from $-100$ (very little) to $+100$ (a
lot). Each slider position maps to specific constraint bounds; the neutral position ($0$) corresponds to the feasible bounds synthesized (and relaxed) from user examples.}
\vspace{-2mm}
\label{fig:slider-example}
\end{figure}

\subsection{Encoding intent into feasible constraints}
\label{sec:intent}
The bound synthesizer (i)~synthesizes initial, potentially infeasible, bounds
from user examples to form constraints, (ii)~relaxes the initial bounds to
ensure feasibility, (iii)~exposes the feasible bounds via the slider interface
for iterative refinement, and (iv)~translates the adjusted sliders back to
feasible bounds after refinement.

\subsubsection{Synthesizing the initial bounds.}\label{synthinitbound} We extend
\squid~\cite{squid}, which learns user intent from example tuples, to a
bundle-level granularity. Given example bundles $\mathbf{E} = \{E_i\}_{i=1}^N$,
we use \texttt{\small SUM} as the aggregate to compute their feature profiles
$\{\mathbf{\featureBundle}(E_i)\}_{i=1}^N$. The choice of aggregate is
application-specific and our framework can support any linear aggregate such as
\texttt{COUNT}, \texttt{MIN}, \texttt{MAX}, etc. However, \texttt{\small SUM} is
often preferred since \texttt{\small AVG} cannot distinguish single- from
multi-tuple bundles, and \texttt{\small MAX/MIN} are outlier-sensitive. We
derive the initial constraint bounds $\Theta_{\text{init}} =
\{\langle\lowerbound_j, \upperbound_j\rangle\}_{j=1}^K$ by taking the topic-wise
minimum and maximum of the feature profiles of bundles in $\mathbf{E}$:
{\begin{center} \resizebox{0.8\columnwidth}{!}{
\begin{tabular}{c@{     }c}
$\lowerbound_j = \min_{1 \le i \le N} F_j(E_{i}),$ &
$\quad \upperbound_j = \max_{1 \le i \le N} F_j(E_{i})$
\end{tabular}}
\end{center}}

\begin{example}
\label{ex:intent}
Consider 3 example snippets $E_{\text{\text{UT}}}\,$, $E_{\text{AZ}}\,$, and
$E_{\text{WA}}$ over the source documents $T_{\text{\text{UT}}}^s\,$,
$T_{\text{AZ}}^s\,$, and $T_{\text{WA}}^s$ that represent Wikipedia pages for
Utah, Arizona, and Washington, respectively. The topic profile of
$E_{\text{\text{UT}}}$, {$\mathbf{\featureBundle}(E_{\text{UT}}) =
\mathbf{f}(t_{\text{UT}}^{1}) + \mathbf{f}(t_{\text{UT}}^{2}) +
\mathbf{f}(t_{\text{UT}}^{3}) = [2.70,1.20,0.10]$}, as shown in the last row of
Table~\ref{tab:topic_distribution}. The topic profiles for the example snippets,
along with the initial constraint bounds  are given below, which results in
$\Theta_{\text{init}} = \{\langle1.50, 2.70\rangle, \langle0.90, 1.20\rangle,
\langle0.10, 0.40\rangle\}$.
{\begin{center}
\footnotesize
\resizebox{0.85\columnwidth}{!}{
\begin{tabular}{@{}lccc@{}}
\toprule
& \textbf{$\featureBundle_{1}$ (\topic{Demographics})} & \textbf{$\featureBundle_{2}$ (\topic{Geography})} & \textbf{$\featureBundle_{3}$ (\topic{Economy})} \\

\midrule
$\mathbf{\featureBundle}(E_{\text{\text{UT}}})$ & $2.70$ $\uparrow$             & $1.20$ $\uparrow$             & $0.10$ $\downarrow$  \\
$\mathbf{\featureBundle}(E_{\text{AZ}})$ & $1.80$ \phantom{X}            & $0.90$ $\downarrow$            & $0.40$ $\uparrow$ \\
$\mathbf{\featureBundle}(E_{\text{WA}})$ & $1.50$ $\downarrow$           & $1.00$ \phantom{$\uparrow$}  & $0.30$ \phantom{$\downarrow$} \\
\midrule
$\lowerbound$ (min)         & $1.50$                        & $0.90$                        & $0.10$ \\
$\upperbound$ (max)         & $2.70$                        & $1.20$                        & $0.40$ \\
\bottomrule
\end{tabular}}
\end{center}}

   

\end{example}

\looseness-1 \noindent\underline{\emph{Remark on constraint expressivity.}}
\label{constraint-expressivity} \system supports any tuple-level constraint that
standard SQL supports (Appendix~\ref{appx:constraint_representation}), such as equality predicates involving categorical
attributes, logical conditions, and constraints over derived (potentially
non-linear) attributes. Techniques for deriving such tuple-level constraints are
addressed in prior work~\cite{squid}. \system also supports more complex
bundle-level constraints, such as interactions among aggregated attributes, and
special cases of bounded constraints, such as equalities (by setting
$\lowerbound = \upperbound$) and one-sided inequalities (by setting
$\lowerbound=-\infty$ or $\upperbound=\infty$).
\system supports any linear aggregate within the bundle-level constraints. The
choice of the aggregate is application-specific and should be specified by a
domain expert. In the absence of domain expertise, an LLM can suggest a suitable
aggregate based on the task context.
We note that practical applications may involve non-linear constraints, such as
Max-Min diversity~\cite{maxmindiv}. However, this is a limitation of PaQL, which
\system relies on, not of \system itself. In fact, the underlying ILP solvers
such as IBM CPLEX~\cite{cplex} support certain classes of quadratic constraints
with specific properties (e.g., convexity and semi-definiteness) and \system can
potentially support such constraints once PaQL provides support for them.

\subsubsection{Bound relaxation to ensure feasibility}
\label{sec:bound_relaxation} 
\looseness-1 Overly tight bounds can yield infeasible queries that return no
valid bundle (Example~\ref{ex:three}), particularly when bounds are misaligned
with the target data's feature distribution under distributional shift
(Example~\ref{ex:three}). Infeasibility arises from two sources. The first is
\emph{insufficient feature coverage}: the target data cannot satisfy a
feature's lower bound even when all tuples are selected. For example, if the
source data (e.g., Utah, Arizona, \& Washington) and example snippets emphasize
\texttt{national parks} (e.g., Bryce, Grand Canyon, Olympic), but the target
document is Kansas, which contains no national parks, its maximum contribution
to that topic is $0$, making any positive lower bound (e.g., $0.10$)
infeasible. The second is \emph{atomicity}: tuples are indivisible, so
selecting any single tuple may violate an upper bound. For instance, if every
target sentence has a topic score of at least $0.15$, an upper bound of $0.10$
is infeasible because no tuple can be partially selected.

\smallskip\noindent\textbf{Bound relaxation algorithm.} \looseness-1
A standard way to resolve infeasibility is
the solver-native FeasOpt~\cite{feasopt}, which minimizes the total deviation
from the numerical bounds, termed ``slack'', added across all constraint
bounds to reach feasibility. However, minimizing total slack tends to spread
small relaxations across many constraints, which hurts usability: each
relaxation shifts a slider, and changes in many sliders are harder for users to
interpret and refine.

We therefore propose \textit{symmetric bound relaxation} (SYM), which
iteratively widens only the violated constraint bounds until the resulting
query becomes feasible (Algorithm~\ref{alg:main},
lines~\ref{identification}--\ref{line14}). If the initial constraint bounds
$\Theta_{\text{init}}$ are infeasible (line~\ref{feasible}), we first identify
the violated constraints $\Theta_{\text{violated}}$ (line~\ref{identification})
using ILP solver signals (e.g., IBM CPLEX~\cite{cplex}). We then decrease
$\lowerbound_j$ toward $0$ (line~\ref{lower}) and increase $\upperbound_j$
toward $F_{j}(\targetdata)$ (line~\ref{upper}) by the same amount $\rho {\cdot}
\epsilon_j$, where the step multiplier $\rho$ (line~\ref{multiplier}) controls
the relaxation rate based on the number of prior attempts, and the step size
$\epsilon_j$ (line~\ref{stepsize}) is a target-data-specific parameter.

We also considered directional relaxation
(DIR), which uses the ILP solver's information about whether the lower or upper
bound causes a violation. Like SYM, DIR limits \#relaxed
constraints, but it introduces a different usability issue: by widening only
one bound, it skews the user intent and shifts the slider's center toward that
side, leaving less room for tuning in that direction. We empirically found
FeasOpt, DIR, and SYM to yield comparable bundle quality, intent alignment, and
runtime (\S\ref{sec:relaxation_mechanism}) when the degree of infeasibility is
low (the most common case). We adopt SYM for its two \emph{usability benefits}:
(1)~it relaxes fewer constraints than FeasOpt, affecting fewer intent sliders; and
(2)~it widens both bounds equally, keeping the relaxed range centered on what
the user-provided examples expressed and preserving equal tuning headroom in
both directions.

\DontPrintSemicolon
\SetKwInOut{Input}{Input}
\SetKwInOut{Output}{Output}
\let\oldWhile\While
\renewcommand{\While}[2]{%
  \oldWhile{#1}{#2}%
  \vspace{-\baselineskip}%
}

\SetCommentSty{mycommfont}
\newcommand\mycommfont[1]{\scriptsize\ttfamily\textcolor{darkgray}{\textbf{#1}}}
\SetKwComment{tcc}{\textcolor{darkgray}{\texttt{// }}}{\textcolor{darkgray}{}}

\DontPrintSemicolon

\setlength{\textfloatsep}{0pt}

\begin{algorithm}[t]
\LinesNumbered
\small{

\Input{
    Initial bounds $\Theta_{\text{init}}$, schema $\schema$, target data $\targetdata$\\
    Relaxation parameters $\mu$ and $\tau$
}
\Output{
    Feasible bounds $\Theta$ to construct the \paql constraints 
}

$\Theta \gets \Theta_{\text{init}}$\\

$\text{\#attempts} \gets 0$ \tcc*{Initialize \#attempts}
\MyWhile{$PQ(\Theta, \targetdata) = \emptyset$\label{feasible}\tcc*{While query is infeasible}}{  
    $\Theta_{\text{violated}} = \text{IdentifyViolations}(\Theta, \targetdata)$\label{identification} \tcc*{Violated constraints}
    $\Theta_{\text{relaxed}} = \emptyset$\label{initConst} \tcc*{Relaxed  constraints}
    $\rho \gets \exp\!\left(\mu {\cdot} \max\left(1, \left\lfloor \frac{\text{\#attempts++}}{\tau} \right\rfloor\right)\right)\!$\label{multiplier} \tcc*{Relaxation step multiplier} 
    \ForEach{$\langle\lowerbound_j, \upperbound_j\rangle \in \Theta_{\text{violated}}$\label{repeatBound}}{        
        $\epsilon_j \gets \frac{\featureBundle_j(\targetdata)}{|\targetdata|}$\label{stepsize} \tcc*{Relaxation step size}
        $\lowerbound_j \gets \max(\lowerbound_j - \rho {\cdot} \epsilon_j,\;\; 0)$ \label{lower}\tcc*{Decrease lower bound}
        $\upperbound_j \gets \min(\upperbound_j + \rho {\cdot} \epsilon_j,\;\; \featureBundle_j(\targetdata))$\label{upper} \tcc*{Increase upper bound}
        $\Theta_{\text{relaxed}} = \Theta_{\text{relaxed}} \cup \{\langle\lowerbound_j, \upperbound_j\rangle\}$\label{line13}
    }
    $\Theta = (\Theta \setminus \Theta_{\text{violated}}) \cup \Theta_{\text{relaxed}}$ \label{line14}\tcc*{Update constraint bounds}
}

\Return $\Theta$
}

\caption{\system bound relaxation algorithm} 
\label{alg:main}

\end{algorithm}

\smallskip\noindent\textbf{Relaxation step multiplier, $\rho$.} It controls the
relaxation rate and is defined as $e^{\mu \cdot \max\left(1, \left\lfloor
\#\text{attempts}/{\tau} \right\rfloor\right)}$. Here, $\mu$ (default $0.5$)
sets the base relaxation rate, while $\tau$ (default $10$) controls how
frequently $\rho$ is boosted. Consequently, $\rho$ increases exponentially
every $\tau$ attempts, enabling faster escape from infeasible regions.

\smallskip\noindent\textbf{Relaxation step size, $\epsilon_j$.}
\label{relaxation-step} While $\rho$ determines the rate of relaxation,
different features exhibit varying score distributions within $\targetdata$. To
account for this, we derive a feature-specific step size from the target data
distribution and set $\epsilon_j = \frac{F_{j}(\targetdata)}{|\targetdata|}$,
which corresponds to the average contribution of feature $f_j$ per tuple in
$\targetdata$.
Since the output bundle is a subset of the target data $\targetdata$, the
appropriate relaxation magnitude depends on the empirical scale of each feature
in $\targetdata$. We therefore derive $\epsilon_j$ from $\targetdata$: a fixed
step can be too small or too large for features with different value ranges.
Computing $\epsilon_j$ per feature automatically adapts the step size to the
scale of that feature, even when feature values are not normalized.

\begin{example} Consider $\targetdata_{\text{CA}}$ over
$5$ sentences ($|\targetdata_{\text{CA}}| = 5$), where
$\featureBundle_2(\targetdata_{\text{CA}}) = 0.60$. The initially synthesized
constraint $0.90 \le \featureBundle_{2}(B) \le 1.20$
(Example~\ref{ex:intent}) is infeasible since the maximum achievable
score over $\targetdata_{\text{CA}}$ is $0.60$, requiring relaxation. For the
first $10$ iterations, $\rho {=} e^{0.5 {\times} 1} {\approx} 1.65$ and
$\epsilon {=} \frac{0.60}{5} {=} 0.12$. Thus, the first iteration relaxes the
bounds to $\max(0.90 {-} 1.65{\times}0.12, 0){=}0.70 \le \featureBundle_{2}(B)
\le$ $min(1.20 {+} \linebreak  1.65{\times}0.12, 0.60){=}0.60$.
This is still infeasible, so the second iteration yields $\max(0.70 {-}
1.65{\times}0.12,0){=}0.50 \le \featureBundle_{2}(B) \le \min(0.60 {+}
1.65{\times}0.12, \linebreak 0.60){=}0.60$, at which point the constraint becomes feasible. The upper bound is not relaxed due to it already being at
max. \label{ex:relaxed_constraints} \end{example}


\subsubsection{Bound relaxation after slider interaction.} Since slider
adjustments may re-introduce infeasibility, we re-relax the bounds using
Algorithm~\ref{alg:main}. Once feasible bounds are obtained, we snap the slider
to the position whose bounds are closest to the feasible bounds and subsume
them. Unlike the initial relaxation, we have an advantage here since we only
need to relax bounds for a single constraint, with all other constraint bounds
fixed. Therefore, if previously feasible bounds $\langle \lowerbound_j^*,
\upperbound_j^* \rangle$ are known, we use them as an additional reference
during relaxation. Specifically, we reduce $\lowerbound_j$ until $\max(0,
\lowerbound_j^*)$ (line~\ref{lower}) and increase $\upperbound_j$ until
$\min(\upperbound_j^*, F_{j}(\targetdata))$ (line~\ref{upper}). If the user
requests termination of relaxation before the process finishes, we move the
slider to the closest feasible bounds.

\subsection{Devising a quality function for optimality}
\label{sec:quality}
The quality function (\S\ref{sec:define-bundle-optimality})
defines the notion of optimality used to break ties among multiple valid
bundles, particularly important when relaxation  yields
multiple valid bundles. The quality function is application-specific and is
typically specified by a domain expert prior to user interaction with an
instance of \system, as shown in step (0) in Fig.~\ref{fig:system-framework}.
Given (i)~a tuple-level scoring function $\scorefunction: \tuple \mapsto
\mathbb{R}$, which assigns a numeric score to a tuple $t$, and (ii)~an
application-specific aggregate function $\aggregate$, the quality of a
candidate bundle $B$, $quality(B) = \aggregate_{\tuple \in B}
\scorefunction(t)$. Below, we present example quality functions for the three
applications discussed in \S\ref{sec:intro}; Appendix~\ref{appx:system_variants_analysis}
compares quality-function variants for FTSE.

\smallskip\noindent\emph{Supplier selection for business.} \emph{Cost} and
\emph{reliability} are common quality functions when selecting a bundle of
suppliers for business expansion into a new country. For cost, a natural
aggregate is \texttt{SUM}, with the objective of minimizing total cost, i.e.,
$\displaystyle \Sigma_{\tuple \in B} \mathtt{cost}(\tuple)$. For reliability,
suitable aggregates include \texttt{AVG} or \texttt{MIN}, depending on the
preference: maximizing average reliability, or maximizing the minimum
reliability among the selected suppliers.

\smallskip\noindent\emph{Focused snippet extraction.} \looseness-1
\emph{Informativeness} and \emph{conciseness} are common quality functions for
text snippet extraction. Sentence-level informativeness can be modeled using
term frequency within a sentence $\tuple$, weighted by inverse document
frequency in the corpus (TF-IDF), which assigns higher scores to sentences
containing rarer words. Informativeness can also be approximated using
structural signals such as the presence of numbers or citations, which often
indicate factual or reference-rich content. Conciseness can be modeled using
the word or character count of a sentence, with the objective of minimizing the
total count in a snippet. In both cases, the most appropriate aggregate is
\texttt{SUM}. Alternatively, conciseness can be modeled by minimizing the
\texttt{COUNT} of sentences in a snippet.

\smallskip\noindent\emph{Playlist recommendation.} \looseness-1
\emph{Popularity} and \emph{diversity} are common quality functions in playlist
recommendation. Popularity can be modeled using the number of weekly plays or
chart rankings of a song, with the objective of maximizing overall popularity
using aggregate \texttt{AVG}, i.e., $\displaystyle \mathtt{AVG}_{\tuple \in
B}\; \mathtt{popularity}(\tuple)$.
Diversity can be modeled using pairwise dissimilarity between songs based on
features such as genre, mood, or artist. This can be captured via a distance
function over song features, with the objective of maximizing diversity within
the playlist. A suitable aggregate is \texttt{MIN} over all pairwise distances,
i.e., $\displaystyle \texttt{MIN}_{\tuple_i, \tuple_j \in B, i \neq j}
\mathtt{distance}(\tuple_i, \tuple_j)$, and maximizing this ensures that even
the most similar pair of songs in the playlist remains sufficiently dissimilar.
While we assume a linear, tuple-wise scoring function, \system can support
quadratic objectives via CPLEX's quadratic programming or MILP linearization.

\subsection{Package query execution}
\label{sec:execution}
For the package query execution engine, we use \paql~\cite{BrucatoBAM16}, a
query engine for declarative package queries. \paql translates queries into
Integer Linear Programs (ILPs), which are then solved using off-the-shelf
solvers such as IBM CPLEX~\cite{cplex}. For large target data with many tuples,
solving the ILP becomes computationally expensive due to the combinatorial
nature of the search space. To address this, \system first \textsc{PreFilters}
$\targetdata$ to at most $10x$ tuples most similar to the examples, where $x$
is the average example size; the ILP then enforces the constraints and selects
the optimal bundle from that pool. As $x$ grows, \textsc{PreFilter} retains
more tuples, making direct ILP slow. To remedy this, \system also applies
\paql's \sketchref~\cite{BrucatoBAM16}, which solves the ILP over cluster
representatives with a guaranteed $(1+\epsilon)^{6}$ approximation of the
optimal bundle. Empirical results are in \S\ref{sec:scalability_analysis}.
%
%

\system formulates a \paql query that encodes the constraint bounds $\Theta$
(\S\ref{sec:intent}) as constraints over packages (analogous to snippets in
FTSE or bundles in general) and defines the optimization objective using the
scoring function $\scorefunction$ together with the aggregate $\aggregate$
(\S\ref{sec:quality}), as shown in \S\ref{sec:paqlsynthesis}. It then executes
the package query, and, if feasible, returns the optimal bundle.
The \paql engine also informs the bound synthesizer (\S\ref{sec:intent}) about any violated constraints when the query is infeasible in line~\ref{identification} of Algorithm~\ref{alg:main}.

\begin{example}\label{ex:seven}
For the constraint bounds $\Theta{=}\{\langle1.50, 2.70\rangle,\langle0.50,
0.60\rangle, \allowbreak \langle0.10, 0.40\rangle\}$ 
(derived in Example~\ref{ex:intent} and relaxed in Example~\ref{ex:relaxed_constraints}) and the target
document $\targetdata_{\text{CA}}$, we want the snippet with the minimum
total word count. The function $\texttt{word\_count}: \tuple \mapsto \mathbb{N}^+$
returns the number of words in a sentence $t$. The synthesized \paql query is as
follows:

\begin{paqllisting}{$\mathrm{PQ}(\targetdata_{\text{CA}}, \Theta)$:}
SELECT PACKAGE(*) AS B FROM |$\targetdata_{\text{CA}}$|
SUCH THAT |$\featureBundle_{\text{\topic{1}}}(B)$| BETWEEN 1.50 AND 2.70
  AND |$\featureBundle_{\text{\topic{2}}}(B)$| BETWEEN 0.50 AND 0.60
  AND |$\featureBundle_{\text{\topic{3}}}(B)$| BETWEEN 0.10 AND 0.40
MINIMIZE SUM|$_{t \in B}$| word_count(|$t$|)
\end{paqllisting}
\end{example}

\noindent\underline{\emph{Remark:}} \looseness-1 Example-based inference
assumes that the provided examples are correct, yet in practice they may be
insufficiently representative or noisy,
causing bounds synthesized from attribute-wise minimum and maximum values to be
overly restrictive or permissive. \system mitigates this by letting users
adjust the intent sliders directly. Alternatively, examples that differ
substantially from the others could be flagged for user verification, as in
prior work on disambiguating user intent in programming by
example~\cite{uistpaper}.

\newcommand{\keytakeaway}[1]{
    \begin{tcolorbox}[
        colback=blue!5!white, colframe=blue!20!gray, boxrule=0.4pt,
        sharp corners, boxsep=1pt, left=1pt, right=1pt,
        top=1pt, bottom=1pt, before skip=2pt, after skip=1pt,
    ]
    #1
    \end{tcolorbox}
}

\section{Experimental Results}\label{sec:experiments} We now present
experimental results addressing two sets of research questions. RQ1--RQ3
evaluate to what extent \system's \paql synthesis algorithm meets the three
\emph{correctness criteria} for \bqbe: constraint satisfaction, alignment with
user intent, and consistency between constraints and bundles (detailed in
\S\ref{correctness-criteria}). RQ4--RQ6 focus on scalability, practical
efficiency, and robustness aspects.

\vspace{1mm}
\noindent
\resizebox{\linewidth}{!}{
\begin{tabular}{|@{ }l@{ }p{9cm}@{ }|}
\hline
\textbf{RQ1} & To what extent do the bundles returned by \system satisfy the
    synthesized constraints? How does it compare to alternatives?
    (\S\ref{sec:cs_performance}).\\
\textbf{RQ2} & To what extent do the bundles returned by \system align with user
    intent? How does it compare to LLM-based baselines?
    (\S\ref{sec:retrieval_summarization_performance}).\\
\textbf{RQ3} & Does moving farther from the synthesized constraints produce 
	bundles that are increasingly less aligned with the intended bundle?
	(\S\ref{sec:consistency_analysis}).\\
\textbf{RQ4} &  How does \system's runtime scale with the size of example bundles and the
    size and dimensionality of the documents?
    (\S\ref{sec:scalability_analysis}).\\

\textbf{RQ5} & How often is constraint relaxation triggered as example size increases,
    and how effective is it for \bqbe? (\S\ref{sec:relaxation_analysis}).\\

\textbf{RQ6} & How robust is \system to varying degrees of skewness in the target
	document's topic distribution? (\S\ref{sec:robustness}).\\ \hline
	\end{tabular}}



\smallskip\noindent\textbf{Implementation.} \system uses a Python backend---with
Gensim for frequency-based topic modeling, SBERT~\cite{reimers-2019-sentence-bert} for
semantics-aware topic modeling, IBM CPLEX 20.1 via docplex for ILP solving, and
a JavaScript/Bootstrap frontend for the slider interface. We ran experiments on
a shared compute cluster, utilizing 4 Intel Emerald Rapids CPU cores, 64 GB RAM, 
and an NVIDIA H200 GPU.

\smallskip\noindent\textbf{Datasets.} We evaluate \system in two applications:
(1)~supplier selection over TPC-H~\cite{tpc-h}, using a synthesized view of
\code{Supplier} (100 rows), \code{Partsupp} (8{,}000 rows), and \code{Nation}
(25 rows) at scale factor 0.01, with five attributes: \code{price},
\code{availability}, \code{balance}, \code{region\_europe}, and
\code{region\_america}; and (2)~focused text snippet extraction (FTSE) over two
datasets: 
(a)~\dataset~\cite{subsume}, containing 275 (user, intent) pairs, each with $8$
manually curated summaries; we use $5$ summaries as examples and the remaining
$3$ for test. We refer to each (user, intent, state) combination as an
instance; and (b)~\cnndm~\cite{hermann2015teaching}, containing human-written highlights
spanning four news categories. Since these highlights are abstractive, we use
ChatGPT-4o to align each with source-article sentences, capping the snippet at
10 sentences or 30\% of the article's sentences (details in
Appendix~\ref{appx:dataset_details}).

\smallskip\noindent\textbf{Baselines.} \looseness-1 We consider the following
baselines, where $k$ is the average example size (details are in
Appendix~\ref{appx:baseline_details}):

\begin{itemize}[nosep, leftmargin=*]

\item \emph{Greedy} greedily selects $k$ tuples by decreasing quality score.

\item \emph{Cluster} uses K-means clustering over the attribute space, and
selects the tuple with highest quality score from each cluster.

\item \emph{Random} selects $k$ tuples uniformly at random.

\item \emph{Top-k} selects the $k$ most similar sentences to the examples by
\SBERT~\cite{reimers-2019-sentence-bert} cosine similarity.

\item \emph{\sudocu}~\cite{sudocu} is our prior FTSE system, using Latent
Dirichlet Allocation~\cite{BleiNJ03} topic distributions.

\item \emph{\presumm}~\cite{liu2019text} is an extractive summarizer adapted
for \bqbe by pre-filtering sentences based on similarity to the
examples.

\item \emph{\memsumm}~\cite{gu2021memsum} is a reinforcement-learning
extractive summarizer for long documents, adapted via the same pre-filtering.

\item \emph{ChatGPT-4o}~\cite{openai2023gpt4} is prompted for FTSE.

\item \emph{LLM-agent} uses Llama-3.3-70B-Instruct-FP8~\cite{llama33} for
agent-based \paql synthesis (including bounds), relaxation, and execution.

\item \emph{LLM-\system} uses the same LLM for \paql synthesis, but with
\system's constraint relaxation and execution.
\end{itemize}


\subsection{Correctness Criteria \& Metrics}\label{correctness-criteria} We
define the correctness criteria for \bqbe along three axes: 

\vspace{0.5mm}\noindent\textbf{Constraint satisfaction} provides a formal
correctness guarantee for feasible queries by ensuring that the returned bundle
satisfies the synthesized constraints. To measure the degree of constraint
satisfaction, we use \emph{Constraint Satisfaction Rate (CSR)}, the fraction of
synthesized constraints satisfied by the retrieved bundle.

\vspace{0.5mm}\noindent\textbf{Intent alignment} is a key correctness criterion
that ensures the result bundle produced by a \bqbe system closely matches the
ground truth in terms of user intent. When
the ground-truth is available, such as for the \dataset dataset, we use
(1)~\emph{ROUGE-1/2/L}~\cite{lin2004rouge}: F$_1$ scores measuring unigram,
bigram, and longest-common-sub\-sequence overlap between result and ground
truth; and (2)~\emph{Semantic Similarity (SS)}: cosine similarity between
average \SBERT embeddings of result and ground truth.
In absence of ground truth, the intent is
fuzzy, which requires use of application-specific metrics as proxies to measure
intent alignment. For the supplier selection problem over TPC-H, we use~the
\emph{quality score} (the optimization objective), defined as the sum of
\code{price}, \code{availability}, and \code{balance}.

\vspace{0.5mm}\noindent\textbf{Constraint-bundle alignment consistency}
evaluates whether deviations from the synthesized \paql $Q$ produce
corresponding deviations in the resulting bundle. Specifically, as a query $Q'$
moves farther from $Q$, its resulting bundle should become increasingly less
aligned with the intended bundle. We measure this using Pearson's correlation
coefficient (PCC) between the Hausdorff distance~\cite{hausdorff} between the
feasible regions induced by two queries and the cosine distance between the
\SBERT embeddings of the resulting bundles.

\subsection{Constraint satisfaction effectiveness} \label{sec:cs_performance}
\looseness-1 We evaluate RQ1 (constraint satisfaction) for the task of supplier
selection on TPC-H. We construct $3$ example bundles representing diverse
procurement strategies (conservative, price-focused, and balanced) and \system
infers constraints from these examples. We compare \system against 3 baselines:
constraint-agnostic \emph{Greedy} (top-$k$ highest-scoring tuples),
\emph{Cluster} (best tuples from each of the $k$ clusters), and \emph{Random}
($k$ tuples, averaged over 5 runs).

\begin{sloppypar}
	Table~\ref{tab:constraint_satisfaction_results} shows that \system satisfies
all constraints (CSR~=~100\%) with a competitive average quality score (11.72).
Greedy achieves the highest quality score (15.61) but with an average CSR of
33.3\%, indicating that objective-only optimization fails to enforce
constraints. Cluster outperforms greedy (CSR = 66.7\%), but fails to satisfy
many constraints. Random has a higher CSR (70.0\%) than Greedy and Cluster, but
with a low quality score (10.48).
\end{sloppypar}
	

\begin{table}[t]
\renewcommand{\arraystretch}{1.1}
\centering
\small
\resizebox{1\columnwidth}{!}{
\begin{tabular}{@{} l S[table-format=3.1] @{} S[table-format=2.2] S[table-format=1.4] @{}}
\toprule
\textbf{System} & \multicolumn{1}{c}{\textbf{CSR (\%)} $\uparrow$} & \textbf{Average Quality
Score $\uparrow$} & \textbf{Runtime (s) $\downarrow$} \\ 
\midrule
Random              & \phantom{0}\underbar{70.0}& 10.48            & \textbf{0.0005} \\
Greedy              & \phantom{0}33.3           & \textbf{15.61}   & \underbar{0.0024} \\
Cluster             & \phantom{0}66.7           & \underbar{14.04} & 0.4712 \\
\textbf{\system}    & \textbf{100.0}            & 11.72            & 0.0207 \\
\bottomrule
\end{tabular}}
\caption{{\small \system achieves full constraint satisfaction with a competitive quality score. \textbf{Bold} = best; \underline{underlined} = second-best.}}
\vspace{-3mm}
\label{tab:constraint_satisfaction_results}
\end{table}

\vspace{1mm}
\noindent\resizebox{\columnwidth}{!}
{
\centering
\keytakeaway{
    \begin{itemize}[nosep, leftmargin=*, label={\footnotesize\textcolor{Bittersweet}{\faLightbulb}}]  

        \item Greedy and Cluster have lower constraint satisfaction rate (CSR)
        while Random has low quality score.
		
        \item \system achieves 100\% CSR with a competitive quality score as it
        accounts for both objective and constraints.
		
    \end{itemize}
}}

\subsection{Alignment with user intent}
\label{sec:user-intent-alignment}
\subsubsection{Comparison with non-LLM baselines}
\label{sec:retrieval_summarization_performance}
For RQ2, we use the \dataset dataset and evaluate \system against Top-$k$, \sudocu,
\presumm, and \memsumm. We report ROUGE-1/2/L F$_1$ scores and SBERT-based
semantic similarity (SS) between retrieved and ground-truth snippets, together
with per-query runtime.

\looseness-1 Fig.~\ref{fig:extractive_performance} shows that \system
consistently outperforms all baselines on ROUGE and SS. \sudocu is a \bqbe
system and outperforms non-\bqbe extractive summarizers \memsumm and \presumm,
but its topic modeling is inferior to \system and Top-$k$, both of which use
semantics-aware topic modeling.

%
\vspace{-1mm}
\subsubsection{Comparison with LLM-based baselines as direct snippet extractor}
\label{sec:llm_summarization_performance}
We compare \system against LLM-based baselines prompted to directly retrieve
snippets from a target document given a few example snippets. We use ChatGPT-4o
as a RAG baseline that performs few-shot learning from the examples. We
evaluate two intents from \dataset---one more specific and one less
specific---to assess how each approach handles varying intent specificity
(Table~\ref{tab:sbert_similarity_chatgpt4o}; Appendix~\ref{appx:llm_baselines_extraction}). \system outperforms ChatGPT-4o on
both intents, but the advantage narrows for the less specific intent. For
generic intents over \cnndm, ChatGPT-4o outperforms \system, which highlights
that generic intent is too broad for example-driven retrieval, making \system
inappropriate. However, the benefit of \system being LLM-free is that its
latency scales with the corpus rather than model size, it applies to private
data where LLM pipelines cannot run, and it incurs no per-query inference cost.


\pgfplotsset{compat=1.18}
\usetikzlibrary{patterns}
\usetikzlibrary{pgfplots.groupplots}

\definecolor{myblue}{RGB}{30, 144, 255}   
\definecolor{mygrey}{RGB}{192, 192, 192}   
\definecolor{myorange}{RGB}{255, 165, 0}   
\definecolor{mygreen}{RGB}{50, 205, 50}    
\definecolor{myred}{RGB}{220, 20, 60}      

\begin{figure}[t]
\centering
\resizebox{1\columnwidth}{!}{%
\begin{tikzpicture}
\begin{groupplot}[
    group style={
        group size=2 by 1,
        horizontal sep=0.2cm, 
    },
    ybar,
    xtick pos=left,
    xticklabel pos=lower,
    xtick style={draw=none},
    enlarge y limits=false,
    enlarge x limits=0.4,
    ymajorgrids=true,
    grid style={dashed, gray!30},
    height=3.5cm,
    legend style={
        at={(0.63,1)}, 
        anchor=south,
        legend columns=5, draw=none, column sep=5pt,
    },
    legend image code/.code={
        \draw[draw=black,fill opacity=1] (0cm,-0.08cm) rectangle (0.3cm,0.1cm);
    },
]
\nextgroupplot[
    ylabel={F$_1$ Score (\%)},    
    ylabel style={yshift=-5pt},
    ymin=0, ymax=105,
    nodes near coords,
    symbolic x coords={ROUGE-1,ROUGE-2,ROUGE-L},
	enlarge x limits=0.25,
    xtick=data,
	width=0.52\textwidth, 
    every node near coord/.append style={
        font=\tiny,                            
        /pgf/number format/fixed,              
        /pgf/number format/precision= 1,
        /pgf/number format/zerofill      
    }
]
\addplot [fill=black, draw=black] coordinates {(ROUGE-1,58.01) (ROUGE-2,44.00) (ROUGE-L,47.01)}; \addlegendentry{\system\phantom{x}}
\addplot [fill=black!50,  draw=black] coordinates {(ROUGE-1,53.80) (ROUGE-2,36.97) (ROUGE-L,41.04)}; \addlegendentry{Top-$k$\phantom{x}}
\addplot [fill=white, draw=black] coordinates {(ROUGE-1,40.01) (ROUGE-2,19.81) (ROUGE-L,25.38)}; \addlegendentry{\sudocu\phantom{x}}
\addplot [fill=white,  postaction={pattern=horizontal lines}, draw=black] coordinates {(ROUGE-1,37.57) (ROUGE-2,15.51) (ROUGE-L,21.14)}; \addlegendentry{\memsumm\phantom{x}}
\addplot [fill=white, postaction={pattern=north west lines},   draw=black] coordinates {(ROUGE-1,22.23) (ROUGE-2,6.58)  (ROUGE-L,14.47)}; \addlegendentry{\presumm\phantom{x}}

\nextgroupplot[
    ymin=0, ymax=105,
    nodes near coords,
    yticklabels=\empty,
    symbolic x coords={Semantic Similarity},
    xtick=data,
    width=0.22\textwidth,
    every node near coord/.append style={
        font=\tiny,                            
        /pgf/number format/fixed,              
        /pgf/number format/precision= 1,
        /pgf/number format/zerofill      
    }
]
\addplot [fill=black, draw=black] coordinates {(Semantic Similarity,86.13)};
\addplot [fill=black!50, draw=black] coordinates {(Semantic Similarity,83.87)};
\addplot [fill=white, draw=black] coordinates {(Semantic Similarity,75.16)};
\addplot [fill=white, postaction={pattern=horizontal lines}, draw=black] coordinates {(Semantic Similarity,72.24)};
\addplot [fill=white, postaction={pattern=north west lines},   draw=black] coordinates {(Semantic Similarity,58.85)};


\end{groupplot}
\end{tikzpicture}
}
\vspace{-4mm}
\caption{\small For focused text snippet extraction on the \dataset dataset, \system outperforms all other baselines across all metrics.}
\vspace{-1mm}
\label{fig:extractive_performance}
\end{figure}

\begin{table}[t]
    \centering
    \small
    \resizebox{1\columnwidth}{!}{
    \begin{tabular}{@{}l l c }
    \toprule
    \multicolumn{1}{c}{\textbf{Intent}} & \multicolumn{1}{c}{\textbf{System}} & \multicolumn{1}{c}{\textbf{\makecell{Semantic Similarity}} $\uparrow$} \\
    \midrule
    \multirow{2}{*}{\parbox{4cm}{Intent 1: What about this state's arts and culture attracts you the most?}} 
     & ChatGPT-4o & 0.64 \\
     & \system & \textbf{0.88} \\
    \midrule
    \multirow{2}{*}{\parbox{4cm}{Intent 2: What are some of the most interesting things about this state?}} 
     & ChatGPT-4o & 0.65 \\
     & \system & \textbf{0.76} \\
    \bottomrule
    \end{tabular}}
    \caption{\system outperforms ChatGPT-4o on both \dataset intents; the advantage narrows on the more general one.}
    \vspace{-1mm}
    \label{tab:sbert_similarity_chatgpt4o}
\end{table}

\vspace{-1mm}
\subsubsection{Comparison with LLM-based baselines as intermediate \paql
generator} \label{llm-paql-comparison}

We additionally compare against two LLM-based
baselines that generate \paql queries as the mechanism for snippet retrieval,
rather than directly retrieving snippets. \emph{LLM-agent} is an agent-based
approach that performs the end-to-end pipeline of \paql construction, bound
relaxation, and query execution, while \emph{LLM-\system} is a hybrid approach
that delegates relaxation and execution to \system after the LLM generates the
\paql query. We evaluate both approaches over three infeasible (user,
intent) pairs on \dataset, each with example bundles from 5 source documents
and 3 target documents, allowing up to three self-critique-based relaxation
steps per target document for LLM-agent. Table~\ref{tab:llm-paql} reports
results averaged over the 9 target documents; the prompts are in
Appendix~\ref{appx:llm_baselines_paql}.

\looseness-1 \system outperforms both baselines across all metrics: (1)~it is
several orders of magnitude faster, (2)~it retrieves more intent-aligned
bundles, (3)~it requires fewer relaxations, and (4)~incurs no token cost.
LLM-\system outperforms LLM-agent in terms of quality metrics, runtime
(32$\times$ faster), and token cost, but requires more relaxations. These
results highlight two advantages of \system: superior initial bound synthesis
and a data-aware relaxation strategy that outperforms LLMs' self-correction
loop. Notably, LLM-\system's advantage over LLM-agent persists even when both
start from LLM-generated bounds, demonstrating the benefit of \system's
data-aware relaxation strategy. In contrast, LLM-generated bounds require more
relaxation rounds, increasing not only runtime but also token cost.

\begin{table}[t]
\centering
\footnotesize \resizebox{\linewidth}{!}{
\begin{tabular}{@{}lrrrrrcr@{}}
\toprule
Method          & R1 & R2 & RL & SS & Time (s) & \#Relaxation & \#Tokens \\
\midrule
LLM-agent        & 0.35 & 0.17 & 0.23 & 0.68 & 923.60  & 2.11 & 104{,}127  \\
LLM-\system     & 0.39 & 0.22 & 0.27 & 0.74 & 29.11   & 2.67 & 6{,}536 \\
\system         & \textbf{0.56} & \textbf{0.43} & \textbf{0.48} & \textbf{0.85} & \textbf{0.84}    & \textbf{1.67} & \textbf{0}         \\
\bottomrule
\end{tabular}}
\caption{\small \system outperforms LLM-agent and LLM-\system.}
\vspace{-2mm}
\label{tab:llm-paql}
\end{table}

\vspace{1mm}
\keytakeaway{
    \begin{itemize}[nosep, leftmargin=*,
        label={\footnotesize\textcolor{Bittersweet}{\faLightbulb}}]

        \item \system achieves the highest intent alignment across traditional
        and LLM-based baselines, especially for focused intents.

        \item LLM-generated \paql queries struggle to produce intent-aligned
        bundles, while LLM-based bound relaxation is less effective than
        \system's data-aware relaxation.
		
    \end{itemize}
}


\subsection{Constraint-bundle alignment consistency}
\label{sec:consistency_analysis} To evaluate RQ3, we test whether larger
deviations from synthesized constraints produce larger deviations in the
resulting bundles. For each query and topic, we slide that topic's bounds by $s$
($\lowerbound'=\lowerbound+s$, $\upperbound'=\upperbound+s$), with
15 log-spaced values of $s/(\upperbound-\lowerbound)$ in $[0.01,20]$. We
retain only feasible, unrelaxed original--shifted pairs, yielding approximately
411 pairs per topic (4{,}115 total); the resulting Hausdorff distances range
from $0.003$ to $12.9$. We compute PCC between the Hausdorff distance of the
induced feasible regions and the cosine distance between the corresponding
bundle embeddings. Across topics, the distances are strongly correlated with an
average PCC of $0.93$ (Appendix~\ref{appx:consistency_analysis}), indicating that larger constraint deviations
consistently produce larger bundle deviations.

\vspace{0.5mm}
\keytakeaway{
    \begin{itemize}[nosep, leftmargin=*, label={\footnotesize\textcolor{Bittersweet}{\faLightbulb}}] 
        
		\item Bounds synthesized by \system consistently align with
		the intended bundles, achieving a PCC of $0.93$.

    \end{itemize}
}


\begin{figure}[t]
\centering
\centering
\resizebox{\linewidth}{!}{
\begin{tikzpicture}
\begin{groupplot}[
    group style={
        group size=3 by 1,
        horizontal sep=0.6cm,
        group name=scalbcdgroup
    },
    every axis plot/.append style={opacity=0.7},
    width=4cm,
    height=3cm,
    grid=both,
    grid style={line width=.1pt, draw=gray!20},
    major grid style={line width=.2pt, draw=gray!40},
    label style={font=\large},
    tick label style={font=\large},
    scaled x ticks=false,
    xticklabel style={
        /pgf/number format/fixed,
        /pgf/number format/1000 sep={,},
    },
	legend style={
	    at={(1.6,1.05)},
	    anchor=south,
	    font=\small,
	    legend columns=3,
	    draw=none,
	    column sep=3pt,
	    /tikz/every odd column/.append style={column sep=0pt},
	    /tikz/every even column/.append style={column sep=8pt},
	},
	legend cell align=left,
]

\nextgroupplot[
    xlabel={Example size},
    xlabel style={yshift=1mm, font=\small},
    ylabel style={font=\small},
    ylabel={Total time (s)},
    xmin=-5, xmax=105,
    xtick={0, 20, 40, 60, 80, 100},
	ymax = 1.8,
	ymin = -0.2,
    ytick={0.0, 0.5, 1.0, 1.5},
    yticklabels={0.0, 0.5, 1.0, 1.5},
    xticklabel style={font=\small},
    yticklabel style={font=\small},
]

\addplot[color=ForestGreen, mark=*, mark size=3pt, line width=1pt, densely dotted] coordinates {
    (5, 0.125) (25, 0.137) (45, 0.175) (60, 0.201) (80, 1.382)
    (100, 1.346)
};
\addlegendentry{\textsc{PreFilter}}

\addplot[color=blue, mark=square*, mark size=2pt, line width=1pt] coordinates {
    (5, 0.370) (25, 0.239) (45, 0.299) (60, 0.303) (80, 0.287)
    (100, 0.380)
};
\addlegendentry{\sketchref}

\addplot[color=orange, mark=triangle*, mark size=3pt, line width=1pt, dashed] coordinates {
    (5, 0.163) (25, 0.199) (45, 0.287) (60, 0.266) (80, 0.234)
    (100, 0.345)
};
\addlegendentry{\system (\textsc{PreFilter} + \sketchref)}

%
%
%

\nextgroupplot[
    xlabel={Target document size (K)},
    xlabel style={yshift=1mm, font=\small},
    xmin=-50, xmax=1100,
    xtick={100,400,700,1000},
    xticklabels={100,400,700,1000},
    xticklabel style={font=\small},
    yticklabel style={font=\small},
]

\addplot[color=ForestGreen, mark=*, mark size=3pt, line width=1pt, densely dotted] coordinates {
    (100, 1.504) (200, 1.511)(400, 1.519)
    (600, 1.536)(800, 1.570) (1000, 1.470) 
};

\addplot[color=orange, mark=triangle*, mark size=3pt, line width=1pt, dashed] coordinates {
    (100, 0.387) (200, 0.421)  (400, 0.442)
    (600, 0.461) (800, 0.474) (1000, 0.390)
};

\addplot[color=blue, mark=square*, mark size=2pt, line width=1pt] coordinates {
    (100, 40.98) (200, 79.09) (400, 160.263)
    (600, 238.91) (800, 320.77) (1000, 422.647) 
};

\nextgroupplot[
    xlabel={\#Topics},
    xlabel style={yshift=1mm, font=\small},
    xmode=log,
    log basis x=2,
	ymax = 18,
	ymin = -2,
    ytick={0, 5, 10, 15},
    xmin=0, xmax=2056,
    xtick={4,16,64,256,1024},
    xticklabel style={font=\small},
    yticklabel style={font=\small},
]

\addplot[color=ForestGreen, mark=*, mark size=3pt, line width=1pt, densely dotted] coordinates {
    (2, 0.20) (4, 0.28) (8, 0.52) (16, 1.34) (32, 2.55) (64, 3.36)
    (128, 5.328) (256, 6.599) (512, 10.285) (1024, 13.691)
};

\addplot[color=blue, mark=square*, mark size=2pt, line width=1pt] coordinates {
    (2, 0.24) (4, 0.29) (8, 0.31) (16, 0.83) (32, 0.94) (64, 0.93)
    (128, 2.96) (256, 3.39) (512, 6.21) (1024, 8.867)
};

\addplot[color=orange, mark=triangle*, mark size=3pt, line width=1pt, dashed] coordinates {
    (2, 0.35) (4, 0.39) (8, 0.44) (16, 0.68) (32, 0.66) (64, 0.75)
    (128, 2.05) (256, 2.53) (512, 4.25) (1024, 7.349)
};

\end{groupplot}
\end{tikzpicture}}
\vspace{-8mm}

\caption{\system achieves scalability w.r.t example size (left), target document
size (center), and topic dimensionality (right).}

\label{fig:scalability}
\end{figure}

\subsection{Scalability analysis}
\label{sec:scalability_analysis}

\setlength{\textfloatsep}{5pt}
\looseness-1 We evaluate \system's scalability (RQ4) w.r.t (a)~the average
\#sentences in the examples $x$; (b)~the size of the target document in
\#sentences; and (c)~the dimensionality of the documents in \#topics. \system
achieves practical scalability via the two mechanisms of \S\ref{sec:execution},
\textsc{PreFilter} and \paql's \sketchref. For small $x$, \textsc{PreFilter}
keeps the ILP size independent of target document size. However, its
effectiveness diminishes as $x$ grows or limiting search to only $10x$
sentences becomes suboptimal, making \sketchref essential.


\looseness-1 Over 400 \dataset instances, we vary the example size from 5 to 100
sentences (with target documents of $\approx$500 sentences) and the target
document size from 100K to 1M sentences (with $x=100$), using 5 source and 45
target documents per instance. We increase \#sentences in each document by paraphrasing the
original sentences (Appendix~\ref{appx:scalability_relaxation_details}). To synthesize the example bundles, we iteratively sample
sentences without replacement to avoid duplicates, randomly selecting sentences
across topics until the required bundle size is reached.
\looseness-1 Fig.~\ref{fig:scalability} (left) shows that for small documents
($\approx$ 500 sentences), \sketchref scales better than \textsc{PreFilter} as
$x$ grows. This is because \textsc{PreFilter} relies on direct ILP: when the
document contains fewer than $10x$ sentences, it retains the full target
document. In contrast, \sketchref always considers the full document but
reduces the search space via clustering and invokes ILP over cluster
representatives. \system applies \textsc{PreFilter} before \sketchref,
benefiting from both. 

Fig.~\ref{fig:scalability} (center) shows that for a
fixed $x=100$, \textsc{PreFilter}'s runtime increases linearly with target
document size due to the single pass required to retain the top $1000$
sentences, while \sketchref's runtime increases more steeply. \system benefits
from both and inherits the scalability of \textsc{PreFilter}, with runtime
under $0.5$ seconds even for documents with 1M sentences (excluding the
one-time embedding). 
\looseness-1 However,
\textsc{PreFilter} may compromise bundle quality. Since no ground truth is
available, we treat the bundle retrieved \textit{without} \textsc{PreFilter} as
the reference bundle with optimal quality, and measure a retrieved bundle's
quality function value relative to it. We observe that the relative bundle
quality decreases only modestly: $93.7\%$ for the 100K-sentence document and
$86.6\%$ for the 1M-sentence document. Note that \textsc{PreFilter} retains only
$1\%$ and $0.1\%$ of the target document sentences, respectively. This modest
quality loss is well justified by the massive runtime reduction it offers
($106\times$ and $1084\times$ speedup w.r.t.\ \sketchref, respectively).

\SBERT embedding generation for \textsc{PreFilter} incurs a one-time
cost per target document, growing linearly from 27.6s to 267.5s as the
\#sentences increases from 100K to 1M. Per-query \textsc{PreFilter} indexing
cost increases from 0.01s to 0.1s over the same range.

Fig.~\ref{fig:scalability} (right) shows that increasing
document dimensionality in \#topics (or constraints) causes linear runtime
growth for all approaches. However, practical workloads rarely exceed $1024$
constraints, and even then, \system keeps its runtime under $9$s.

\vspace{0.5mm}
\keytakeaway{
    \begin{itemize}[nosep, leftmargin=*, label={\footnotesize\textcolor{Bittersweet}{\faLightbulb}}] 
		
        \item \system scales nearly independently of document size.
		
        \item \system's runtime grows linearly with \#topics, and remains under
        $9$ seconds even for practically large \#constraints.
		
    \end{itemize}
}

\smallskip \noindent \emph{Remark.} When
higher quality is required, \textsc{PreFilter} can retain more sentences and
\sketchref keeps the runtime manageable over the larger retained set. The
trade-off between runtime and quality can be tuned based on application-specific
requirements.


\subsection{Relaxation behavior and alternatives}

\subsubsection{Relaxation behavior analysis}
\label{sec:relaxation_analysis}
We analyze relaxation behavior (RQ5) over an augmented version of \dataset,
where we add synthetic data to the original dataset (details in
Appendix~\ref{appx:scalability_relaxation_details}). We vary the example size from 10 to 10K sentences and the
target document size from 10 to 1M sentences. Fig.~\ref{fig:scal_relaxation}
reports only infeasible queries, which make up 22\% when varying example size
and 47\% when varying document size. Fig.~\ref{fig:scal_relaxation} (left)
shows that \#relaxations grows with example size, as more examples yield
higher lower bounds of the constraints, making them tighter. Conversely,
\#relaxations decreases with increasing target document size, since larger
documents are more likely to contain sentences satisfying the example-derived
bounds (Fig.~\ref{fig:scal_relaxation} (right)).

\vspace{-1mm}
\subsubsection{Alternative relaxation mechanisms}
\label{sec:relaxation_mechanism} 
We compare our symmetric relaxation (SYM), which relaxes both bounds of each
violated constraint equally, with two alternatives: (1)~\emph{directional}
relaxation (DIR), which uses CPLEX's conflict refiner to choose relaxation
direction, and (2)~\emph{FeasOpt}~\cite{feasopt}, which
minimizes the total slack to ensure feasibility. Since only $5/825$ ($0.6\%$)
queries over \dataset are infeasible, we induce infeasibility by tightening
each constraint's bounds by $\mathit{pad}$, i.e., $(\lowerbound,
\upperbound)\mapsto((1+\mathit{pad})\times\lowerbound,
(1-\mathit{pad})\times\upperbound)$; larger $\mathit{pad}$ yields greater
infeasibility. We apply $\mathit{pad}\in\{30\%,40\%,50\%,60\%\}$ to $80$
queries, of which 45, 66, 77, and 80 become infeasible, respectively, and
report results over these infeasible queries (more in
Appendix~\ref{appx:relaxation_mechanisms}).

\looseness-1 Figure~\ref{fig:relaxation_quality} shows that (a)~SYM,
DIR, and FeasOpt yield comparable semantic similarity, and (b)~their runtimes
are comparable at low $\mathit{pad}$, although FeasOpt converges faster at high
$\mathit{pad}$.  However, (c)~shows
that SYM and DIR relax fewer constraints than FeasOpt at every padding level.
Thus, in the common case of mild infeasibility, SYM matches FeasOpt in
convergence and intent alignment while relaxing fewer constraints, justifying
our choice of SYM for the two usability benefits
(\S\ref{sec:bound_relaxation}). FeasOpt converges faster with minimal total
relaxation, albeit by altering more constraints, and may be preferable when
severe infeasibility is expected. However, we never observed severe
infeasibility on the original \dataset dataset; only 0.6\% of queries were infeasible,
all with a low degree of infeasibility.


\definecolor{myblue}{RGB}{30, 144, 255}

\begin{figure}[t]
\centering
\resizebox{\columnwidth}{!}{
\begin{tikzpicture}
\begin{groupplot}[
    group style={
        group size=4 by 1,
        horizontal sep=1cm,
        vertical sep=1.6cm,
        group name=scalrelaxgroup
    },
    width=4cm,
    height=2.7cm,
    xtick pos=left,
    ymajorgrids=true,
    xmajorgrids=true,
    enlarge y limits=false,
    grid style={dashed, gray!30},
    label style={font=\small},
    tick label style={font=\footnotesize},
    title style={at={(0.5,-0.75)}, anchor=north, font=\small},
    every axis plot post/.append style={mark size=2pt},
]

\nextgroupplot[
    ymin=0, ymax=55,
    xmin=0.5, xmax=6.5,
    xlabel={Example Size},
    xlabel style={yshift=1mm},
    ylabel={\#Relaxations},
    ylabel style={xshift=-1.5mm},
    xtick={1,2,3,4,5,6},
    xticklabels={10,50,100,500,1K,10K},
]
\addplot+[boxplot/draw direction=y, boxplot prepared={lower whisker=1, lower quartile=4.0, median=6.0, upper quartile=8.0, upper whisker=16}, boxplot/draw position=1, draw=myblue, solid, fill=myblue!20, thick] coordinates {};
\addplot+[boxplot/draw direction=y, boxplot prepared={lower whisker=1, lower quartile=1.0, median=2.5, upper quartile=5.75, upper whisker=15}, boxplot/draw position=2, draw=myblue, solid, fill=myblue!20, thick] coordinates {};
\addplot+[boxplot/draw direction=y, boxplot prepared={lower whisker=1, lower quartile=4.0, median=5.0, upper quartile=8.25, upper whisker=18}, boxplot/draw position=3, draw=myblue, solid, fill=myblue!20, thick] coordinates {};
\addplot+[boxplot/draw direction=y, boxplot prepared={lower whisker=1, lower quartile=5.25, median=7.5, upper quartile=14.75, upper whisker=34}, boxplot/draw position=4, draw=myblue, solid, fill=myblue!20, thick] coordinates {};
\addplot+[boxplot/draw direction=y, boxplot prepared={lower whisker=1, lower quartile=11.75, median=14.5, upper quartile=21.5, upper whisker=39}, boxplot/draw position=5, draw=myblue, solid, fill=myblue!20, thick] coordinates {};
\addplot+[boxplot/draw direction=y, boxplot prepared={lower whisker=5, lower quartile=40.0, median=51.0, upper quartile=51.0, upper whisker=51}, boxplot/draw position=6, draw=myblue, solid, fill=myblue!20, thick] coordinates {};

\nextgroupplot[
    width=6cm,
    height=2.7cm,
    ymin=0, ymax=55,
    xmin=0.5, xmax=7.5,
    xlabel={Target Document Size},
    xlabel style={yshift=1mm},
    ylabel={},
    xtick={1,2,3,4,5,6,7},
    xticklabels={10,50,100,1K,10K,100K,1M},
]
\addplot+[boxplot/draw direction=y, boxplot prepared={lower whisker=30, lower quartile=51, median=51, upper quartile=51, upper whisker=51}, boxplot/draw position=1, draw=myblue, solid, fill=myblue!20, thick] coordinates {};
\addplot+[boxplot/draw direction=y, boxplot prepared={lower whisker=2, lower quartile=40, median=49, upper quartile=51, upper whisker=51}, boxplot/draw position=2, draw=myblue, solid, fill=myblue!20, thick] coordinates {};
\addplot+[boxplot/draw direction=y, boxplot prepared={lower whisker=1, lower quartile=1, median=3, upper quartile=4.25, upper whisker=10}, boxplot/draw position=3, draw=myblue, solid, fill=myblue!20, thick] coordinates {};
\addplot+[boxplot/draw direction=y, boxplot prepared={lower whisker=1, lower quartile=1, median=3, upper quartile=4.25, upper whisker=9}, boxplot/draw position=4, draw=myblue, solid, fill=myblue!20, thick] coordinates {};
\addplot+[boxplot/draw direction=y, boxplot prepared={lower whisker=1, lower quartile=1, median=3, upper quartile=4.25, upper whisker=9}, boxplot/draw position=5, draw=myblue, solid, fill=myblue!20, thick] coordinates {};
\addplot+[boxplot/draw direction=y, boxplot prepared={lower whisker=1, lower quartile=1, median=3, upper quartile=4.25, upper whisker=10}, boxplot/draw position=6, draw=myblue, solid, fill=myblue!20, thick] coordinates {};
\addplot+[boxplot/draw direction=y, boxplot prepared={lower whisker=1, lower quartile=1, median=3, upper quartile=4.25, upper whisker=10}, boxplot/draw position=7, draw=myblue, solid, fill=myblue!20, thick] coordinates {};

\end{groupplot}
\end{tikzpicture}
}
\vspace{-8mm}
\caption{Number of relaxation rounds for infeasible queries as (left)
example size and (right) target document size grow. Queries still infeasible
at the 50-round cap are shown at 51.}
\vspace{-2mm}
\label{fig:scal_relaxation}
\end{figure}


\usetikzlibrary{pgfplots.groupplots,calc,patterns.meta}

\begin{figure}[t]
\centering
\resizebox{\columnwidth}{!}{%
\begin{tikzpicture}
\begin{groupplot}[
    group style={
        group size=3 by 1,
        horizontal sep=1.2cm,
    },
    width=4cm,
    height=3.24cm,
    xmin=24, xmax=66,
    xtick={30,40,50,60},
    xticklabels={30,40,50,60},
    xlabel style={align=center, yshift=4pt},
    xtick pos=left,
    x tick style={draw=none},
    ytick pos=left,
    ymajorgrids=true,
    grid style={dashed, gray!30},
    label style={font=\small},
    ylabel style={font=\footnotesize, yshift=-3pt},
    tick label style={font=\footnotesize},
    yticklabel style={/pgf/number format/fixed},
    every axis plot post/.append style={mark size=2.2pt},
    legend style={draw=none, legend columns=1, column sep=3pt, font=\small, cells={anchor=west}},
]

\nextgroupplot[
    xlabel={$\mathit{pad}$ (\%)\\(a)}, ylabel={SS},
    ybar=0pt, bar width=4.2pt, tick align=inside,
    ymin=0, ymax=100, ytick={0,20,40,60,80},
]
\addplot[draw=black!80, fill=black!80] coordinates {(30,83.68) (40,84.04) (50,83.98) (60,82.40)};
\addplot[draw=black!15, fill=black!15, postaction={pattern={Dots[distance=0.9pt, radius=0.18pt]}, pattern color=black!70}] coordinates {(30,83.68) (40,83.87) (50,84.10) (60,82.54)};
\addplot[draw=black!55, fill=black!55] coordinates {(30,83.13) (40,82.89) (50,83.35) (60,82.56)};

\nextgroupplot[
    xlabel={$\mathit{pad}$ (\%)\\(b)}, ylabel={Runtime (s)},
    ymode=log, ymin=0.2, ymax=300,
    ytick={1,10,100}, yticklabels={$10^0$,$10^1$,$10^2$}, yminorticks=false,
    boxplot/draw direction=y,
    every axis plot post/.style={mark size=0.6pt},
    legend to name=relaxqualitylegend,
    legend image code/.code={\draw[#1] (0cm,-0.1cm) rectangle (0.25cm,0.1cm);},
]
\addlegendimage{draw=black!80, fill=black!80} \addlegendentry{SYM}
\addlegendimage{draw=black!15, fill=black!15, postaction={pattern={Dots[distance=0.9pt, radius=0.18pt]}, pattern color=black!70}} \addlegendentry{DIR}
\addlegendimage{draw=black!55, fill=black!55} \addlegendentry{FeasOpt}
\addplot[forget plot, draw=black, fill=black!80, mark=*, mark options={draw=black, fill=black},
    boxplot prepared={draw position=27.2, box extend=2.4, lower whisker=0.31, lower quartile=0.58, median=1.58, upper quartile=3.60, upper whisker=7.90, every median/.style={draw=white, thick}},
] table[row sep=\\, y index=0] {10.32 \\ 11.28 \\ 12.71 \\ 60.36 \\ 60.44 \\};
\addplot[forget plot, draw=black, fill=black!15, boxplot/every box/.style={postaction={pattern={Dots[distance=0.9pt, radius=0.18pt]}, pattern color=black!70}}, mark=*, mark options={draw=black, fill=black},
    boxplot prepared={draw position=30.0, box extend=2.4, lower whisker=0.35, lower quartile=0.58, median=1.56, upper quartile=3.86, upper whisker=7.79, every median/.style={draw=black, thick}},
] table[row sep=\\, y index=0] {9.64 \\ 10.85 \\ 12.54 \\ 60.43 \\ 61.04 \\};
\addplot[forget plot, draw=black, fill=black!55, mark=*, mark options={draw=black, fill=black},
    boxplot prepared={draw position=32.8, box extend=2.4, lower whisker=0.38, lower quartile=0.66, median=1.14, upper quartile=2.13, upper whisker=2.56, every median/.style={draw=black, thick}},
] table[row sep=\\, y index=0] {5.21 \\ 11.33 \\ 13.79 \\ 22.15 \\ 25.01 \\ 56.54 \\ 60.78 \\ 61.08 \\};
\addplot[forget plot, draw=black, fill=black!80, mark=*, mark options={draw=black, fill=black},
    boxplot prepared={draw position=37.2, box extend=2.4, lower whisker=0.40, lower quartile=0.94, median=1.35, upper quartile=5.22, upper whisker=11.55, every median/.style={draw=white, thick}},
] table[row sep=\\, y index=0] {12.02 \\ 12.79 \\ 12.81 \\ 12.99 \\ 17.79 \\ 30.87 \\ 60.35 \\ 60.36 \\ 60.69 \\ 61.65 \\ 78.69 \\};
\addplot[forget plot, draw=black, fill=black!15, boxplot/every box/.style={postaction={pattern={Dots[distance=0.9pt, radius=0.18pt]}, pattern color=black!70}}, mark=*, mark options={draw=black, fill=black},
    boxplot prepared={draw position=40.0, box extend=2.4, lower whisker=0.35, lower quartile=0.93, median=1.29, upper quartile=5.32, upper whisker=11.68, every median/.style={draw=black, thick}},
] table[row sep=\\, y index=0] {12.54 \\ 12.83 \\ 18.04 \\ 30.54 \\ 60.42 \\ 60.42 \\ 60.51 \\ 61.84 \\ 78.39 \\};
\addplot[forget plot, draw=black, fill=black!55, mark=*, mark options={draw=black, fill=black},
    boxplot prepared={draw position=42.8, box extend=2.4, lower whisker=0.45, lower quartile=0.88, median=1.24, upper quartile=1.81, upper whisker=3.15, every median/.style={draw=black, thick}},
] table[row sep=\\, y index=0] {3.33 \\ 5.50 \\ 5.76 \\ 6.30 \\ 8.05 \\ 8.89 \\ 12.19 \\ 14.34 \\ 14.86 \\ 21.73 \\ 35.86 \\ 60.44 \\ 60.93 \\};
\addplot[forget plot, draw=black, fill=black!80, mark=*, mark options={draw=black, fill=black},
    boxplot prepared={draw position=47.2, box extend=2.4, lower whisker=0.57, lower quartile=1.08, median=1.48, upper quartile=3.81, upper whisker=7.12, every median/.style={draw=white, thick}},
] table[row sep=\\, y index=0] {8.15 \\ 10.19 \\ 10.36 \\ 10.49 \\ 11.49 \\ 12.10 \\ 14.27 \\ 16.84 \\ 18.30 \\ 23.11 \\ 57.57 \\ 60.56 \\ 60.66 \\};
\addplot[forget plot, draw=black, fill=black!15, boxplot/every box/.style={postaction={pattern={Dots[distance=0.9pt, radius=0.18pt]}, pattern color=black!70}}, mark=*, mark options={draw=black, fill=black},
    boxplot prepared={draw position=50.0, box extend=2.4, lower whisker=0.59, lower quartile=1.01, median=1.49, upper quartile=3.79, upper whisker=7.75, every median/.style={draw=black, thick}},
] table[row sep=\\, y index=0] {9.12 \\ 9.85 \\ 10.49 \\ 10.66 \\ 12.57 \\ 14.33 \\ 16.30 \\ 20.12 \\ 20.56 \\ 57.76 \\ 60.59 \\ 60.97 \\};
\addplot[forget plot, draw=black, fill=black!55, mark=*, mark options={draw=black, fill=black},
    boxplot prepared={draw position=52.8, box extend=2.4, lower whisker=0.62, lower quartile=0.99, median=1.39, upper quartile=1.99, upper whisker=3.45, every median/.style={draw=black, thick}},
] table[row sep=\\, y index=0] {4.96 \\ 7.60 \\ 11.12 \\ 11.40 \\ 11.44 \\ 12.70 \\ 19.35 \\ 30.66 \\ 60.48 \\ 60.55 \\ 60.57 \\ 60.59 \\};
\addplot[forget plot, draw=black, fill=black!80, mark=*, mark options={draw=black, fill=black},
    boxplot prepared={draw position=57.2, box extend=2.4, lower whisker=0.41, lower quartile=0.96, median=2.23, upper quartile=8.86, upper whisker=18.59, every median/.style={draw=white, thick}},
] table[row sep=\\, y index=0] {21.03 \\ 22.34 \\ 22.51 \\ 34.08 \\ 40.77 \\ 60.99 \\ 62.28 \\ 89.31 \\ 180.89 \\};
\addplot[forget plot, draw=black, fill=black!15, boxplot/every box/.style={postaction={pattern={Dots[distance=0.9pt, radius=0.18pt]}, pattern color=black!70}}, mark=*, mark options={draw=black, fill=black},
    boxplot prepared={draw position=60.0, box extend=2.4, lower whisker=0.46, lower quartile=0.98, median=2.16, upper quartile=8.92, upper whisker=18.31, every median/.style={draw=black, thick}},
] table[row sep=\\, y index=0] {20.99 \\ 21.97 \\ 22.08 \\ 32.25 \\ 40.59 \\ 61.89 \\ 61.92 \\ 85.15 \\ 181.28 \\};
\addplot[forget plot, draw=black, fill=black!55, mark=*, mark options={draw=black, fill=black},
    boxplot prepared={draw position=62.8, box extend=2.4, lower whisker=0.33, lower quartile=0.62, median=0.95, upper quartile=2.15, upper whisker=4.23, every median/.style={draw=black, thick}},
] table[row sep=\\, y index=0] {4.82 \\ 5.95 \\ 6.42 \\ 6.44 \\ 10.95 \\ 12.32 \\ 16.29 \\ 20.92 \\ 21.81 \\ 23.97 \\ 35.59 \\ 42.82 \\ 50.99 \\ 60.45 \\ 60.95 \\};

\nextgroupplot[
    xshift=-0.25cm,
    xlabel={$\mathit{pad}$ (\%)\\(c)}, ylabel={\#Constraints},
    ylabel style={xshift=-3pt},
    ybar=0pt, bar width=4.2pt, tick align=inside,
    ymin=0, ymax=9,
]
\addplot[draw=black!80, fill=black!80] coordinates {(30,1.49) (40,1.89) (50,1.36) (60,3.54)};
\addplot[draw=black!15, fill=black!15, postaction={pattern={Dots[distance=0.9pt, radius=0.18pt]}, pattern color=black!70}] coordinates {(30,1.49) (40,1.89) (50,1.35) (60,3.55)};
\addplot[draw=black!55, fill=black!55] coordinates {(30,2.93) (40,3.86) (50,5.38) (60,7.69)};

\end{groupplot}
\node[anchor=west] at ($(group c3r1.east)+(0cm,0)$)
    {\pgfplotslegendfromname{relaxqualitylegend}};
\end{tikzpicture}
}
\vspace{-8mm}

\caption{(a)~Mean semantic similarity, (b)~runtime, and
(c)~\#constraints relaxed, by SYM, DIR, \& FeasOpt on infeasible queries.}

\label{fig:relaxation_quality}
\end{figure}


\begin{figure}[t]
\centering
\resizebox{\columnwidth}{!}{
\begin{tikzpicture}
\begin{groupplot}[
    group style={
        group size=3 by 1,
        horizontal sep=1.2cm,
        vertical sep=1.6cm,
        group name=skewgroup
    },
    width=4cm,
    height=2.7cm,
    xtick pos=left,
    ymajorgrids=true,
    xmajorgrids=true,
    enlarge y limits=false,
    grid style={dashed, gray!30},
    label style={font=\small},
    tick label style={font=\footnotesize},
    title style={at={(0.5,-0.75)}, anchor=north, font=\small},
    every axis plot post/.append style={mark size=2pt},
]

\nextgroupplot[
    symbolic x coords={50,60,70,80,90},
    xtick={50,60,70,80,90},
    ymin=0, ymax=13,
    xlabel={Degree of Skewness (\%)},
    xlabel style={yshift=1mm},
    ylabel={\#Relaxations},
    ylabel style={xshift=-1mm},
    title={(a)},
]
\addplot+[thick, mark=*] coordinates {
    (50,10.905)
    (60,9.756)
    (70,10.750)
    (80,8.313)
    (90,10.614)
};

\nextgroupplot[
    symbolic x coords={50,60,70,80,90},
    xtick={50,60,70,80,90},
    ymin=0.0, ymax=1,
    xlabel={Degree of Skewness (\%)},
    xlabel style={yshift=1mm},
    ylabel={Runtime (s)},
    yticklabel style={
        /pgf/number format/fixed,
        /pgf/number format/fixed zerofill,
        /pgf/number format/precision=2
    },
    title={(b)},
]
\addplot+[thick, mark=*] coordinates {
    (50,0.361)
    (60,0.348)
    (70,0.361)
    (80,0.335)
    (90,0.370)
};

\nextgroupplot[
    symbolic x coords={50,60,70,80,90},
    xtick={50,60,70,80,90},
    ymin=0, ymax=1,
    ytick={0,0.2,0.4,0.6,0.8,1.0},
    yticklabel style={
        /pgf/number format/fixed,
        /pgf/number format/fixed zerofill,
        /pgf/number format/precision=1
    },
    xlabel={Degree of Skewness (\%)},
    xlabel style={yshift=1mm},
    ylabel={SS},
    title={(c)},
]
\addplot+[thick, mark=*] coordinates {
    (50,0.880)
    (60,0.888)
    (70,0.883)
    (80,0.885)
    (90,0.870)
};

\end{groupplot}
\end{tikzpicture}
}
\vspace{-8mm}
\caption{\system's (a)~number of relaxations, (b)~runtime, and (c)~intent
alignment as the topic skew in the target document varies.}

\label{fig:skew}
\end{figure}

\vspace{0.5mm}
\keytakeaway{
    \begin{itemize}[nosep, leftmargin=*, label={\footnotesize\textcolor{Bittersweet}{\faLightbulb}}] 

		\item More examples result in stricter (less feasible) constraints while larger documents provide more options to satisfy constraints.
		
         \item SYM, DIR, and FeasOpt differ in \#constraints relaxed and in runtime under severe infeasibility.
		
    \end{itemize}
}

\subsection{Robustness to data skew}
\label{sec:robustness}
To evaluate robustness to varying degrees of skewness in the target document's
topic distribution (RQ6), we designate one topic as the \emph{focus topic} and
vary the fraction of sentences drawn from it (50\%, 60\%, 70\%, 80\%, and
90\%), with the remaining sentences drawn uniformly from the other topics, over
$100$ instances of the \dataset dataset (details are in
Appendix~\ref{appx:robustness_analysis}). Fig.~\ref{fig:skew} shows the effect
of increasing topic skew: (a) shows that the number of relaxations stays stable
despite skew; (b) shows that runtime remains largely unaffected by target data
skew; and (c) shows that the resulting bundles maintain a consistent level of
alignment with the ground truth across all skew levels.

\vspace{3mm}
\keytakeaway{
\begin{itemize}[nosep, leftmargin=*, label={\footnotesize\textcolor{Bittersweet}{\faLightbulb}}]

    \item \system's runtime, intent alignment, and \#relaxations remain robust
    to skew in the target document's topic distribution.

\end{itemize}
}

\section{Intent Sliders: A User study} \label{sec:userstudy} \looseness-1 To
evaluate \system's slider-based intent-refinement interface for FTSE, we ran a
within-subjects study (Appendix~\ref{appx:userstudyappendix}) with 20 Amazon
Mechanical Turk participants (fluent in English, experienced with document
search), who identified their preferred US state for relocation from Wikipedia
articles using \system versus \SBERT (the Top-$k$ baseline). We exclude
non-\bqbe baselines (e.g., conversational LLMs), which underperform on FTSE
(Table~\ref{tab:sbert_similarity_chatgpt4o}). We set the slider parameter
$\alpha=0.1$ and, following a mixed-methods approach~\cite{greene1989toward},
randomized each participant's starting tool (anonymized) to mitigate transfer
effects. Each participant was assigned 25 randomly selected states per tool and
asked to provide snippets for at least 2, with no upper limit. With \system,
participants could select sentences, search by keyword, and refine intent via
sliders; with \SBERT, they could select sentences and search by keyword too, but
had to adjust intent by manually adding or removing example sentences.

\looseness-1 We quantitatively analyzed the data by comparing the participants'
ratings of summary quality, ease of use, and preference for both \SBERT and
\system. We also compared the number of interactions with the interface (e.g.,
selecting examples, adjusting sliders, requesting snippet retrieval, etc.)
between the two tools to understand how participants engaged with them.
Participants' open-ended feedback on their experience additionally surfaced
signals relevant to the perceived interpretability of the sliders. The majority
used the sliders to refine their intents, and their comments suggest an
interpretability benefit, for instance: \emph{``I used them [sliders] to reduce
importance of certain things [\dots] to see how that would affect generated
summaries''} and \emph{``I raised the slider up so that the output summaries
would prioritize [\dots] economy and climate.''}

\begin{sloppypar}
\vspace{0.5mm}\noindent\textbf{Participants' satisfaction.} On task completion,
participants rated each tool on quality, ease of use, and daily usage
preference. \system received higher ratings than SBERT on all three: (i) 12/20
rated \system as higher quality (8/20 for SBERT), (ii) 16/20 found \system
easier to use (only 4/20 for SBERT), and (iii) 11/20 preferred \system for
daily use (9/20 for SBERT).
\end{sloppypar}

\vspace{0.5mm}\noindent\textbf{Participants' interactions.} \looseness-1 To
understand how participants interacted with the tools, we categorized
interactions into six types: \textit{Search}: keyword queries,
\textit{Selection}: example selection, \textit{Update}: example modification,
\textit{Learn}: intent learning, \textit{Retrieval}: backend retrieval, and
\textit{Slider}: interactive parameters. Averaging interaction counts per user,
\SBERT users performed about 3.1 times more \textit{Search} interactions than
\system users. In contrast, \system users had more \textit{Slider} interactions
and consequently about 2.1 times more \textit{Retrieval} interactions, as
\system automatically retrieves bundles upon slider adjustments. \system's
interactive sliders encouraged more iterative refinement of results, while
\SBERT's lack of interactivity led to more manual searching.

Although the user study involved only 20
participants, each provided multiple summaries (2--25). Nevertheless, the
evaluation should be viewed as preliminary rather than definitive.
            

\section{Related Work}

\looseness-1 \textbf{Programming by example}~\cite{DBLP:conf/popl/Gulwani11,
DBLP:conf/aplas/GulwaniJ17, DBLP:conf/cade/Gulwani16, uistpaper} is a paradigm
where users provide examples to express their intent and has seen significant
success in data discovery~\cite{DBLP:journals/pvldb/RezigBFPVGS21,
DBLP:journals/pvldb/KhatriMR25, 10.1145/3786682} and querying
(\qbe)~\cite{DBLP:conf/vldb/Zloof75, QPlain, DBLP:conf/sigmod/ShenCCDN14,
DBLP:conf/sigmod/PsallidasDCC15, DBLP:journals/tods/BonifatiCS16, squid,
DBLP:journals/is/FarihaCMM26}. Some works prune the search space for efficient
query synthesis~\cite{DBLP:conf/sigmod/ShenCCDN14,
DBLP:journals/tods/BonifatiCS16} or generalize the
output~\cite{DBLP:conf/sigmod/PsallidasDCC15}, and \squid~\cite{squid,
DBLP:journals/is/FarihaCMM26} semantically synthesizes SQL queries from
user-provided examples. However, all existing \qbe systems are designed to
retrieve individual items and do not account for the combinatorial nature of
bundle retrieval. \system addresses the bundle retrieval problem through the
lens of \qbe, where users express their intent through example bundles, as
opposed to example tuples.

\smallskip\noindent\textbf{Focused text snippet extraction}, also known as
query-focused extractive summarization, allows users to specify keywords or
queries to guide snippet selection~\cite{daume2009bayesian, li-li-2014-query,
litvak2017query}, but requires users to explicitly formulate their intents.
RAG-based approaches~\cite{NEURIPS2020_6b493230, REALM, langchain,
Liu_LlamaIndex_2022} can also support snippet extraction via few-shot
prompting~\cite{NeurIPS20TomLanguage}, but are not designed to jointly satisfy
multiple constraints. \emph{Faceted query-by-example} approaches leverage
representative documents as an example query, additionally conditioned on
explicit facets (constraints) to extract snippets across multiple
constraints~\cite{mysore-etal-2022-multi, DBLP:journals/corr/abs-2310-04678,
DBLP:conf/acl/DoRKL25}. However, existing systems either require users to
specify constraints via keywords~\cite{DBLP:conf/acl/DoRKL25} or natural
language~\cite{DBLP:journals/corr/abs-2310-04678}, or rely on opaque vector
matching that lacks interpretability~\cite{mysore-etal-2022-multi}. 


\smallskip\noindent\textbf{Bundle retrieval} \looseness-1 involves selecting an
optimal set of items subject to a set of constraints~\cite{BrucatoBAM16,
DBLP:journals/pvldb/MaiWABHM24, DBLP:journals/pvldb/DrosouP12,
DBLP:journals/pvldb/WangMM18, sage}. It has also been studied extensively in
recommender systems across various domains, such as
e-commerce~\cite{DBLP:conf/sigir/WeiLYWZ22, DBLP:conf/recsys/SunXZZLW23,
DBLP:conf/kdd/LiuWTMWC25}, entertainment~\cite{DBLP:conf/sigir/PathakGM17,
DBLP:journals/tkde/ChangGHJL23}, and travel
planning~\cite{DBLP:conf/mdm/FangLT14, DBLP:journals/tsc/GuCL22,
DBLP:journals/tsc/ZhangMGC24}. However, these systems rely on greedy heuristics
or approximate generative models to generate bundles and struggle to enforce
strict constraints. Package queries increase bundle query
efficiency~\cite{BrucatoBAM16, DBLP:journals/pvldb/MaiWABHM24}, but require
users to formulate the query. Diversification-based
retrieval~\cite{DBLP:journals/pvldb/DrosouP12, DBLP:journals/pvldb/WangMM18,
sage} retrieves bundles that collectively satisfy user constraints, but targets
simpler constraints involving cardinality or diversity. In contrast, \system
addresses the challenge of automatically synthesizing multi-constraint package
queries from user examples.

\smallskip\noindent\textbf{Constraint relaxation.}
\label{sec:related_work_relaxation} 
FeasOpt~\cite{feasopt} relaxes infeasible constraints to obtain a feasible
solution, potentially at the cost of solution quality. Related work in
provenance and explanation addresses similar problems but in different settings.
CAPE~\cite{DBLP:conf/sigmod/MiaoZGR19} explains surprising aggregate results by
mining counterbalancing patterns from data outside the query's provenance, while
QFix~\cite{DBLP:conf/sigmod/WangM017} diagnoses data errors by computing minimal
repairs to historical queries that could cause the observed discrepancy.
However, these approaches do not directly apply to our \paql constraint
relaxation problem.


\section{Summary and Future Directions}
\label{sec:conclusion}
\looseness-1 We presented \system, a framework that enables users to specify
bundle query intents through examples, synthesizes package queries from
examples, and employs constraint relaxation to ensure query feasibility.
Several directions extend the framework. 
\system currently optimizes linear objectives over per-tuple scores; native
\paql support for pairwise or quadratic objectives (e.g., bundle diversity for
playlists) would broaden expressiveness. Specification of the quality function
requires domain expertise; inferring both objective and constraints from
example bundles---a problem related to inverse
optimization~\cite{Heuberger04}---would remove the domain-expertise
requirement. Choosing the best aggregate to use in the constraints is another
open problem. LLMs can potentially suggest the best aggregate based on the task
context, which will lower the configuration burden currently on the domain
expert. Though our initial experimental
results are limited to the \dataset and TPC-H datasets, we hope that further
research will produce more datasets with ground-truth for evaluating \bqbe
problems.



\bibliographystyle{ACM-Reference-Format}
\bibliography{paper}


\appendix

\setlength{\textfloatsep}{5pt}

\renewcommand{\thesection}{\Alph{section}}
\makeatletter
\renewcommand{\@seccntformat}[1]{%
  \ifnum\pdfstrcmp{#1}{section}=\z@
    APPENDIX~\csname the#1\endcsname:\hspace{-2mm}\quad%
  \else
    \csname the#1\endcsname\quad
  \fi
}
\makeatother
\renewcommand{\thesubsection}{\Alph{section}.\arabic{subsection}}
\renewcommand{\thesubsubsection}{\Alph{section}.\arabic{subsection}.\arabic{subsubsection}}
\renewcommand{\thefigure}{A\arabic{figure}}
\setcounter{figure}{0}
\renewcommand{\thetable}{A\arabic{table}}
\setcounter{table}{0}
\newcommand{\stepCounter}[1]{{\Large\textcircled{{\small#1}}}}

\section{Example Use Cases}\label{appx:appendix_applications} \looseness-1 We now
present three real-world applications where example-driven intent specification
lowers the barrier for data bundle retrieval: \textit{supplier selection} for
business, \textit{focused snippet extraction} from text documents, and
\textit{playlist recommendation} in streaming services.

\subsubsection*{Supplier selection for business expansion into a new country.}
\looseness-1 Bundle retrieval by example is valuable in business settings where
an owner seeks to expand into a new country and must identify a set of new
\textit{suppliers} to establish relationships with. The objective is to select a
bundle of about 100 suppliers from a large candidate pool over 10{,}000
suppliers (similar to the TPC-H~\cite{tpc-h} dataset). Manually constructing an
optimal supplier set here is tedious and error-prone, as the selected suppliers
must collectively satisfy a range of constraints. Formulating these requirements
as an explicit package query is equally challenging: business owners often lack
precise knowledge of the exact parameter values for their desired constraints in
an unfamiliar market (e.g., acceptable price ranges, inventory availability, or
financial stability thresholds). However, they can readily provide example
bundles of suppliers drawn from countries where their business already operates
successfully. \system can leverage these examples to infer the underlying
implicit constraints, adjust them to the new supplier database of the target
country, and retrieve an optimal set of suppliers that collectively satisfies
the constraints.
 
\subsubsection*{Focused text snippet extraction}
\looseness-1 Another use case of \system is focused snippet extraction from text
documents, also termed as personalized extractive summarization~\cite{subsume}.
Here, the goal is to select (extract) a subset of sentences (snippet) from a
text document that, collectively, best aligns with the user's information need
(focus). For example, when a user wishes to extract sentences from a technical
paper that are most relevant to their own research interests. While such intent
can, in principle, be expressed in natural language, prior work~\cite{sudocu,
subsume} shows that articulating subjective intents this way is difficult and
often leads to user fatigue~\cite{DBLP:conf/sigir/ZamaniMCLDBCD20} due to
iterative, back-and-forth intent refinement via conversations. Instead, users
can provide a few example $\langle$document, snippet$\rangle$ pairs to
implicitly communicate their focus~\cite{DBLP:conf/emnlp/YanNL11,
takatsu-etal-2021-personalized, hu-etal-2012-context,
DBLP:conf/emnlp/SlobodkinNASD23}. For instance, a journalist summarizing reports
of all U.S.\ states with a particular focus---e.g., ``basic'' economic
information, ``moderate'' coverage of education, and ``strong'' emphasis on
technology---can highlight sentences from a few states' reports to form example
snippets. From these exemplars, \system can infer the journalist's focus,
transparently encode it as \paql constraints, and apply them to new state
reports to extract snippets with the same focus. Notably, for reports with
substantially different structure or content, \system can minimally adjust these
inferred constraints to ensure feasible snippet extraction.

\subsubsection*{Playlist recommendation for streaming services}
\looseness-1
In streaming services, users seek recommendations for \textit{bundles}
of items that collectively satisfy their preferences. Spotify, YouTube
Music, and Netflix exemplify this through playlist/watchlist
recommendations (e.g., Discover Weekly~\cite{spotify_music}, Your Daily
Discover~\cite{youtube_music}, and Top Picks for
You~\cite{netflix}), where the goal is to generate a set of songs/movies aligned with user interest. Existing approaches typically
rely on item-level similarity---recommending songs similar to those in a
user's listening history---or on similarity to a single reference
playlist. For instance, a user who frequently listens to \texttt{\small AC/DC}
may receive playlists dominated by rock songs, or playlists resembling
those curated for users with similar histories. While effective in
homogeneous preference settings, such approaches struggle to capture
diverse tastes and are notoriously known to cause user frustration, as evident from complaints about lack of variety in Spotify's Discover Weekly playlist~\cite{spotify-frustration}.
Consider a user whose taste spans multiple
genres---predominantly energetic rock, with a mix of heavy metal, some
soft country songs, alongside occasional calm classical music. A playlist based only on rock similarity or a single reference list cannot capture such diversity. Moreover, explicitly specifying such nuanced preferences (e.g., how to balance genre, artist, mood, and tempo) is difficult, even with natural language or LLM-based interfaces. Example-driven bundle recommendation addresses this: users provide a few example playlists that reflect their desired diversity across genres, artists, tempo, and mood, and \system infers implicit constraints to generate playlists that collectively satisfy the inferred constraints.

\section{Constraint Expressivity}
\label{appx:constraint_representation}

Here, we provide additional details on how we represent different types of
constraints in the integer linear program (ILP) formulation of \paql. We
illustrate three example queries that demonstrate categorical,
nonlinear/interaction, and  logical operators. Even with linear representations,
PaQL can accommodate richer expressions. To show that these extensions are
usable in practice, we ran three case studies on the real-world Spotify Tracks
dataset\footnote{The Spotify tracks dataset:
\url{www.kaggle.com/datasets/maharshipandya/-spotify-tracks-dataset}.}. For each
case study, we write a package query with the constraint extension for a
natural-language playlist intent and solve it with \system's ILP formulation.
We remove duplicate tracks, which the dataset lists once per genre.

\smallskip\noindent\textbf{Q1: Categorical.} Categorical information can be represented
through indicator variables in the integer linear program. To represent this
extension, we created a PaQL query about ``retrieving a workout playlist that's
mostly dance tracks to keep users energy up, but throw in a couple of hip-hop
tracks for variety.'' This is a typical example of a categorical constraint as
it involves selecting items based on their category (genre=`dance'). The query
below bounds the count of dance tracks with a range, while constraining the
count of hip-hop tracks to an exact value:

\begin{paqllisting}{\texttt{\textbf{Q1}:}}
SELECT PACKAGE(*) FROM SPOTIFY
WHERE genre IN ('dance', 'hip-hop')
SUCH THAT COUNT(*) BETWEEN 10 AND 14
  AND COUNT(genre = 'dance') BETWEEN 8 AND 11
  AND COUNT(genre = 'hip-hop') = 3
MAXIMIZE SUM(popularity);
\end{paqllisting}

The result shows \code{COUNT(*)}=14, \code{COUNT(dance)}=11,
\code{COUNT(hip\\-hop)}=3, \code{SUM(popularity)}=1296. All bounds were
feasible and no relaxation was needed.

\smallskip\noindent\textbf{Q2: Nonlinear/Interaction.} Feature interactions and
nonlinear relationships can be modeled by computing a derived attribute once as
a preprocessing step and then bounding it like any other attribute. The intent,
``retrieve a quick hits mix that includes tracks that are popular given how
short they are, not old epics that only racked up plays because they're long.
About 10 songs, where the average duration is around 3.5 minutes'', includes
non-linear attribute hit density. The hit density derived attribute that
expresses non-linear relationships is computed as $hit\_density = popularity /
duration\_m$, where $duration\_m$ is the track's duration in minutes and
$popularity$ is its popularity score.

\begin{paqllisting}{\texttt{\textbf{Q2}:}}
SELECT PACKAGE(*) FROM SPOTIFY
SUCH THAT COUNT(*) BETWEEN 8 AND 12
  AND SUM(popularity / duration_m)
      BETWEEN 112.0 AND 240.0
  AND AVG(duration_m) BETWEEN 3.49 AND 3.51
MAXIMIZE SUM(popularity);
\end{paqllisting}

We encode ``around 3.5 minutes'' as a range of $\pm 0.01$ minutes. The result
shows \code{COUNT(*)}=10, \code{AVG(duration\_m)}=3.51, \code{SUM(hit
density)}=239.97, \code{SUM(popularity)}=843. All bounds were feasible; the
hit-den\-sity upper bound is tight (239.97 of 240) and is what limits the
playlist's total popularity.

\smallskip\noindent\textbf{Q3: Logical Operators.} Logical conditions can be expressed with
AND, OR, and NOT. The intent for Q3 is ``retrieve a road-trip playlist that
covers a mix of moods: it must include at least one relaxed chill track AND at
least one upbeat pop track, but keep every track under 5 minutes. About 10 songs
total.'' With this intent, we add a logical constraint to ensure the playlist
meets the specified criteria.

\begin{paqllisting}{\texttt{\textbf{Q3}:}}
SELECT PACKAGE(*) FROM SPOTIFY
WHERE genre IN ('chill', 'pop')
SUCH THAT COUNT(*) BETWEEN 8 AND 12
  AND COUNT(genre = 'chill') >= 1
  AND COUNT(genre = 'pop') >= 1
  AND NOT (COUNT(duration_m > 5) >= 1)
MAXIMIZE SUM(popularity);
\end{paqllisting}

\begin{sloppypar}
The result shows \code{COUNT(*)}=12, \code{COUNT(chill)}=1,
\code{COUNT(pop)}=11, \code{SUM(popularity)}=1132. All bounds were feasible and
no relaxation was needed.
\end{sloppypar}

\section{Quality Function Variants}
\label{appx:system_variants_analysis}

\pgfplotsset{compat=1.18}
\usetikzlibrary{patterns}
\usetikzlibrary{pgfplots.groupplots}

\begin{figure*}[t]
\centering
\resizebox{0.9\textwidth}{!}{%
\begin{tikzpicture}
\begin{groupplot}[
    group style={
        group size=2 by 1,
        horizontal sep=2cm, 
    },
    ybar,
    xtick pos=left,
    xticklabel pos=lower,
    xticklabel style={font=\footnotesize},    
    yticklabel style={font=\footnotesize},
    xtick style={draw=none},
    enlarge y limits=false,
    enlarge x limits=0.4,
    ymajorgrids=true,
    grid style={dashed, gray!30},
    height=3cm,
    legend style={
        font=\footnotesize,
        at={(0.7,1)}, 
        anchor=south,
        legend columns=5, draw=none, column sep=5pt,
    },
    legend image code/.code={
        \draw[draw=black,fill opacity=1] (0cm,-0.08cm) rectangle (0.3cm,0.1cm);
    },
]

\nextgroupplot[
    ylabel={F$_1$ Score (\%)},  
    ylabel style={yshift=-5pt, font=\footnotesize},
    ymin=0, ymax=105,
    nodes near coords,
    symbolic x coords={ROUGE-1,ROUGE-2,ROUGE-L},
	enlarge x limits=0.25,
    xtick=data,
	width=0.52\textwidth, 
    every node near coord/.append style={
        font=\tiny,                            
        /pgf/number format/fixed,              
        /pgf/number format/precision= 1,
        /pgf/number format/zerofill,
        text=black      
    }
]
\addplot+[ybar, fill=black, draw=black] coordinates {(ROUGE-1,58.01) (ROUGE-2,44.00) (ROUGE-L,47.01)};
\addlegendentry{C-FREQ\phantom{xxx}}

\addplot+[ybar, fill=black!60, draw=black] coordinates {(ROUGE-1,55.31) (ROUGE-2,39.78) (ROUGE-L,43.28)};
\addlegendentry{P-FREQ\phantom{xxx}}

\addplot+[ybar, fill=black!20, postaction={pattern=dots}, draw=black] coordinates {(ROUGE-1,54.37) (ROUGE-2,38.53) (ROUGE-L,42.13)};
\addlegendentry{BS\phantom{xxx}}

\addplot+[ybar, fill=white,  postaction={pattern=dots}, draw=black] coordinates {(ROUGE-1,50.74) (ROUGE-2,31.93) (ROUGE-L,36.05)};
\addlegendentry{TS}

\nextgroupplot[
    ymin=0, ymax=105,
    nodes near coords,
    symbolic x coords={Semantic Similarity},
    xtick=data,
    ylabel={Similarity (\%)},
    ylabel style={yshift=-5pt, font=\footnotesize},
    ymin=0, ymax=105,
    width=0.22\textwidth,
    every node near coord/.append style={
        font=\tiny,                            
        /pgf/number format/fixed,              
        /pgf/number format/precision= 1,
        /pgf/number format/zerofill,
        text=black     
    }
]
\addplot+[ybar, fill=black, draw=black] coordinates {(Semantic Similarity,86.13)};
\addplot+[ybar, fill=black!60, draw=black] coordinates {(Semantic Similarity,85.57)};
\addplot+[ybar, fill=black!20, postaction={pattern=dots},  draw=black] coordinates {(Semantic Similarity,85.25)};
\addplot+[ybar, fill=white, postaction={pattern=dots}, draw=black] coordinates {(Semantic Similarity,84.17)};

\end{groupplot}
\end{tikzpicture}
} 
\vspace{-4mm}
\caption{\small Frequency-based quality functions (C-FREQ \& P-FREQ) outperform similarity-based functions (BS \& TS) on \dataset.}
\vspace{-4mm}

\label{fig:ablation}
\end{figure*}
Fig.~\ref{fig:ablation} demonstrates how individual quality functions
affect the performance of \system in retrieving sentences on the \dataset.
In the comparison with non-LLM baselines (Fig.~\ref{fig:extractive_performance}),
\system uses C-FREQ; in all other experiments, it uses the combined quality
function \textbf{ALL}, which sums four per-sentence scores, BS, TF-IDF, LPR,
and TS, defined below. Here, we instead use each quality function on its own as the
objective of the \system ILP formulation.
\textbf{C-FREQ} uses only the term-frequency quality function: the sum of a
sentence's TF-IDF weights, with the TF-IDF model fit on the user examples.
\textbf{BS} uses only the \SBERT cosine-similarity quality function, which
maximizes the similarity between selected sentences and the user examples.
\textbf{P-FREQ} combines TF-IDF with the Log Probability Ratio (LPR), which
scores a sentence by how much more frequent its words are in the examples
than in the target document.
\textbf{TS} uses only a topic-similarity quality function based on cosine
similarity, thereby maximizing topic similarity between selected sentences and
the examples.

Among the single quality functions, C-FREQ achieves the best performance across
all metrics, demonstrating its ability to identify sentences that are
semantically aligned with the user examples. Despite its simplicity, it
outperforms both the similarity-based functions (BS and TS) and P-FREQ in terms
of both ROUGE scores and semantic similarity.
The variant incorporating topic similarity (TS) shows the worst performance
across all metrics. This indicates that explicitly maximizing topic similarity
is not a reliable signal for selecting sentences that are important and
semantically relevant to the examples.
Ultimately, while integrating sentence-level semantic similarity or topic
similarity seems intuitively promising, they are not effective signals for
sentence selection compared to frequency-based objectives. This finding is
highly consistent with prior work~\cite{DBLP:conf/nips/Thakur0RSG21,
kedzie-etal-2018-content}, which observes that frequency-based methods often
outperform pure semantic similarity-based methods in retrieval tasks.

\vspace{2mm}
{
\centering
\keytakeaway{
    \begin{itemize}[nosep, leftmargin=*, label={\footnotesize\textcolor{Bittersweet}{\faLightbulb}}]
        \item Users can choose different quality functions to optimize for;
        among single quality functions, the term-frequency quality function
        (C-FREQ) achieves the best performance across all metrics.
        \item Frequency-based quality functions outperform semantic similarity-based
        objectives, suggesting that frequency is a more reliable signal for
        selecting sentences aligned with user intent in this setting.
    \end{itemize}
}}

\vspace{2mm}

\section{Dataset Details}
\label{appx:dataset_details}
\noindent{\textbf{TPC-H}.} For the package query retrieval task, we use
the TPC-H benchmark datasets~\cite{tpc-h}, which consists of a set of
business-oriented queries and concurrent data modification. For our experiments,
we synthesize a \texttt{\small SupplierFeatures} view derived from three tables
using a 0.01 scale factor: \texttt{\small Supplier} with 100 tuples,
\texttt{\small Partsupp} with 8{,}000 rows, and \texttt{\small Nation} with 25
rows. Each supplier is represented by five normalized features: \textit{price
competitiveness}, \textit{inventory availability}, \textit{financial stability}
(account balance), and two geographic indicators (\textit{America} and
\textit{Europe}). \system infers sum-based bounds for these features by
observing three manually provided example packages representing distinct
strategies: Conservative, Price-Focused, and Balanced. The optimization
objective is to maximize the aggregate utility of cost, availability, and
stability: $Obj = \sum (f_{price} + f_{avail} + f_{bal})$. We compare \system
against a \textit{Greedy} baseline, which selects the top-$k$ suppliers by raw
objective score while ignoring constraints, a \textit{Cluster} baseline, which
uses K-means clustering over the attribute space and selects the
highest-scoring supplier from each cluster, and a \textit{Random} baseline. 

\smallskip\noindent{\textbf{\dataset}~\cite{subsume}.} \dataset is a dataset for
the evaluation of subjective summary extraction systems. The dataset contains
\numtriplets \textit{(document, intent, summary)} triplets over the 50 Wikipedia
pages for US states, with \numintents intents of varying subjectivity, provided
by \numpeople individuals over Mechanical Turk. ``Intent'' is the underlying
question motivating the creation of a summary. In order to evaluate the
effectiveness of a \emph{personalized} automatic summarization system, given an
intent, we need a few snippets where a subset of the snippets are used as
examples provided to the system and the rest are used to evaluate the snippets
that the system produces. Accordingly, \dataset includes 275 unique (user,
intent) pairs, each contributing 8 different, manually curated snippets. Five of
the snippets are used as example user-inputs for the system to derive a
user-intent model, and the remaining three are held-out as a test-set to
evaluate the system's ability to capture user-summarization-intent using the
five examples. In this work, we explore ten random splits between the example
and test sets for each user in the dataset. The ChatGPT-4o comparison
(Table~\ref{tab:sbert_similarity_chatgpt4o}) instead splits each of its two
intents into three example states and five test states, used by both \system
and ChatGPT-4o (Table~\ref{tab:appendix_gpt4o_per_state}).

\smallskip\noindent{\textbf{\cnndm}.} The \cnndm dataset provides
snippets in the form of human-written highlights~\cite{hermann2015teaching}.
However, because these highlights are often abstractive they cannot be directly
used for extractive tasks. To adapt this dataset for sentence-level extractive
summarization, we employed ChatGPT-4o to align the abstractive snippets with
their source articles. The model was tasked with identifying the indices of the
article sentences that most closely correlate with the reference highlights. To
ensure the resulting extractive snippets remained concise and useful, we imposed
two structural constraints: the selection could not exceed 10 sentences or 30\%
of the original article's sentence count. These constraints prevent the
inclusion of excessive or redundant text, forcing the model to prioritize
high-density semantic information.

\section{Baseline Details}
\label{appx:baseline_details}
\noindent{\textbf{\SBERT}}~\cite{reimers-2019-sentence-bert} is a widely used
sentence embedding model that generates semantically meaningful vector
representations of sentences. We use \SBERT as a simple baseline for
example-driven text snippet extraction for selecting sentences that are
semantically similar to the example snippets provided by the user based on
cosine similarity. We select a snippet using top-$k$ high-scoring sentences in
the document, where $k$ is simply the average number of sentences across the
examples.

\smallskip\noindent{\textbf{\sudocu}}~\cite{sudocu} is our prior example-driven
snippet extraction system. It models topics with Latent Dirichlet
Allocation~\cite{BleiNJ03} instead of \system's topic model: we fit a 10-topic
LDA model on the \dataset sentences with \sudocu's original settings and use each
sentence's topic distribution. \sudocu synthesizes per-topic bounds and a length
bound ($\pm 25\%$ of the examples' average length) from the examples, solves an
ILP over the whole target document that maximizes a word-frequency sentence
score, and relaxes infeasible topic bounds with a fixed-step multiplier.

\smallskip\noindent{\textbf{\presumm}}~\cite{liu2019text} is a widely adopted
extractive summarization model. By default, \presumm solely considers the input
document and extracts important sentences based on general document content,
without accommodating specific user inputs. During its training, it learns to
identify important sentences by optimizing for word overlap (e.g., ROUGE scores)
with gold-standard snippets. To adapt \presumm for our example-driven text
snippet extraction task, we first pre-filter the target document, retaining only
the candidate sentences that are semantically similar to the user-provided
example snippets. We then feed this pre-filtered document into \presumm. The
intuition behind this approach is that by restricting the input space to
sentences semantically aligned with the examples, we effectively bias \presumm
towards extracting a snippet that reflects the user's specific intent. In our
implementation, we use the \presumm model pre-trained on the \cnndm
dataset~\cite{hermann2015teaching}.

\smallskip\noindent{\textbf{\memsumm}}~\cite{gu2021memsum} is a state-of-the-art
reinforcement-learning model for long-document extractive summarization. Unlike
\presumm, \memsumm is history-aware: it conditions each extraction on the
sentences already selected, producing snippets with greater semantic consistency
and reduced redundancy. Sentence selection is trained to maximize word overlap
with the gold-standard snippet. In this work, we use the model pre-trained on
the GovReport dataset~\cite{govreport}. Similarly to \presumm, we first
pre-filter each target document using the same \SBERT-based approach.

\smallskip\noindent{\textbf{ChatGPT-4o}}~\cite{openai2023gpt4} is a large
language model from OpenAI. We use it as a retrieval-augmented baseline for
example-driven text snippet extraction to evaluate how well it can understand
user intent and learn extraction patterns from a few examples. We use a file
attachment method to provide the model with the target document and example
snippets, and we prompt the model to generate a snippet that captures the intent
expressed in the examples.

\smallskip\noindent{\textbf{LLM-agent}} is a fully LLM-driven baseline that
performs every stage of the pipeline---bound synthesis, retrieval, and
relaxation---itself, with no ILP solver invoked. Since a target document is
generally too long to fit in context, we first retrieve the top-$k$ most
relevant chunks via SBERT cosine similarity between the user's intent and each
chunk's mean sentence embedding, with each candidate sentence tagged by its real
sentence ID. The LLM is then shown these retrieved passages as a target document
(and the example bundles) and, in a single response, must (1) identify the most
salient recurring themes as topics and write a \code{[min, max]} bound for each,
plus a bound on the number of sentences, and (2) directly select the sentence
IDs---only from those shown to it---that jointly satisfy those bounds and
maximize the same quality function \system's ILP objective uses. We check the
LLM's own selection against the bounds it just wrote; on a violation, we tell it
exactly which bound(s) it broke and ask it to relax appropriately and re-select,
up to 3 self-critique rounds, mirroring \system's relax-and-retry loop (each
round allows up to 2 retries for invalid JSON or unseen sentence IDs). If the
model's selection is still infeasible after the final round, we keep that
round's answer rather than discard it and mark it infeasible.

\begin{sloppypar}
\smallskip\noindent{\textbf{LLM-\system}} is a hybrid baseline that replaces
only \system's bound-synthesis step (Section~\ref{sec:intent}) with an LLM call,
while retrieval and bound relaxation remain unchanged, handled by \system's ILP.
It runs in three stages. First, it reuses the same retrieved chunks as
LLM-agent, so both LLM baselines see the same view of each target document.
Second, the LLM (Llama-3.3-70B-Instruct-FP8) is shown the 5 example bundles
together with the retrieved passages of the target document and asked to output
a per-topic \code{[min, max]} bound in a single call per target document, with
up to 2 retries if the output is not valid JSON (retries count toward its
latency and token cost); the bound on the number of sentences is \system's own
(Appendix~\ref{appx:implementation}). Unlike LLM-agent, it neither selects
sentences nor relaxes its own bounds. Third, these frozen bounds are handed to
\system unmodified (\code{bounds=<LLM output>}), which executes the resulting
PaQL query with its ILP and relaxes it via its usual symmetric-relaxation loop
(Section~\ref{sec:bound_relaxation}) if infeasible. We report the LLM's
bound-generation latency as an added per-query cost on top of \system's own
retrieval time.
\end{sloppypar}

\section{LLM-based baselines as direct snippet extractors}
\label{appx:llm_baselines_extraction}

\newtcolorbox{promptbox}[1]{
    colback=gray!5,       
    colframe=gray!50,     
    arc=2mm,              
    boxrule=0.5pt,        
    left=6pt, right=6pt, top=8pt, bottom=8pt,
    fontupper=\small,
    title=#1,             
    colbacktitle=gray!20, 
    coltitle=black,       
    fonttitle=\bfseries,
    enhanced,
    breakable,            
    attach boxed title to top left={yshift=-2mm, xshift=2mm},
    boxed title style={sharp corners=south, size=small, colframe=gray!50}
}

\subsubsection*{SubSumE}
For the initial set of ChatGPT-4o few-shot experiments, we conducted inference
on each user input file, where each file corresponds to a single intent
question. The process involved uploading the relevant Wikipedia text files for
the states alongside a structured prompt to instruct the model. For instance,
one of the prompts used is as follows:
\begin{promptbox}{Few-Shot Prompt for SubSumE}
    \texttt{\small I am providing you with eight text documents. For three of
		them, Delaware, Idaho, and Wisconsin, I will give you example
		summaries. From these, you need to understand my summarization intent
		and learn the summarization pattern from the given examples. Then
		summarize the remaining five documents based on the learned pattern.
		Make sure to use extractive summarization: select only the original
		sentences from the given text without generalizing. You can perform
		detailed summarization by selecting relevant sentences from the text
		files. Return the final summaries as a combined paragraph.}
\end{promptbox}

\begin{table}[t]
	\centering
    \resizebox{\columnwidth}{!}{
	\begin{tabular}{p{4cm} p{6cm}}
		\toprule
		\textbf{Intent in \dataset} & \textbf{ChatGPT-4o Predicted Intent} \\
		\midrule
		How is the government structured in this state? & What is the governmental and legislative structure of \{state\}? \\[6mm]
		How is the weather of the state? & What is the climate like in \{state\}? \\[3mm]
		What drives the economy in this state? & What are the economic drivers and geographical characteristics of different U.S. states? \\
		\bottomrule
	\end{tabular}}
	\vspace{2mm}
	\caption{Comparison of original intents in the \dataset dataset and intents predicted by ChatGPT-4o using file attachment for context.}

	\label{tab:appendix_intent_comparison_chatgpt4o}
\end{table}

\begin{table}[t]
    \centering
    \small
    \begin{tabular}{@{} l l r}
        \toprule
        \textbf{Example States} & \textbf{Test State} & \textbf{Semantic Similarity} \\ 
        \midrule
        \multicolumn{3}{@{}l}{\textbf{Task 1:} \textit{What about this state's arts and culture attracts you the most?}} \\
        \hline
        \addlinespace[0.3em]
        \multirow{5}{*}{\parbox{3cm}{Delaware,\\ Idaho,\\Wisconsin}} & Colorado & 0.662 \\
                                   & New Jersey & 0.633 \\
                                   & Utah & 0.742 \\
                                   & Virginia & 0.560 \\
                                   & West Virginia & 0.598 \\
        \cmidrule(lr){2-3}
                                   & \textbf{Average} & \textbf{0.639} \\
        
        \midrule
        \multicolumn{3}{@{}l}{\textbf{Task 2:} \textit{What are some of the most interesting things about this state?}} \\
        \hline
        \addlinespace[0.3em]
        \multirow{5}{*}{\parbox{3cm}{New Hampshire,\\ Hawaii,\\Washington}} & Arkansas & 0.693 \\
                                   & California & 0.737 \\
                                   & Georgia & 0.694 \\
                                   & Kansas & 0.533 \\
                                   & Michigan & 0.587 \\
        \cmidrule(lr){2-3}
                                   & \textbf{Average} & \textbf{0.649} \\
        \bottomrule
    \end{tabular}
    \caption{Semantic similarity for ChatGPT-4o across focused intent text snippet extraction tasks, per test state.}
    \label{tab:appendix_gpt4o_per_state}

	\centering
    \resizebox{\columnwidth}{!}{
	\begin{tabular}{@{}cl>{\centering\arraybackslash}p{1.7cm}>{\centering\arraybackslash}p{2cm}>{\centering\arraybackslash}p{2cm}@{}}
		\toprule
		\multicolumn{2}{c}{} & \textbf{\system} & \textbf{ChatGPT-4o (prompt)} & \textbf{ChatGPT-4o (few-shot)} \\
		\midrule
		\multirow{3}{*}{\textbf{ROUGE-1}} 
		& \textbf{Precision} & 0.2714 & 0.6699 & 0.6655 \\
		& \textbf{Recall}    & 0.4769 & 0.5155 & 0.5108 \\
		& \textbf{F1}        & 0.3235 & 0.5504 & 0.5483 \\
		\midrule
		\multirow{3}{*}{\textbf{ROUGE-2}} 
		& \textbf{Precision} & 0.1364 & 0.5389 & 0.5304 \\
		& \textbf{Recall}    & 0.2318 & 0.4106 & 0.4021 \\
		& \textbf{F1}        & 0.1590 & 0.4403 & 0.4332 \\
		\midrule
		\multirow{3}{*}{\textbf{ROUGE-L}} 
		& \textbf{Precision} & 0.1977 & 0.5873 & 0.5769 \\
		& \textbf{Recall}    & 0.3445 & 0.4504 & 0.4405 \\
		& \textbf{F1}        & 0.2335 & 0.4814 & 0.4734 \\
		\midrule
		\multicolumn{2}{c}{\textbf{Semantic Similarity}} & 0.6272 & 0.7965 & 0.7853 \\
		\bottomrule
	\end{tabular}}
	\caption{Focused text snippet extraction with generic intent on \system, ChatGPT-4o (prompt), and ChatGPT-4o (few-shot) with ground truth snippets from the dataset's highlights on \cnndm dataset politics articles.
	}
	\label{tab:appendix_generic_snippet}
\end{table}

Here we first report the performance of ChatGPT-4o for predicting intents given
example snippets. In Table~\ref{tab:appendix_intent_comparison_chatgpt4o}, we
show the original intents in the \dataset dataset and the intents predicted by
ChatGPT-4o via an additional query such as ``Given these example summaries, what
is the summarization intent?''. The original intents are more specific and
focused on particular aspects of the states, while the ChatGPT-4o predicted
intents tend to be broader or more general. For example, the original intent of
``How is the government structured in this state?'' is focused on governmental
structure, whereas the ChatGPT-4o predicted ``What is the governmental and
legislative structure of \{state\}?'', which is more general and encompasses
both governmental and legislative aspects.

Table~\ref{tab:appendix_gpt4o_per_state} shows the semantic similarity for ChatGPT-4o
across focused text snippet extraction tasks, per test state. For Task 1 (``What
about this state's arts and culture attracts you the most?''), the average
semantic similarity across the five test states is 0.639, with Utah having the
highest similarity of 0.742 and Virginia having the lowest similarity of 0.560.
For Task 2 (``What are some of the most interesting things about this state?''),
the average semantic similarity across the five test states is 0.649, with
California having the highest similarity of 0.737 and Kansas having the lowest
similarity of 0.533. The results give us insight into how ChatGPT-4o's understanding of
user intent generally varies across different states, which may be influenced by
the specific content and characteristics of each state document.

\subsubsection*{\cnndm}
For the \cnndm dataset, we used a similar few-shot prompting approach
with ChatGPT-4o as \dataset. The prompt was designed to guide the model in learning
snippet extraction patterns from a few examples and applying them to new articles.
The prompt structure is as follows:

\begin{promptbox}{Few-Shot Prompt for \cnndm}
    \textbf{System Instructions:}\\
    \texttt{\small You are an advanced summarization assistant. I will provide some examples of original articles with its summarization. Please try to learn the pattern of summarization and use it to generate a similar summarization on new given content.} \\
    
    \textbf{Few-Shot Examples:}\\
    \texttt{\small \{\{example\_article\}\}}
    
    \hrulefill 
    
    \textbf{User Input:}\\
    \texttt{\small The given new article is \{\{test\_article\}\}. Please return a list of zero-based sentence indices from the original text that best align with the intent of the summary. Like [1,2].} \\
    
    \textbf{Output Format:}\\
    \texttt{\small [index\_0, index\_1, ...]}
\end{promptbox}

\begin{table}[t]
\centering
\small
\centering
\resizebox{0.99\columnwidth}{!}{
\begin{tabular}{@{}llcccc@{}}
\toprule
\textbf{Category} & \textbf{System} & \multicolumn{1}{c}{\textbf{ROUGE-1}$\uparrow$} & \multicolumn{1}{c}{\textbf{ROUGE-2}$\uparrow$} & \multicolumn{1}{c}{\textbf{ROUGE-L}$\uparrow$} & \multicolumn{1}{c@{}}{\makecell{\textbf{Semantic}\\\textbf{Similarity}}$\uparrow$} \\
\midrule
\multirow{2}{*}{\textbf{Politics}} 
& ChatGPT-4o  & \textbf{0.5483} & \textbf{0.4332} & \textbf{0.4734} & \textbf{0.7853} \\
& \system  & 0.3235 & 0.1590 & 0.2335 & 0.6272 \\
\midrule
\multirow{2}{*}{\textbf{Crime}} 
& ChatGPT-4o  & \textbf{0.4844} & \textbf{0.3513} & \textbf{0.4019} & \textbf{0.7601} \\
& \system  & 0.3246 & 0.1499 & 0.2204 & 0.5829 \\
\midrule
\multirow{2}{*}{\textbf{Sports}} 
& ChatGPT-4o  & \textbf{0.5170} & \textbf{0.4219} & \textbf{0.4548} & \textbf{0.8022} \\
& \system  & 0.3148 & 0.1743 & 0.2330 & 0.6255 \\
\midrule
\multirow{2}{*}{\textbf{Lifestyle}} 
& ChatGPT-4o  & \textbf{0.5060} & \textbf{0.3812} & \textbf{0.4294} & \textbf{0.7949} \\
& \system  & 0.3136 & 0.1304 & 0.1982 & 0.6016 \\
\bottomrule
\end{tabular}}
\vspace{2mm}
\caption{\small ChatGPT-4o outperforms \system in generic intent alignment
across CNN/DailyMail news categories.
}
\label{tab:cnn_daily_mail}
\end{table}

\setcounter{dbltopnumber}{1}
\begin{table*}[tp]
	\centering
    \resizebox{0.9\textwidth}{!}{
	\begin{tabular}{%
			cl
			>{\centering\arraybackslash}p{1.5cm}
			>{\centering\arraybackslash}p{2.5cm}
			>{\centering\arraybackslash}p{1.5cm}
			>{\centering\arraybackslash}p{2.5cm}
			>{\centering\arraybackslash}p{1.5cm}
			>{\centering\arraybackslash}p{2.5cm}
		}
		\toprule
		\multicolumn{2}{c}{} 
		& \multicolumn{2}{>{\centering\arraybackslash}m{4.2cm}}{\textbf{``Summarize the crime mentioned in this article.''}}
		& \multicolumn{2}{>{\centering\arraybackslash}m{4.2cm}}{\textbf{``Who are the suspects and victims in this event?''}}
		& \multicolumn{2}{>{\centering\arraybackslash}m{4.2cm}}{\textbf{``Summarize the locations where the crime was committed.''}} \\
		\cmidrule(lr){3-4} \cmidrule(lr){5-6} \cmidrule(lr){7-8}
		\multicolumn{2}{c}{} 
		& \textbf{\system} 
		& \textbf{ChatGPT-4o (few-shot)}
		& \textbf{\system} 
		& \textbf{ChatGPT-4o (few-shot)}
		& \textbf{\system} 
		& \textbf{ChatGPT-4o (few-shot)}\\
		\midrule
		\multirow{3}{*}{\textbf{ROUGE-1}} 
		& \textbf{Precision} & 0.5126 & 0.6831 & 0.4562 & 0.6901 & 0.4016 & 0.6825 \\
		& \textbf{Recall}    & 0.4489 & 0.6043 & 0.4215 & 0.5855 & 0.3783 & 0.4260 \\
		& \textbf{F1}        & 0.4608 & 0.6206 & 0.4207 & 0.5999 & 0.3686 & 0.5021 \\
		\midrule
		\multirow{3}{*}{\textbf{ROUGE-2}} 
		& \textbf{Precision} & 0.3009 & 0.5586 & 0.2493 & 0.5781 & 0.2110 & 0.5010 \\
		& \textbf{Recall}    & 0.2523 & 0.4928 & 0.2198 & 0.4780 & 0.1862 & 0.3116 \\
		& \textbf{F1}        & 0.2644 & 0.5066 & 0.2235 & 0.4946 & 0.1861 & 0.3684 \\
		\midrule
		\multirow{3}{*}{\textbf{ROUGE-L}} 
		& \textbf{Precision} & 0.3526 & 0.5944 & 0.3168 & 0.6079 & 0.2904 & 0.5641 \\
		& \textbf{Recall}    & 0.3004 & 0.5256 & 0.2850 & 0.5088 & 0.2660 & 0.3498 \\
		& \textbf{F1}        & 0.3124 & 0.5397 & 0.2872 & 0.5242 & 0.2619 & 0.4135 \\
		\midrule
		\multicolumn{2}{c}{\textbf{Semantic Similarity}}
		& 0.7435 
		& 0.8307 
		& 0.7021 
		& 0.8089 
		& 0.6117 
		& 0.7552 \\
		\bottomrule
	\end{tabular}}
	\caption{Focused text snippet extraction by \system and ChatGPT-4o (few-shot) on three focused intents for crime articles of \cnndm, using LLM-selected ground-truth snippets.}
    \vspace{-3mm}
	\label{tab:focused_system_llm}
\vspace{6mm}
    \centering
    \resizebox{0.95\textwidth}{!}{
    \setlength{\tabcolsep}{3pt}
	\begin{tabular}{cl>{\centering\arraybackslash}p{1.5cm}>{\centering\arraybackslash}p{4cm}>{\centering\arraybackslash}p{4cm}>{\centering\arraybackslash}p{4.9cm}}
		\toprule
		& & \textbf{Generic} & \textbf{``Summarize the crime mentioned in this article.''} & \textbf{``Who are the suspects and victims in this event?''} & \textbf{``Summarize the locations where the crime was committed.''} \\
		\midrule
		\multirow{3}{*}{\textbf{ROUGE-1}}
		& \textbf{Precision} & 0.3068 & 0.5126 & 0.4562 & 0.4016 \\
		& \textbf{Recall}    & 0.4029 & 0.4489 & 0.4215 & 0.3783 \\
		& \textbf{F1}        & 0.3246 & 0.4608 & 0.4207 & 0.3686 \\
		\midrule
		\multirow{3}{*}{\textbf{ROUGE-2}}
		& \textbf{Precision} & 0.1395 & 0.3009 & 0.2493 & 0.2110  \\
		& \textbf{Recall}    & 0.1853 & 0.2523 & 0.2198 & 0.1862 \\
		& \textbf{F1}        & 0.1499  & 0.2644 & 0.2235 & 0.1861 \\
		\midrule
		\multirow{3}{*}{\textbf{ROUGE-L}}
		& \textbf{Precision} & 0.2065 & 0.3526 & 0.3168 & 0.2904 \\
		& \textbf{Recall}    & 0.2746 & 0.3004 & 0.2850  & 0.2660  \\
		& \textbf{F1}        & 0.2204 & 0.3124 & 0.2872 & 0.2619 \\
		\midrule
		\multicolumn{2}{c}{\textbf{Semantic Similarity}}
		& 0.5829 & 0.7435 & 0.7021 & 0.6117 \\
		\bottomrule
	\end{tabular}}
	\caption{\system's performance on the generic intent and three category-level focused intents for the crime category of \cnndm.}
    \vspace{-5mm}
	\label{tab:crime_summary}
\end{table*}

\begin{table*}[t]
    \centering
    \resizebox{\textwidth}{!}{
	\begin{tabular}{cl>{\centering\arraybackslash}p{1.5cm}>{\centering\arraybackslash}p{4.5cm}>{\centering\arraybackslash}p{4.5cm}>{\centering\arraybackslash}p{4.5cm}}
		\toprule
		& & \textbf{Generic} & \textbf{``Summarize the politicians and celebrities mentioned.''} & \textbf{``Summarize the contributions or impacts of this political event or statement.''} & \textbf{``What are the key issues or debates highlighted in this political event?''} \\
		\midrule
		\multirow{3}{*}{\textbf{ROUGE-1}}
		& \textbf{Precision} & 0.2714 & 0.4699 & 0.4480  & 0.5055 \\
		& \textbf{Recall}    & 0.4769 & 0.3970  & 0.5131 & 0.4703 \\
		& \textbf{F1}        & 0.3235 & 0.3920  & 0.4626 & 0.4684 \\
		\midrule
		\multirow{3}{*}{\textbf{ROUGE-2}}
		& \textbf{Precision} & 0.1364 & 0.2807 & 0.2686 & 0.2975 \\
		& \textbf{Recall}    & 0.2318 & 0.2076 & 0.3019 & 0.2617 \\
		& \textbf{F1}        & 0.1590 & 0.2125 & 0.2751 & 0.2669 \\
		\midrule
		\multirow{3}{*}{\textbf{ROUGE-L}}
		& \textbf{Precision} & 0.1977 & 0.3422 & 0.3126 & 0.3511 \\
		& \textbf{Recall}    & 0.3445 & 0.2675 & 0.3600   & 0.3146 \\
		& \textbf{F1}        & 0.2335 & 0.2699 & 0.3233 & 0.3186 \\
		\midrule
		\multicolumn{2}{c}{\textbf{Semantic Similarity}}
		& 0.6272 & 0.6710  & 0.7454 & 0.7658 \\
		\bottomrule
	\end{tabular}}
	\caption{\system's performance on the generic intent and three category-level focused intents for the politics category of \cnndm.}
    \vspace{-3mm}
	\label{tab:politics_summary}
\vspace{6mm}
    \centering
    \resizebox{\textwidth}{!}{
	\begin{tabular}{cl>{\centering\arraybackslash}p{1.5cm}>{\centering\arraybackslash}p{4.2cm}>{\centering\arraybackslash}p{4.3cm}>{\centering\arraybackslash}p{4.8cm}}
		\toprule
		& & \textbf{Generic} & \textbf{``Summarize the sports teams or athletes involved.''} & \textbf{``Summarize the match results or the event details.''} & \textbf{``Where is the stadium or location hosting the sports match event.''} \\
		\midrule
		\multirow{3}{*}{\textbf{ROUGE-1}}
		& \textbf{Precision} & 0.3148 & 0.4544 & 0.5007 & 0.3875 \\
		& \textbf{Recall}    & 0.3886 & 0.5049 & 0.4372 & 0.2503 \\
		& \textbf{F1}        & 0.3148 & 0.4675 & 0.4488 & 0.2828 \\
		\midrule
		\multirow{3}{*}{\textbf{ROUGE-2}}
		& \textbf{Precision} & 0.1823 & 0.2903 & 0.3285 & 0.2407 \\
		& \textbf{Recall}    & 0.2098 & 0.3095 & 0.2670  & 0.1407 \\
		& \textbf{F1}        & 0.1743 & 0.2932 & 0.2833 & 0.1658 \\
		\midrule
		\multirow{3}{*}{\textbf{ROUGE-L}}
		& \textbf{Precision} & 0.2405 & 0.3410  & 0.3872 & 0.3115 \\
		& \textbf{Recall}    & 0.2852 & 0.3732 & 0.3273 & 0.1863 \\
		& \textbf{F1}        & 0.2330  & 0.3486 & 0.3410  & 0.2176 \\
		\midrule
		\multicolumn{2}{c}{\textbf{Semantic Similarity}}
		& 0.6255 & 0.7552 & 0.7250  & 0.6476 \\
		\bottomrule
	\end{tabular}}
	\caption{\system's performance on the generic intent and three category-level focused intents for the sports category of \cnndm.}
    \vspace{-3mm}
	\label{tab:sports_summary}
\end{table*}

\begin{table*}[t]
    \centering
    \resizebox{\textwidth}{!}{
	\begin{tabular}{cl>{\centering\arraybackslash}p{1.2cm}>{\centering\arraybackslash}p{4cm}>{\centering\arraybackslash}p{4.8cm}>{\centering\arraybackslash}p{5.4cm}}
		\toprule
		& & \textbf{Generic} & \textbf{``What are the main lifestyle trends or topics highlighted in the article?''} & \textbf{``Summarize the key tips or advice offered for enhancing personal well-being and daily life.''} & \textbf{``What significant insights or stories does the article provide about cultural practices or personal experiences?''} \\
		\midrule
		\multirow{3}{*}{\textbf{ROUGE-1}}
		& \textbf{Precision} & 0.3056 & 0.5585 & 0.4078 & 0.6109 \\
		& \textbf{Recall}    & 0.3554 & 0.4031 & 0.3706 & 0.4331 \\
		& \textbf{F1}        & 0.3136 & 0.4533 & 0.3638 & 0.4931 \\
		\midrule
		\multirow{3}{*}{\textbf{ROUGE-2}}
		& \textbf{Precision} & 0.1240  & 0.3291 & 0.2250  & 0.4043 \\
		& \textbf{Recall}    & 0.1519 & 0.2235 & 0.1885 & 0.2735 \\
		& \textbf{F1}        & 0.1304 & 0.2596 & 0.1925 & 0.3178 \\
		\midrule
		\multirow{3}{*}{\textbf{ROUGE-L}}
		& \textbf{Precision} & 0.1919 & 0.3868 & 0.2798 & 0.4454 \\
		& \textbf{Recall}    & 0.2268 & 0.2726 & 0.2450  & 0.3032 \\
		& \textbf{F1}        & 0.1982 & 0.3105 & 0.2445 & 0.3514 \\
		\midrule
		\multicolumn{2}{c}{\textbf{Semantic Similarity}}
		& 0.6016 & 0.7495 & 0.6730 & 0.7832 \\
		\bottomrule
	\end{tabular}}
	\caption{\system's performance on the generic intent and three category-level focused intents for the lifestyle category of \cnndm.}
    \vspace{-5mm}
	\label{tab:lifestyle_summary}
\end{table*}

Here, we provide the detailed results for the generic text snippet extraction
and customized focused text snippet extraction tasks on the \cnndm dataset, as
well as the semantic similarity scores for each focused intent text snippet
extraction task. For ChatGPT-4o, we report the results for both prompt-only and
few-shot prompting approaches. The prompt-only approach uses a single prompt to
guide the model's generation, while the few-shot prompting approach includes
three examples for each intent, which are randomly selected from the training
set of the \cnndm dataset.

\paratitle{Generic intent} 
Table~\ref{tab:appendix_generic_snippet} specifically shows in politics articles
that ChatGPT-4o (both prompt and few-shot) significantly outperforms \system
across all ROUGE metrics and Semantic Similarity, demonstrating its general
capability in understanding user intent and generating snippets for generic text
snippet extraction tasks. Interestingly, the prompt-only approach performs
slightly better than the few-shot approach in terms of ROUGE scores and Semantic
Similarity, which may suggest that providing multiple examples for few-shot
prompting does not necessarily lead to better performance in this case.

In Table~\ref{tab:cnn_daily_mail}, we compare against ChatGPT-4o (few-shot) across four categories
(Politics, Crime, Sports, \& Life\-style). ChatGPT-4o leads on all ROUGE and
Semantic Similarity metrics in every category (Table~\ref{tab:cnn_daily_mail});
on Crime, for instance, ChatGPT-4o achieves ROUGE-L 0.4019 and Semantic
Similarity 0.7601, while 0.2204 and 0.5829 for \system (similar gaps elsewhere).
This is expected since generic news snippets have broad intent that
example-driven retrieval cannot capture~\cite{DBLP:journals/tacl/ZhangLDLMH24},
marking the regime where \system is not the appropriate tool.

\looseness-1 Table~\ref{tab:focused_system_llm} reports the detailed results for
the focused text snippet extraction tasks on the \cnndm dataset contrasting
\system and ChatGPT-4o (few-shot). Crime articles from \cnndm contain
multi-aspect descriptions, so we use three focused crime intents: ``Summarize
the crime mentioned in this article.'', ``Who are the suspects and victims in
this event?'', and ``Summarize the locations where the crime was committed.''
Although focused, these intents remain relatively generic compared with the
\dataset intents in the main paper (e.g., ``What about this state's arts and
culture attracts you the most?''). We observe that ChatGPT-4o
(few-shot) generally outperforms \system across all ROUGE metrics and semantic
similarity on these intents, suggesting that ChatGPT-4o
is more effective at retrieving snippets aligned with broad user intents than
\system. However, \system still shows competitive performance relative to its
own generic-intent baseline for crime articles (Table~\ref{tab:crime_summary}),
suggesting that \system remains effective at retrieving snippets for specific
intents, even though those intents are still relatively broad.

Tables~\ref{tab:crime_summary} to~\ref{tab:lifestyle_summary} report the
detailed results for the focused text snippet extraction tasks on the \cnndm
dataset with our manually crafted focused intents for crime, politics, sports,
and lifestyle categories. These intents are written at the category level and,
like those in Table~\ref{tab:focused_system_llm}, remain relatively generic
compared with the \dataset intents in the main paper. Across all categories,
the results show that \system's performance is generally better for these
focused intents than for the generic intent, suggesting that \system becomes
more effective as the intent becomes more specific, even at this relatively
coarse level.

\section{LLM-based baselines as PaQL generators}
\label{appx:llm_baselines_paql}
\subsubsection*{LLM-\system (bound synthesis)}
For LLM-\system, the LLM is only asked to synthesize the bounds from the 5
example bundles and the retrieved passages of the target document; retrieval and
relaxation are handled entirely by \system afterward. The prompt is as follows:

\begin{promptbox}{LLM-\system prompt for bound synthesis}
    \textbf{System Instructions:}\\
    \texttt{\small You are standing in for a PaQL (Package Query Language)
	query-authoring engine. PaQL specifies a desired package (bundle) of
	sentences as a set of constraints instead of an explicit selection: SELECT
	PACKAGE(*) FROM \{\{target\_doc\_name's sentences\}\} WHERE \{\{your
	per-topic bounds\}\} MAXIMIZE \{\{a fixed per-sentence quality score,
	summed over the bundle\}\}. Each sentence's per-topic relevance score is a
	value between 0 and 1 that behaves like a soft topic-membership
	distribution. The bundle's size is set by the solver, not by you. The
	MAXIMIZE clause is fixed and scores each candidate sentence s as score(s) =
	BS(s) + TFIDF(s) + LPR(s) + TS(s), where BS(s) = cosine(SBERT\_embedding(s),
	mean(SBERT\_embedding(examples))); TFIDF(s) = sum of TF-IDF weights of s's
	words, divided by the maximum (TF-IDF model fit on the examples); LPR(s) =
	sum over words w in s of log10(P(w\textbar{}examples) / P(w\textbar{}target
	doc)) (negative terms only, then min-max scaled); and TS(s) =
	cosine(topic\_vector(s), mean(topic\_vector(examples))). A separate solver---not you---takes the WHERE bounds you
	write, searches the target document under the fixed MAXIMIZE objective, and
	widens whichever bounds turn out to be unsatisfiable; you are not asked to
	select sentences or to relax anything yourself. You are not told in advance
	what the topics are---find them yourself from the recurring themes you
	notice below, and order them by prominence (topic\_0 = most prominent,
	topic\_1 = next, and so on).}

    \textbf{Examples:}\\
    \texttt{\small \{\{example\_bundles\}\}}

    \hrulefill

    \textbf{Retrieved Passages:}\\
    \texttt{\small The target document ``\{\{target\_doc\_name\}\}'' is too long
	to include in full here, so below are the passages retrieved from it as most
	relevant to the intended bundle---they are your only view into this
	document's actual content. \{\{retrieved\_context\}\}}

    \hrulefill

    \textbf{Instructions:}\\
    \texttt{\small 1. Using the example bundles above and the retrieved
	passages, identify the most salient recurring themes and decide what topical
	focus the returned bundle should have. 2. For each theme, in prominence
	order, write a [min, max] bound on that theme's total relevance score summed
	over the bundle's sentences. 3. Output ONLY a single valid JSON
	object---no markdown, no explanation.}

    \textbf{Output Format:}\\
    \texttt{\small \{"topic\_0": [min, max], ..., "topic\_9": [min, max]\}}
\end{promptbox}

\subsubsection*{LLM-agent (bound synthesis \& selection)}
For LLM-agent, the LLM stands in for the entire PaQL pipeline: it is shown
examples, must invent its own topic themes, write bounds for them, and directly
select sentence IDs satisfying those bounds while maximizing \system's quality
function. On a self-detected bound violation, it is re-prompted to relax and
re-select, up to 3 rounds:

\begin{promptbox}{LLM-agent prompt for PaQL synthesis and execution}
    \textbf{System Instructions:}\\
    \texttt{\small You are standing in for a PaQL (Package Query Language)
	 query-authoring-AND-execution engine. PaQL specifies a desired package
	 (bundle) of sentences as a set of constraints instead of an explicit
	 selection: SELECT PACKAGE(*) FROM \{\{target\_doc\_name's sentences\}\}
	 WHERE \{\{your per-topic and length bounds\}\} MAXIMIZE \{\{a fixed
	 per-sentence quality score, summed over the bundle\}\}. The WHERE clause
	 bounds each topic's total relevance score over the chosen sentences, and
	 separately bounds the bundle's size (sentence count). Each sentence's
	 per-topic relevance score is a value between 0 and 1 that behaves like a
	 soft topic-membership distribution (one sentence's scores across all topics
	 roughly sum to 1). Normally a solver takes your bounds and, under a fixed
	 objective, returns whichever feasible bundle scores highest---not just any
	 bundle that satisfies the bounds. That objective scores each candidate
	 sentence s as score(s) = BS(s) + TFIDF(s) + LPR(s) + TS(s), where BS(s) =
	 cosine(SBERT\_embedding(s), mean(SBERT\_embedding(examples))); TFIDF(s) =
	 sum of TF-IDF weights of s's words, divided by the maximum (TF-IDF model fit
	 on the examples); LPR(s) = sum over words w in s of
	 log10(P(w\textbar{}examples) / P(w\textbar{}target doc)) (negative terms
	 only, then min-max scaled); and TS(s) = cosine(topic\_vector(s),
	 mean(topic\_vector(examples))). Here, instead of writing bounds for a
	 separate solver to execute, you must play every role at once: write the
	 WHERE bounds yourself, AND directly select the sentences that both satisfy
	 them and are the ones a MAXIMIZE step would have picked. You are not told
	 in advance what the topics are---find them yourself from the recurring
	 themes you notice below, and order them by prominence (topic\_0 = most
	 prominent, topic\_1 = next, and so on).}

    \textbf{Examples:}\\
    \texttt{\small \{\{example\_bundles\}\}}

    \hrulefill

    \textbf{Retrieved Passages:}\\
    \texttt{\small The target document ``\{\{target\_doc\_name\}\}'' is too long
	to include in full here, so below are the passages retrieved from it as most
	relevant to the intended bundle---they are your only view into this
	document's actual content, and the ONLY sentences you may select from. Each
	sentence is tagged with its own sentence ID in [brackets].
	\{\{retrieved\_context\}\}}

    \hrulefill

    \textbf{Instructions:}\\
    \texttt{\small 1. Using the example bundles above (if any) and the retrieved
	passages, identify the most salient recurring themes and decide what topical
	focus, length, and style the returned bundle should have. 2. For each theme
	you identified, in prominence order, write a [min, max] bound on that
	theme's total relevance score summed over the bundle's sentences. Also write
	a [min, max] bound on the NUMBER OF SENTENCES the bundle should contain. 3. Select
	the sentence IDs---and ONLY sentence IDs shown in brackets above---that
	together form a bundle satisfying the bounds you just wrote. Do not invent
	sentence IDs. Your selection will be checked against the bounds you provide;
	if it does not satisfy them, you will be told which bound(s) were violated
	and asked to relax (widen) exactly those bounds and select again.}

    \textbf{Output Format:}\\
    \texttt{\small \{"topic\_0": [min, max], ..., "length": [min, max],
    "selected\_sids": [id\_0, id\_1, ...]\}}
\end{promptbox}

\setcounter{dbltopnumber}{2}
\section{Constraint-bundle alignment consistency}
\label{appx:consistency_analysis}

\smallskip\noindent\textbf{Why Hausdorff distance?} \looseness-1 Two \paql
queries that share the same objective but differ in their constraint bounds
$\Theta = \{\langle\lowerbound_j, \upperbound_j\rangle\}_{j=1}^K$ induce
different feasible regions. Because $\Theta$ bounds every topic independently,
its feasible region is an axis-aligned box: the Cartesian product of $K$
per-topic intervals, $\{z \in \mathbb{R}^K \mid \lowerbound_j \le z_j \le
\upperbound_j,\ 1 \le j \le K\}$. Given two such boxes, one per constraint
set, we want a single number that quantifies how far apart they are. We use
the Hausdorff distance~\cite{hausdorff}, which for two regions $Z$ and $Y$ is
defined as
\begin{equation*}
    d_H(Z, Y) = \max\Big\{ \sup_{z \in Z} \inf_{y \in Y} d(z,y),\ \ \sup_{y \in Y} \inf_{z \in Z} d(z,y) \Big\}.
\end{equation*}
For each point in $Z$, this measures the distance to its nearest point in
$Y$ ($\inf_{y \in Y} d(z,y)$), then takes the largest such value over every
point of $Z$ ($\sup_{z \in Z} \inf_{y \in Y} d(z,y)$); since this quantity is
not symmetric in $Z$ and $Y$, we compute it in both directions and take the
larger of the two.

\begin{sloppypar}
Since the feasible regions we compare are axis-aligned boxes rather than
arbitrary shapes, we choose the Manhattan distance for the point-to-point
distance $d(z,y) = \sum_{j=1}^K |z_j - y_j|$, as it is separable across
coordinates, matching the per-topic product structure of the boxes. This lets
us compute $d_H$ one topic at a time. Fix a topic $j$ and a point $z_j \in
[\lowerbound_j, \upperbound_j]$, and let $[\lowerbound_j', \upperbound_j']$
denote the other query's bounds for that topic. Its nearest point in
$[\lowerbound_j', \upperbound_j']$ depends on whether $z_j$ falls below,
inside, or above that interval, giving
\begin{equation*}
    \inf_{y_j \in [\lowerbound_j', \upperbound_j']} |z_j - y_j| = \max(0,\ \lowerbound_j' - z_j) + \max(0,\ z_j - \upperbound_j').
\end{equation*}
Taking the supremum of this expression over $z_j \in [\lowerbound_j,
\upperbound_j]$ is maximized at one of the interval's own endpoints, since
each term is piecewise linear and increases only as $z_j$ moves away from
$[\lowerbound_j', \upperbound_j']$. Writing $\delta_j(\Theta, \Theta') =
\max\big(\max(0, \lowerbound_j' {-} \lowerbound_j),\ \max(0, \upperbound_j {-}
\upperbound_j')\big)$ for how far $\Theta$'s bound on topic $j$ extends beyond
$\Theta'$'s, this supremum is
\begin{equation*}
    \sup_{z_j \in [\lowerbound_j, \upperbound_j]} \inf_{y_j} |z_j - y_j| = \delta_j(\Theta, \Theta'),
\end{equation*}
i.e., whichever side of $[\lowerbound_j, \upperbound_j]$ extends further
beyond $[\lowerbound_j', \upperbound_j']$, if either does. Summing this
per-topic quantity across all $K$ topics, by separability of the Manhattan
distance, gives $\sup_{z \in Z} \inf_{y \in Y} d(z,y)$; swapping the roles of
the two constraint sets gives $\sup_{y \in Y} \inf_{z \in Z} d(z,y) =
\sum_{j=1}^K \delta_j(\Theta', \Theta)$. The Hausdorff distance between two
\paql constraint sets therefore reduces to
\begin{equation*}
    d_H(\Theta, \Theta') = \max\Big\{ \textstyle\sum_{j=1}^K \delta_j(\Theta, \Theta'),\ \ \textstyle\sum_{j=1}^K \delta_j(\Theta', \Theta) \Big\}.
\end{equation*}
Intuitively, this reduces the Hausdorff distance to comparing the declared
bounds directly: for each topic, we measure how far one query's interval extends
beyond the other's on each side, and sum these overhangs across topics without
needing to reason about the boxes as geometric objects or enumerate points
inside them.
\end{sloppypar}

\begin{figure}[h]
  \centering
  \resizebox{0.60\linewidth}{!}{\includegraphics[width=0.8\linewidth]{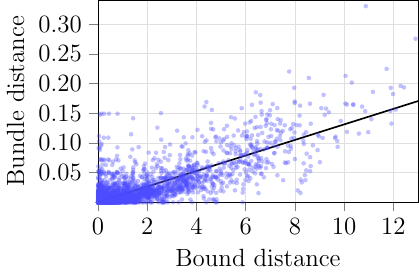}}
  \vspace{-4mm}
  \caption{Consistency of constraint shifts and bundle deviations.}
    \vspace{-4mm}
  \label{fig:consistency}
\end{figure}

\smallskip\noindent\textbf{Consistency evaluation.} \looseness-1 To evaluate
whether \system's synthesized bounds produce consistent output, we test whether
shifting the synthesized constraint bounds consistently shifts the returned
output bundle. Each query uses the first 5 summaries of a (user, intent) pair in
\dataset as its example bundles, and all queries share one synthetic target
document: for each of the 10 topics, it contains the 20 sentences that score
highest on that topic across all 46 state documents (200 sentences in total).
This gives every topic enough strongly scoring content that the shifted queries
have a fair chance of remaining feasible. For each query and topic, we shift
only that topic's constraint bounds by the same additive amount $s$ (i.e.,
$\lowerbound' = \lowerbound + s$ and $\upperbound' = \upperbound + s$), with 15
log-spaced values of $s/(\upperbound-\lowerbound)$ in $[0.01, 20]$, avoiding
linear spacing since it tends to produce infeasible constraints especially at
the extremes. We restrict this experiment to queries whose original and shifted
constraints are both feasible and unrelaxed, since a relaxed or infeasible
query's ILP solver falls back to a similar feasible region regardless of the
shift and would therefore not test consistency. The Hausdorff distance between
the original and shifted bounds is observed to be in the range $[0.003, 12.9]$;
shifts producing a larger distance were infeasible. This restriction yields
about 411 clean pairs of original and shifted constraints per topic, 4{,}115
pairs in total across topics. For each query and topic, we compute the Pearson
correlation between Hausdorff distances of original and shifted constraints and
cosine distances of the corresponding output bundles, and average it over
queries and topics. Across topics, the two distances are strongly correlated
(Fig.~\ref{fig:consistency}), with an average PCC of $0.93$, indicating that
different constraint shifts consistently produce different changes in the
returned bundle.

\section{Scalability}
\label{appx:scalability_relaxation_details}
Here we provide additional detail on the two mechanisms underlying
Section~\ref{sec:scalability_analysis}'s scalability results.

\smallskip\noindent\textbf{\textsc{PreFilter}.} \looseness-1 Before invoking the
ILP, \system restricts the candidate pool to the top-$10x$ sentences of the
target document by SBERT cosine similarity to the mean example embedding of the
example bundle, where $x$ is the example bundle's average length. When the
target document has fewer than $10x$ sentences, this pool covers the entire
document and \textsc{PreFilter} is effectively a no-op; its benefit only appears
once the document exceeds this threshold, at which point the ILP's input size is
capped independent of how much larger the document grows.

\smallskip\noindent\textbf{\sketchref.} \looseness-1 When the
\textsc{PreFilter}-ed candidate pool is still large, \system retrieves via
\sketchref instead of solving one ILP directly over the whole pool. \sketchref
first \textsc{\sketchref Partition}s the pool into groups no larger than a
threshold $\tau$, splitting one clustering dimension (topic) at a time---the
dimension with the largest spread within the group being split---rather than
all dimensions at once, so that group size shrinks gradually as $\tau$ is
lowered instead of collapsing directly from the whole pool to near-singleton
groups. It then \textsc{Sketch}es an initial solution by solving one ILP over
only each group's representative (its centroid), and \textsc{Refine}s it by
greedily replacing each group's representative with real sentences from that
group, one group at a time (largest groups first), backtracking and
reprioritizing whichever group most recently failed if a later group's
constraints cannot be satisfied. Because a package's required length can exceed
the candidate pool's own distinct-sentence count (e.g., a large example bundle
against a small document), both the \textsc{Sketch} and \textsc{Refine} stages
allow a sentence to be selected up to $K{+}1$ times, where $K$ is computed per
query as the smallest value that makes the requested length reachable from that
query's actual candidate pool. 

\smallskip\noindent\textbf{Varying dimensionality.} \looseness-1 \dataset's
topic model has only 10 topics, too few to test how \system scales with the
number of constraints. For Fig.~\ref{fig:scalability} (right), we therefore
replace each sentence's topic scores with its sentence embedding from a
Matryoshka Representation Learning (MRL) model~\cite{kusupati2022matryoshka}.
MRL trains the embedding so that its first $d$ dimensions form a usable
embedding on their own. Truncating the same embedding to $d \in \{2, 4, \dots,
1024\}$ dimensions therefore gives feature spaces of increasing size that
describe the same sentences, rather than outputs of unrelated models. Each of
the $d$ dimensions plays the role of a topic: \system synthesizes one bound per
dimension from the example bundles exactly as it does for topics
(\S\ref{sec:intent}), so the resulting query has $d$ constraints. We run this
over the first 100 (user, intent) pairs of \dataset, each with 5 example and 3
target documents.

\smallskip\noindent\textbf{Synthesizing document/example sizes.} \looseness-1
\dataset's real document sizes cap out at 752 sentences (average $\approx$452),
too small to directly test scalability up to 1M sentences. We synthesize larger
target documents by paraphrasing real sentences with an LLM
(Llama-3.3-70B-Instruct-FP8). We paraphrase each real sentence multiple times in
different writing styles and with different seed sentences as context. Each
target document keeps its real sentences and is grown to the desired size by
sampling a topic from the document's topic distribution, then a seed sentence
for that topic, and adding one of that sentence's unused paraphrases, so that
the synthetic document's topic distribution remains close to the original
without duplicate sentences. Experiments on paraphrased documents score all
sentences with a CTM refit on the original \dataset sentences, so that
paraphrases and originals share one topic space. This ensures that the synthetic
documents remain realistic, while allowing us to test \system's scalability
beyond the real corpus' size. For Fig.~\ref{fig:scalability} (center), we
paraphrase the target documents up to each size; each of the 400 instances draws
its example bundles from 5 source documents and queries the remaining 45 target
documents. For the objective-quality comparison of PreFilter against no
PreFilter, we use a single paraphrased target document, whose bundles under
PreFilter reach $93.7\%$ of the unfiltered objective at 100K sentences and
$86.6\%$ at 1M sentences. The relaxation analysis
(Fig.~\ref{fig:scal_relaxation}) builds its example bundles (10 to 10K
sentences) from real sentences: each chunk of 5 picks a topic at random and adds
the source document's 5 highest-scoring unused sentences for that topic, until
the bundle reaches the desired size; sentences repeat only once a bundle is
longer than its source document. Its target documents (10 to 1M sentences) are
the paraphrased documents of Fig.~\ref{fig:scalability} (center), queried with
the same example bundles; a size below a document's real length uses a random
subset of its real sentences.

\usetikzlibrary{pgfplots.groupplots,calc,patterns.meta}

\begin{figure*}[t]
\centering
\resizebox{0.8\linewidth}{!}{%
\begin{tikzpicture}
\begin{groupplot}[
    group style={
        group size=3 by 2,
        horizontal sep=1.2cm,
        vertical sep=1.5cm,
    },
    width=4cm,
    height=3.06cm,
    xmin=24, xmax=66,
    xtick={30,40,50,60},
    xticklabels={30,40,50,60},
    xlabel style={align=center, yshift=4pt},
    xtick pos=left,
    x tick style={draw=none},
    ytick pos=left,
    ymajorgrids=true,
    grid style={dashed, gray!30},
    label style={font=\small},
    ylabel style={font=\footnotesize},
    tick label style={font=\footnotesize},
    yticklabel style={/pgf/number format/fixed},
    legend style={draw=none, legend columns=3, column sep=6pt, font=\small},
    legend image code/.code={
        \draw[#1] (0cm,-0.08cm) rectangle (0.3cm,0.1cm);
    },
]

\nextgroupplot[
    ybar=0pt, bar width=3.2pt,
    xlabel={$\mathit{pad}$ (\%)\\(a)}, ylabel={ROUGE-1 F$_1$ (\%)},
    ymin=0, ymax=60,
    legend to name=relaxappendixlegend,
    area legend,
]
\addplot[draw=black!80, fill=black!80] coordinates {(30,48.07) (40,49.60) (50,50.25) (60,48.10)};
\addplot[draw=black!15, fill=black!15, postaction={pattern={Dots[distance=0.9pt, radius=0.18pt]}, pattern color=black!70}] coordinates {(30,48.07) (40,49.55) (50,50.37) (60,48.10)};
\addplot[draw=black!55, fill=black!55] coordinates {(30,47.38) (40,49.28) (50,49.17) (60,48.40)};
\legend{SYM, DIR, FeasOpt}

\nextgroupplot[
    ybar=0pt, bar width=3.2pt,
    xlabel={$\mathit{pad}$ (\%)\\(b)}, ylabel={ROUGE-2 F$_1$ (\%)},
    ymin=0, ymax=60,
]
\addplot[draw=black!80, fill=black!80] coordinates {(30,26.26) (40,30.26) (50,31.99) (60,29.01)};
\addplot[draw=black!15, fill=black!15, postaction={pattern={Dots[distance=0.9pt, radius=0.18pt]}, pattern color=black!70}] coordinates {(30,26.26) (40,30.15) (50,32.19) (60,28.88)};
\addplot[draw=black!55, fill=black!55] coordinates {(30,25.86) (40,30.03) (50,30.71) (60,30.05)};

\nextgroupplot[
    ybar=0pt, bar width=3.2pt,
    xlabel={$\mathit{pad}$ (\%)\\(c)}, ylabel={ROUGE-L F$_1$ (\%)},
    ymin=0, ymax=60,
]
\addplot[draw=black!80, fill=black!80] coordinates {(30,31.78) (40,34.96) (50,36.95) (60,34.28)};
\addplot[draw=black!15, fill=black!15, postaction={pattern={Dots[distance=0.9pt, radius=0.18pt]}, pattern color=black!70}] coordinates {(30,31.78) (40,34.97) (50,37.10) (60,34.21)};
\addplot[draw=black!55, fill=black!55] coordinates {(30,31.11) (40,34.88) (50,35.45) (60,35.49)};

\nextgroupplot[
    ybar=0pt, bar width=3.2pt,
    xlabel={$\mathit{pad}$ (\%)\\(d)}, ylabel={\#Relaxations},
    ymin=0, ymax=14,
]
\addplot[draw=black!80, fill=black!80] coordinates {(30,3.27) (40,4.82) (50,7.94) (60,12.48)};
\addplot[draw=black!15, fill=black!15, postaction={pattern={Dots[distance=0.9pt, radius=0.18pt]}, pattern color=black!70}] coordinates {(30,3.27) (40,4.82) (50,7.96) (60,12.50)};
\addplot[draw=black!55, fill=black!55] coordinates {(30,1.00) (40,1.00) (50,1.00) (60,1.00)};

\nextgroupplot[
    ybar=0pt, bar width=3.2pt,
    xlabel={$\mathit{pad}$ (\%)\\(e)}, ylabel={\#Constraints},
    ymin=0, ymax=9,
]
\addplot[draw=black!80, fill=black!80] coordinates {(30,1.49) (40,1.89) (50,1.36) (60,3.54)};
\addplot[draw=black!15, fill=black!15, postaction={pattern={Dots[distance=0.9pt, radius=0.18pt]}, pattern color=black!70}] coordinates {(30,1.49) (40,1.89) (50,1.35) (60,3.55)};
\addplot[draw=black!55, fill=black!55] coordinates {(30,2.93) (40,3.86) (50,5.38) (60,7.69)};

\nextgroupplot[
    ybar=0pt, bar width=3.2pt,
    xlabel={$\mathit{pad}$ (\%)\\(f)}, ylabel={Bound shift},
    ymin=0, ymax=7,
]
\addplot[draw=black!80, fill=black!80] coordinates {(30,1.47) (40,2.23) (50,3.21) (60,5.98)};
\addplot[draw=black!15, fill=black!15, postaction={pattern={Dots[distance=0.9pt, radius=0.18pt]}, pattern color=black!70}] coordinates {(30,1.38) (40,2.11) (50,3.11) (60,5.48)};
\addplot[draw=black!55, fill=black!55] coordinates {(30,0.41) (40,0.84) (50,1.70) (60,3.03)};

\end{groupplot}
\node[anchor=south] at ($(group c2r1.north)+(0,0.15cm)$) {\pgfplotslegendfromname{relaxappendixlegend}};
\end{tikzpicture}
}
\vspace{-2mm}
\caption{(a)~Mean ROUGE-1, (b)~ROUGE-2, and (c)~ROUGE-L F$_1$, (d)~\#relaxations,
(e)~\#constraints relaxed, and (f)~bound shift, by SYM, DIR, \&
FeasOpt~\cite{feasopt} on infeasible queries.}
\label{fig:relaxation_appendix}
\end{figure*}

\section{Implementation Details} \label{appx:implementation}
Here we list the settings our implementation adds to the bound synthesizer of
\S\ref{sec:intent}.

\smallskip\noindent\textbf{Bound synthesis.} \looseness-1 For FTSE, the
initial bounds pad the topic-wise minimum and maximum of
\S\ref{synthinitbound} by $10\%$, i.e., $\lowerbound_j = 0.9 \cdot \min_i
F_j(E_i)$ and $\upperbound_j = 1.1 \cdot \max_i F_j(E_i)$. The query also
bounds the snippet length to within $\pm 25\%$ of the examples' average
length $x$, i.e., $0.75x \le \texttt{COUNT}(B) \le 1.25x$; this length
constraint is never relaxed. For supplier selection over TPC-H, the bounds
are the unpadded minimum and maximum over the example bundles, and the
cardinality bound is the minimum and maximum example bundle size.

\smallskip\noindent\textbf{Bound relaxation.} \looseness-1 Our
implementation follows the bound relaxation algorithm of
\S\ref{sec:bound_relaxation}, with an additional constant multiplier in the
step size: $\epsilon_j = \frac{F_j(\targetdata)}{\delta\,|\targetdata|}$ with
$\delta = 0.5$, i.e., the average contribution of two tuples of
$\targetdata$ rather than one. In addition, the step multiplier is rounded
down to an integer, $\rho = \max\big(1, \lfloor e^{\mu \cdot \max(1, \lfloor
\#\text{attempts}/\tau \rfloor)} \rfloor\big)$, which with $\mu = 0.5$ and
$\tau = 10$ gives $\rho = 1$ for the first $2\tau$ attempts and $\rho = 2, 4,
7$ thereafter. Hence, each early relaxation step moves a violated bound by
$2F_j(\targetdata)/|\targetdata|$ instead of the
$1.65\,F_j(\targetdata)/|\targetdata|$ of
Example~\ref{ex:relaxed_constraints}. Further, (i)~upper bounds are widened
without the cap at $F_j(\targetdata)$; (ii)~violated constraints are
identified with CPLEX's conflict refiner, and when it reports no topic
constraint (e.g., when only the length constraint conflicts), we relax every
topic whose bound width is below the average width across topics; and
(iii)~relaxation stops after $50$ rounds, and a query that is still
infeasible then returns no bundle. All FTSE experiments use this
implementation, and DIR uses the same step.

\section{Alternative Relaxation Mechanisms} \label{appx:relaxation_mechanisms}
Here we detail the comparison between our symmetric relaxation (SYM) and two
alternatives, \emph{directional} (DIR) relaxation and
\emph{Feas\-Opt}~\cite{feasopt}, summarized in \S\ref{sec:relaxation_mechanism}.

We contrast our symmetric relaxation (SYM), which relaxes both bounds of each
violated constraint equally, against two alternatives: (1)~\emph{directional}
(DIR) relaxation, which uses CPLEX's conflict refiner to find a minimal set of
conflicting constraints and determine each bound's relaxation direction, and
(2)~\emph{FeasOpt}~\cite{feasopt}, which minimizes the total slack required to
achieve feasibility. Unlike SYM and DIR, which keep the bundle-length
constraint fixed, FeasOpt may also relax it; its bound shift and number of
relaxed constraints count only the topic constraints. For SYM and DIR, which
relax iteratively, the number of relaxed constraints is the number of topic
constraints in the conflict set of the final relaxation round, i.e., the
constraints whose relaxation made the query feasible, rather than the union
over all rounds; FeasOpt, which relaxes once, counts every topic constraint
with nonzero slack. Each method's averages are taken over the queries for
which that method triggered relaxation.

\looseness-1 Over the \dataset query set---275 (document-set, intent)
user-summary sets, each expanded to 3 held-out target documents, for 825
queries total---each query synthesizes its intent bounds from 5 example
summaries via a per-topic min/max that we pad by $\pm10\%$. However, only
$5/825$ ($0.6\%$) queries are infeasible under this default bound synthesis.
Therefore, as described in \S\ref{sec:relaxation_mechanism}, we force
infeasibility by tightening the bounds instead: in place of the $\pm10\%$
padding, we increase each per-topic lower bound and decrease each upper bound
by $\mathit{pad}$, i.e., $(\lowerbound,
\upperbound)\mapsto((1+\mathit{pad})\times\lowerbound,
(1-\mathit{pad})\times\upperbound)$. Larger values of $\mathit{pad}$ yield
tighter bounds and thus a higher degree of infeasibility. We apply
$\mathit{pad}\in\{30\%,40\%,50\%,60\%\}$ to 80 queries, of which 45, 66, 77,
and 80 become infeasible, respectively, requiring on average $3.3$
($\mathit{pad}=30\%$) to $12.5$ ($\mathit{pad}=60\%$) SYM relaxation rounds. We
compare SYM, DIR, and FeasOpt on these infeasible queries.


Figure~\ref{fig:relaxation_appendix} (a)--(c) shows that the three relaxation
mechanisms produce similar ROUGE-1, ROUGE-2, and ROUGE-L F1 scores, consistent
with the Semantic Similarity shown in \S\ref{sec:relaxation_mechanism}. SYM and
DIR slightly outperform FeasOpt at most 1.7 ROUGE points for 30\%, 40\%, and
50\% padding, whereas FeasOpt shows a slight advantage at 60\% padding by less
than 1.3 ROUGE points. Overall, as expected, because the SYM and DIR mechanisms
are only relaxing the violated constraints, they produce slightly better ROUGE
scores, while FeasOpt's relaxation shows a slight advantage when the
infeasibility is heavily severe.

The bound shift needed for feasibility differs across methods: $\text{SYM} >
\text{DIR} > \text{FeasOpt}$ (Figure~\ref{fig:relaxation_appendix} (f)), with
SYM's bound shift on average about 2.5 times bigger than FeasOpt's, since
FeasOpt minimizes total slack rather than minimizing the number of constraints
altered. The number of relaxation rounds needed for feasibility also follows
$\text{DIR} \approx \text{SYM} > \text{FeasOpt}$
(Figure~\ref{fig:relaxation_appendix} (d)), with SYM requiring about 7.1 times
more rounds than FeasOpt on average. On the other hand, the number of relaxed
constraints is $\text{FeasOpt} > \text{DIR} \approx \text{SYM}$
(Figure~\ref{fig:relaxation_appendix} (e)), with FeasOpt relaxing on average
about 2.5 times as many constraints as SYM.

Figure~\ref{fig:relaxation_quality} shows that SYM, DIR, and FeasOpt yield
comparable semantic similarity, differing by at most 1.2 points at any
$\mathit{pad}$ level.
Runtime is comparable at low $\mathit{pad}$, with a median of $1.58$s for SYM
and $1.14$s for FeasOpt at $\mathit{pad}=30\%$, whereas FeasOpt is faster at
high $\mathit{pad}$, taking 0.95s compared to 2.23s for SYM at
$\mathit{pad}=60\%$, since it resolves infeasibility in a single
slack-minimization solve, while SYM and DIR relax iteratively, $12.5$ rounds on
average at $\mathit{pad}=60\%$.

We also observe that SYM relaxes fewer constraints than Feas\-Opt at every
padding level, relaxing 1.4--3.5 constraints on average compared to 2.9--7.7 for
FeasOpt. This is because FeasOpt spreads smaller relaxations across more
constraints, while SYM concentrates larger relaxations on fewer constraints. SYM
takes only $3.3$ relaxation rounds on average at $\mathit{pad}=30\%$, rising to
$12.5$ at $\mathit{pad}=60\%$, whereas FeasOpt always relaxes the constraints
once. Therefore, in the common case of
less severe infeasibility (low pad), SYM is the preferred relaxation mechanism
due to its usability benefits, since it (i)~requires fewer constraints to be
relaxed at comparable quality function, semantic similarity, and runtime, and
(ii)~ensures symmetric relaxation, leaving equal headroom for further tuning of
the corresponding slider. In contrast, Feas\-Opt is ideal when faster
convergence is desired and severe infeasibility (high pad) is expected. However,
across our datasets, we never observed a very high degree of infeasibility. In
fact, on \dataset only 0.6\% of queries were infeasible with a low degree of
infeasibility.

\section{Data Skew Robustness}
\label{appx:robustness_analysis}
Here we provide additional detail on how the skewed target documents in
Section~\ref{sec:robustness} were constructed and evaluated.

\smallskip\noindent\textbf{Constructing skewed documents.} \looseness-1 For each
skew ratio (50\%, 60\%, 70\%, 80\%, 90\%), we build 100 synthetic target
documents of 200 sentences each, each for one (intent, target state, dominant
topic) combination drawn from \dataset. The target states are three \dataset
states: Massachusetts, Minnesota, and New Mexico. The random dominant topic's sentences are first drawn, from an LLM
paraphrase pool generated for that (state, topic) pair; once that pool is
exhausted, the remaining dominant-topic budget is filled with the target state's
own original, top-scored sentences for that topic, since near-duplicate
paraphrases of a small seed set measurably degrade summary quality at high skew.
Every other (non-dominant) topic receives an equal share of the remaining
sentences, drawn uniformly at random from the target state's own original
sentences for that topic, with a floor of 3 sentences per non-dominant topic so
no topic is starved entirely. In addition, the target state's ground-truth
sentences (from the recorded human summary) are always included in the synthetic
document; when a non-dominant topic's share is too small to hold its
ground-truth sentences, that share is enlarged and the dominant topic's share is
reduced accordingly. Since all sentences come from the target state's own text
or its paraphrases, and the ground-truth sentences are present in the document,
the recorded human summary is a valid ground truth for the synthetic document.
The example bundle used to query \system for a given synthetic document is that
same person and that same intent's own example bundle, drawn from the trial's
other states, unchanged from a normal (non-skewed) evaluation.

\smallskip\noindent\textbf{Achieved skew.} \looseness-1 We label each skew level
by the ratio used to construct its documents. At 90\%, the floor raises the
non-dominant topics' $20$ sentences ($2$ or $3$ each) to $27$, capping the
dominant topic at $173$ of $200$ sentences ($86.5\%$). The floor keeps the query
testing skew rather than missing content: \dataset's example summaries ($11.3$
sentences on average) place about $4.5$ sentences outside their main topic, and
with $2$ random sentences per non-dominant topic, $30\%$ of the candidate
documents would have a secondary topic whose own sentences cannot reach its
lower bound, versus $13\%$ with $3$. The intent's main topic, when not dominant,
is kept by the included ground-truth sentences. Over all candidate documents,
the achieved dominant share averages $84.7\%$, $78.9\%$, $69.5\%$, $59.8\%$, and
$49.9\%$ for the nominal 90--50\% levels, so the levels keep their orderr; only
the 90\% level falls notably short of its label, by about $5$ points.

\begin{figure*}[t]
    \centering
    \includegraphics[width=\linewidth]{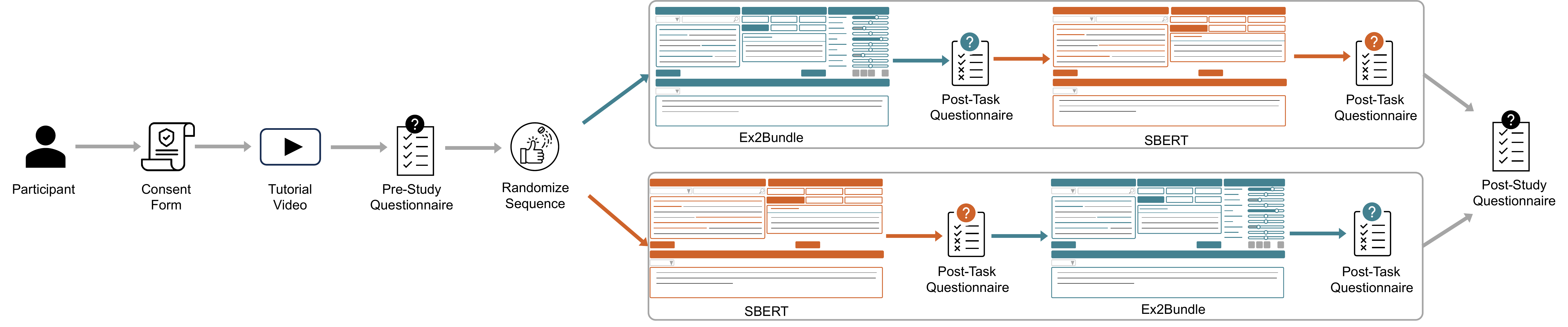}
    \caption{User study protocol. Participants first completed a consent form
    and watched a tutorial video on how to use the assigned tool. Then, they
    first performed a pre-study questionnaire, followed by a decision-making
    task using either \SBERT or \system. After completing the first task and its
    post-task questionnaire, they performed the same task with the
    other tool. Finally, they completed a post-study questionnaire comparing
    both tools and providing feedback on their experience.
	}
    \label{fig:protocol}
\end{figure*}

\begin{figure*}[t]
    \centering
    \includegraphics[width=1\linewidth]{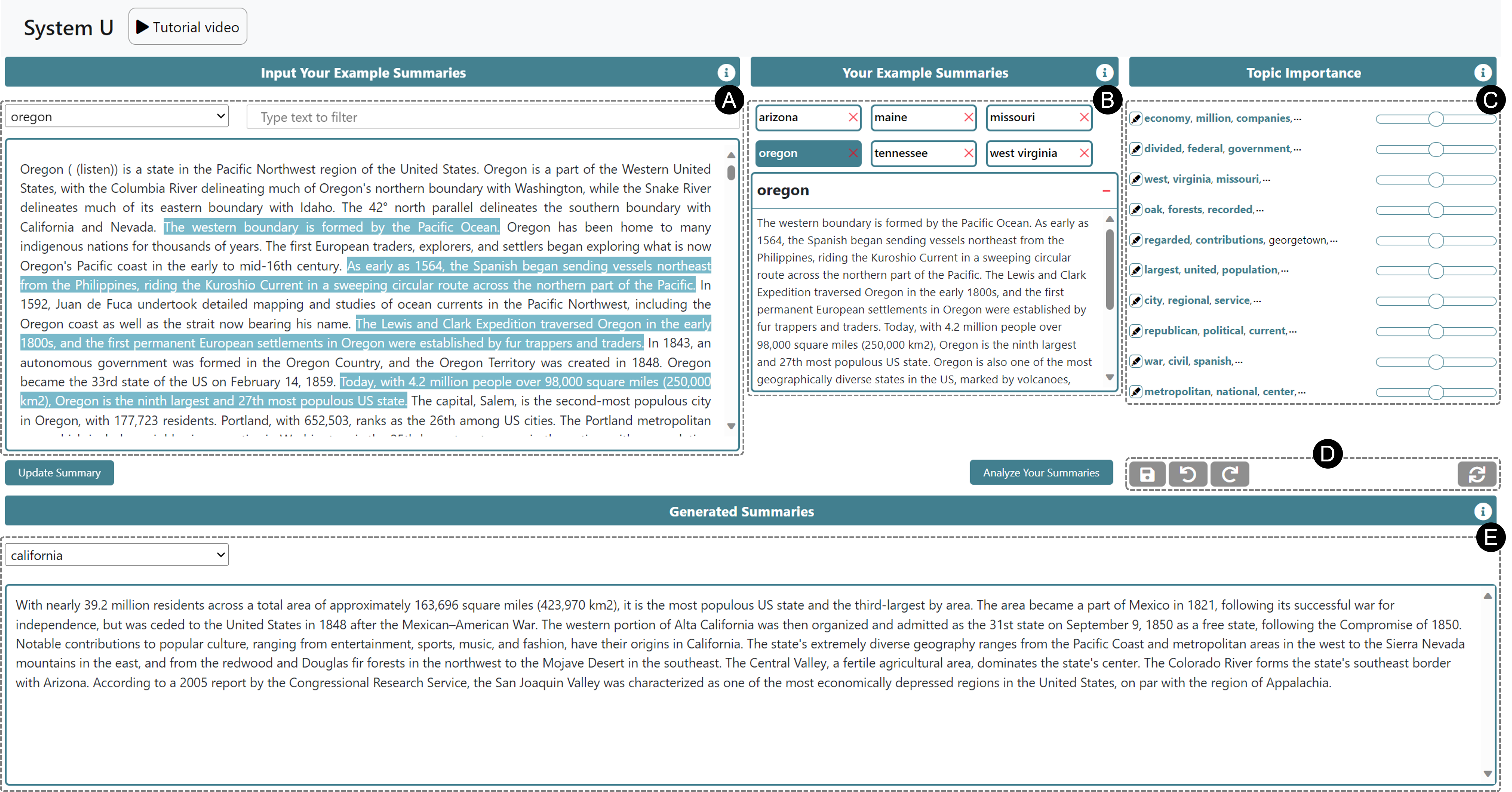}

    \caption{A snapshot of \system's interface. \stepCounter{A} Users can input
    examples by clicking on the interface, and search for keywords that they are
    interested in the documents. \stepCounter{B} Users can review their examples
    and modify them. \stepCounter{C} On the right side, users can adjust the
    importance of different topics to tune the retrieved snippets using
    sliders. \stepCounter{D} Users can save their retrieved snippets.
    \stepCounter{E} The retrieved snippets are displayed under the examples,
    the figure showing the snippets in California state. }

    \Description{Screenshot of the \system\ web interface showing five labelled
    regions: (A) the input panel where users add example sentences and run
    keyword searches, (B) the example list with edit controls, (C) the
    topic-importance slider panel, (D) save controls, and (E) the
    retrieved-snippet display.}
	
    \label{fig:interface}
\end{figure*}

\section{User Study}
\label{appx:userstudyappendix}
To evaluate \system, we conducted a within-subjects study with 20 participants
recruited via Mechanical Turk. Participants were tasked with identifying their
preferred state for relocation based on Wikipedia articles, using both a
baseline (\SBERT) and \system. We followed a mixed-methods
approach~\cite{greene1989toward} to analyze quantitative and qualitative data
from the study. The study protocol was approved by the institutional review
board (IRB) of the University of Massachusetts Amherst.

\begin{figure*}[t]
\centering
\begin{tikzpicture}
\begin{groupplot}[
    group style={
        group size=4 by 1,
        horizontal sep=1.6cm,
        ylabels at=edge left,
    },
    ybar,
    ymin=0, ymax=12.5,
    ytick={0,3,6,9,12},
    ylabel={\# Participants},
    ylabel style={font=\small},
    enlarge y limits=false,
    ymajorgrids=true,
    grid style={dashed, gray!30},
    width=4.5cm,
    height=4cm,
    nodes near coords,
    nodes near coords align={vertical},
    every node near coord/.append style={font=\scriptsize},
    x tick label style={rotate=90, anchor=east, font=\small},
    y tick label style={font=\small},
    xlabel style={font=\small\bfseries, align=center},
    xtick style={draw=none},
    ytick style={draw=none},
    enlarge x limits=0.25,
]

\nextgroupplot[
    xlabel={(a) Search Frequency},
    symbolic x coords={Infrequent, Neutral, Frequent, Always},
    xtick=data,
]
\addplot[fill=gray!100, draw=black] coordinates {
    (Infrequent, 4)
    (Neutral,  3)
    (Frequent, 11)
    (Always,   2)
};

\nextgroupplot[
    xlabel={(b) Time Spent},
    xlabel style={yshift=-11pt},
    symbolic x coords={{<15m}, {15-30m}, {30-45m}, {45-60m}, {60+m}},
    xtick=data,
    enlarge x limits=0.15,
    bar width=0.35cm,
]
\addplot[fill=gray!100, draw=black] coordinates {
    ({<15m},    2)
    ({15-30m},  10)
    ({30-45m},  5)
    ({45-60m},  2)
    ({60+m},    1)
};

\nextgroupplot[
    xlabel={(c) Unexpected Results},
    symbolic x coords={Rarely, Neutral, {Most often}},
    xtick=data,
]
\addplot[fill=gray!100, draw=black] coordinates {
    (Rarely,      10)
    (Neutral,      7)
    ({Most often}, 3)
};

\nextgroupplot[
    xlabel={(d) Search Quality},
    symbolic x coords={Average, {Above avg.}, Good},
    xtick=data,
]
\addplot[fill=gray!100, draw=black] coordinates {
    (Average,      8)
    ({Above avg.}, 6)
    (Good,         6)
};

\end{groupplot}
\end{tikzpicture}
\vspace{-3mm}
\caption{Pre-study questionnaire results (N=20): (a) search frequency from documents or websites, (b) time spent per session (min), (c) frequency of encountering unexpected results, and (d) self-rated search quality.}
\vspace{-2mm}
\label{fig:prestudy_results}
\end{figure*}

\subsection{User Study Settings}

\smallskip\noindent\textbf{Participants.} \looseness-1 We recruited 20
participants (P01-P20) through Amazon Mechanical Turk~\cite{crump2013evaluating}.
Our inclusion criteria stipulated that the participants must reside in North
America, where this study was conducted, and be fluent in English. All
participants were qualified Amazon Master Workers, demonstrating a high success
level across a wide range of tasks. Participants were compensated with \$25. 

\smallskip\noindent{\textbf{Procedure.}} \looseness-1 We followed a
single-session within-subject study protocol (Fig.~\ref{fig:protocol}).
Each study session began with collecting participants' consent. Participants
then watched a 5-minute unskippable video tutorial on using the tool before
performing their tasks.
After the tutorial, participants completed a pre-study questionnaire to gather
their information-seeking purpose, practice, tool usage, time spent, strategy,
and ratings of the quality of information. Then, participants were randomly
assigned to start with either \SBERT or \system to negate the transfer effect.
During the study, participants interacted with an interface,
Fig.~\ref{fig:interface}, to perform their tasks. The interface was designed
to allow users to input examples, review and modify them, adjust the importance
of different topics using sliders only for \system, and view the retrieved
snippets. We anonymized the tools' names to obscure which tool each was using. 

Participants were asked to perform a decision-making task: to identify which
state in the United States (US) they would relocate to and live in if they were
offered their dream job there. We provided them with Wikipedia articles on 25
randomly generated US states. Therefore, each participant worked with two
different sets of 25 states across the two tools.
We instructed them to read and input at least 2 state snippets with no upper
limit. They were also free to use the assigned tool however they wanted.

After working with one tool, we asked the participants to perform the same task
with the other tool, but this time, Wikipedia articles were provided on 25 other
states. This was done to negate potential unwanted bias and transfer effect from
one tool to the other. After completing tasks with both tools, we asked them a
set of post-study questions to compare \SBERT and \system regarding their
summary quality, ease of use, and extracted snippets. We also asked them if
given the choice, which tool they would use to perform their usual
information-seeking tasks. 

\smallskip\noindent{\textbf{Data Collection and Analysis.}} \looseness-1 We
collected both quantitative and qualitative data. Quantitative data include
usage logs, interactions, and time spent working with the tool components. We
also collected responses to the Likert scale as well as qualitative answers to
open-ended questions pre-study, post-task, and post-study questionnaires. We
analyzed the quantitative data by comparing the participants' ratings such as
the summary quality, ease of use, and daily usage of the tools. We analyzed the
qualitative data using thematic analysis~\cite{braun2012thematic}. Two members
of the research team independently coded data from five randomly selected
participants using the open coding method~\cite{khandkar2009open} using
spreadsheets. The coders discussed, resolved disagreements, and consolidated
their codes into a codebook containing the representative set of codes. Then
the coders used this codebook to individually code the data from all 20
participants, adding new codes as they emerged. Finally, they compiled the
codes into themes across five discussion sessions. The themes were discussed
and finalized by the research team.

\balance
\subsection{Pre-study Questionnaire}
\label{pre_study_questionnaire}
In the pre-study questionnaire, we asked participants about their
information-seeking purpose, practice, tool usage, time spent, strategy, and
ratings of the quality of information. We show bar charts of the results for
four of the questions in Fig.~\ref{fig:prestudy_results}. The majority of
participants (11/20) reported frequently searching across multiple documents or
websites. In an information-searching session, the majority of participants
(10/20) spent 15-30 minutes, while 2/20 spent $<$15 minutes, and 5/20 spent
30-45 minutes and 2/20 spent 45-60 minutes, or 1/20 spent 60+ minutes.
When asked about the unexpected results, 10/20 said rarely, 7/20 said
neutral, and 3/20 said often. This suggests that most participants do not often
encounter unexpected results during their information-seeking process. Finally,
6/20 rated the quality of the information gathered after their search process
as good. Another 8/20 rated average, while 6/20 rated above average. In
summary, the pre-study questionnaire results suggest that most participants
frequently search across multiple documents or websites, spend a moderate
amount of time on information-seeking sessions, and do not often encounter
unexpected results. The quality of information gathered is generally rated as
good or above average by most participants.

\begin{figure*}[h]
\centering

\begin{minipage}{\linewidth}
\centering
\resizebox{1\textwidth}{!}{
\begin{tabular}{p{12cm}c}
\toprule
\textbf{Question and Response} & \textbf{Count} \\
\midrule
\multicolumn{2}{l}{\textit{What was your overall experience of searching for information using summaries?}} \\
\hline
\phantom{space}Excellent & 1 \\
\phantom{space}Good & 13 \\
\phantom{space}Average & 3 \\
\phantom{space}Very poor & 3 \\
\midrule
\multicolumn{2}{l}{\textit{While creating example input summaries, did you identify any interesting information that you had not previously thought about?}} \\
\hline
\phantom{space}Yes & 9 \\
\phantom{space}No & 11 \\
\bottomrule
\end{tabular}}
\vspace{1pt}
\par\small(a) 
\end{minipage}

\vspace{10pt}

\begin{minipage}{\linewidth}
\centering
\resizebox{1\textwidth}{!}{
\begin{tabular}{lcc}
\toprule
\textbf{Which tool did you prefer in each category?} & \textbf{\SBERT} & \textbf{\system} \\
\midrule
Which tool generated better summaries? & 8 & \textbf{12} \\
Which tool did you find more useful? & 9 & \textbf{11} \\
Which tool did you find easier to use? & 4 & \textbf{16} \\
Which tool helped you better in making a decision? & 8 & \textbf{12} \\
Which tool did you find more flexible? & 8 & \textbf{12} \\
Which tool did you find more comfortable to use? & 5 & \textbf{15} \\
If you had to choose a tool to search for information from multiple documents/websites, which one would you choose? & 9 & \textbf{11} \\
\bottomrule
\end{tabular}}
\vspace{1pt}
\par\small(b) 
\end{minipage}
\vspace{-3mm}
\caption{User study satisfaction results. (a) Overall experience of the user
study and discovery of new information. Most participants had a good experience
with the system, and nearly half of the participants (9/20) discovered new
information. (b) Tool preference responses. The majority of participants
preferred \system over \SBERT across all categories, showing that \system was
generally favored for extracting text snippets for focused intent tasks.}
\vspace{-2mm}
\label{fig:appendix_user_results}
\end{figure*}

\bigskip

\subsection{Post-study Questionnaire}
We asked participants to answer the same set of questions after completing their
task with each tool: \SBERT and \system. The questions were designed to evaluate
the participants' experience with each tool, including their interactions, the
quality of the extracted snippets, and their preferences for future use.

\smallskip\noindent{\textbf{Post-Task Questionnaire (\SBERT)}.} \looseness-1
After completing their task with \SBERT, participants answered questions about
their experience. Of the 20 participants, one did not submit the post-task
questionnaire for \SBERT, so we analyzed the responses from 19 participants. When asked about
modifying their example input snippets, 8 modified their example input snippets
during the task while 11 did not. Among these participants, 11/19 rated modifying
examples as useful or very useful, 5/19 were neutral, and the remaining 3/19
found it somewhat useful or not useful. Regarding ease of modification, 10/19
rated it as easy or very easy, while 5/19 were neutral. For the quality of the
extracted snippets, 14/19 participants rated them as good or excellent (9 good,
5 excellent), while 3/19 rated them average and 2/19 rated them poor. When asked
whether the extracted snippets helped them make a decision, 14/19 said yes. Only
4/19 reported encountering unexpected information in the snippets, one
participant stating that the snippets showed some information that is not
relevant to their input snippets. Regarding future use, 13/19 agreed or strongly
agreed that they would use \SBERT to search information across multiple
documents, while 2/19 disagreed. Participants who liked \SBERT highlighted its
ease of use and ability to filter relevant content. Common complaints included
slow performance, occasional browser crashes, and the absence of importance
sliders for tuning results.

\smallskip\noindent{\textbf{Post-Task Questionnaire (\system)}.} \looseness-1
After completing their task with \system, participants answered questions about
their experience. We asked them the same questions as in the post-task
questionnaire for \SBERT, but we also asked them additional questions about
their experience with the topic importance sliders, which is a distinctive
feature of \system. Of the 20 participants with valid responses, 10 modified
their example input snippets while 10 did not. 
Among all participants, 8/20 found modifying examples useful or very useful,
9/20 were neutral, and 3/20 found it not useful. Regarding ease of modification,
11/20 rated it as easy or very easy, while 7/20 were neutral and 2/20 found it
not easy.

\looseness-1 A distinctive feature of \system is its relative topic importance sliders. The
majority of participants (18/20) used the sliders during the task, while 2/20
did not, citing technical issues. 
Of those who used the sliders, 7/18 observed a change in the extracted snippet
after adjusting them. 
Regarding usefulness of the sliders, 10/20 rated them as useful or very useful,
6/20 were neutral, and 4/20 found them not useful. In terms of ease of use,
15/20 rated the sliders as easy or very easy.

\looseness-1 For the quality of the extracted snippets, 12/20 participants rated them as
good or excellent (9 good, 3 excellent), while 6/20 rated them average and 2/20 rated them poor or very poor.
When asked whether the snippets helped them make a decision, 14/20 said yes.
For participants who said no, they mentioned that the snippets were not fully
covering the information they wanted.
Only 3/20 reported finding unexpected information in the snippets, mentioning
that some information was not what they expected based on their input snippets. 
Regarding future use, 11/20 agreed or strongly agreed that they would use
\system to search information across multiple documents, while 2/20 strongly
disagreed, 3/20 disagreed, and 4/20 were neutral. 
Participants who liked \system highlighted the topic importance sliders, easy
navigation, and the ability to customize snippet content. Common complaints
included slow performance, freezing, and occasional confusion because of the
tool's technical issues.

\subsection{Quantitative Results after Using Both Tools}
\label{quantitative_results}
We first analyzed the quantitative data by comparing the participants' ratings
such as the summary quality, ease of use, and daily usage preference for both
\SBERT and \system. We also compared the number of interactions of each type
between the two tools to understand how participants engaged with them and how
much time they spent working with the tools to support their decision-making
process.

\smallskip\noindent{\textbf{Participants' satisfaction.}} \looseness-1 After
using both tools, we asked participants to rate the summary quality and ease of
use of each tool. In Fig.~\ref{fig:appendix_user_results}, overall participants
evaluated the experience with retrieving snippets with tools as good (13/20;
Fig.~\ref{fig:appendix_user_results} (a)). When comparing \SBERT and \system,
many participants preferred \system over \SBERT for all dimensions
(Fig.~\ref{fig:appendix_user_results} (b)). For example, 12/20 participants
preferred \system over \SBERT for generating better snippets, 11/20 preferred
\system for usefulness%
, and 16/20 preferred \system for ease of use. When asked which tool they would
choose for their usual information-seeking tasks, 11/20 participants preferred
\system, while 9/20 \SBERT. 

\pgfplotsset{compat=1.18}
\usetikzlibrary{patterns}
\usetikzlibrary{calc}
\usetikzlibrary{decorations.pathmorphing}

\begin{figure}
    \centering
\begin{tikzpicture}
\begin{axis}[
    ybar=0pt,
    xtick style={draw=none},
    enlarge y limits=false,
    height=4.2cm,
    width=\linewidth,
    ymin=0,
    ymax=160,
    font=\small,
    ymajorgrids=true,
    grid style={dashed, gray!30},
    enlarge x limits=0.1,
    ylabel={Average \#Interactions},
    symbolic x coords={Search,Selection,Update,Learn,Retrieval,Slider},
    xtick=data,
    bar width=0.3cm,
    legend style={
        at={(0.98,0.98)},
        anchor=north east,
        legend columns=1,
        draw=none,
        fill=white,
        nodes={text width=2.5cm, align=left},
    },
    legend image code/.code={
        \draw[draw=black,fill opacity=1] (0cm,-0.08cm) rectangle (0.3cm,0.1cm);
    },
]
\addplot+[ybar, fill=white, draw=black] coordinates {(Search,151.26) (Selection,9.63) (Update,8.42) (Learn,2.26) (Retrieval,10.05) (Slider,0.00)};
\addplot+[ybar, fill=black, draw=black] coordinates {(Search,47.89) (Selection,5.00) (Update,2.89) (Learn,1.84) (Retrieval,21.21) (Slider,78.16)};
\legend{SBERT,\system}
\end{axis}
\end{tikzpicture}
    \caption{Comparison between SBERT and \system in terms of interaction type counts by averaging the number of interactions per user. Here, 
    \textit{Search}: keyword queries,
    \textit{Selection}: example selection,
    \textit{Update}: example modification, 
    \textit{Learn}: intent learning,
    \textit{Retrieval}: bundle retrieval, 
    \textit{Slider}: slider adjustments.
    }
    \label{fig:event_types_comparison}
\end{figure}
\smallskip\noindent{\textbf{Participants' interactions.}} To understand how
participants interacted with the tools, we categorized interactions into six
types: \textit{Search}: keyword queries, \textit{Selection}: example selection,
\textit{Update}: example modification, \textit{Learn}: intent learning,
\textit{Retrieval}: backend retrieval, and \textit{Slider}: interactive
parameters. \textit{Search} includes keyword queries that participants used to
find relevant information in the Wikipedia articles such as ``climate''.
\textit{Selection} includes interactions where participants selected specific
sentences or sections from the Wikipedia articles to include in the input
snippets. \textit{Update} includes interactions where participants modified
their selected examples, such as adding or removing sentences from the input
snippets. \textit{Learn} includes \system's learning process, where it learns
the user's intent based on their example selections. \textit{Retrieval} includes
interactions where participants retrieved snippets with \paql after providing
examples or adjusting sliders. \textit{Slider} includes interactions where
participants adjusted the importance sliders to tune the retrieval results. We
averaged the count of each interaction type per user and compared the counts
between \SBERT and \system.

Fig.~\ref{fig:event_types_comparison} shows that \SBERT users have more manual
interactions: they performed about 3.1 times more \textit{Search} interactions
than \system users, and selected and updated more examples than \system users by
about 1.9 and 2.9 times, respectively. However, since \system users can adjust
the importance sliders unlike \SBERT users, they have more \textit{Slider}
interactions than \SBERT users. Naturally, \system users also have about 2.1
times more \textit{Retrieval} interactions than \SBERT users, since \system
retrieves bundles based on slider adjustments, while \SBERT does not have such
capabilities.

\begin{figure}[h]
    \centering
\begin{tikzpicture}
\begin{axis}[
    width=3.2cm,
    height=2.3cm,
    scale only axis=true,
    font=\fontsize{6.5}{7.5}\selectfont,
    tick label style={font=\fontsize{6.5}{7.5}\selectfont},
    label style={font=\fontsize{6.5}{7.5}\selectfont},
    xtick style={draw=none},
    ylabel={Total request time (s)},
    xmin=0.5,
    xmax=2.5,
    xtick={1,2},
    xticklabels={SBERT,\system},
    ymin=0,
    ymax=250,
    ymajorgrids=true,
    grid style=dashed,
    legend pos=north west,
]

\draw[thick] (axis cs:1,6.2908) -- (axis cs:1,14.7089);
\draw[thick] (axis cs:1,31.2806) -- (axis cs:1,72.6811);
\draw[thick] (axis cs:0.92,6.2908) -- (axis cs:1.08,6.2908);
\draw[thick] (axis cs:0.92,72.6811) -- (axis cs:1.08,72.6811);
\draw[thick] (axis cs:0.85,14.7089) rectangle (axis cs:1.15,31.2806);
\draw[thick] (axis cs:0.85,17.6061) -- (axis cs:1.15,17.6061);

\draw[thick] (axis cs:2,6.3507) -- (axis cs:2,14.1374);
\draw[thick] (axis cs:2,90.9806) -- (axis cs:2,209.3841);
\draw[thick] (axis cs:1.92,6.3507) -- (axis cs:2.08,6.3507);
\draw[thick] (axis cs:1.92,209.3841) -- (axis cs:2.08,209.3841);
\draw[thick] (axis cs:1.85,14.1374) rectangle (axis cs:2.15,90.9806);
\draw[thick] (axis cs:1.85,42.0503) -- (axis cs:2.15,42.0503);

\end{axis}
\end{tikzpicture}
\caption{Total backend request time per participant (s) for \SBERT and \system. \system has a higher median request time than \SBERT, which is expected due to the additional processing steps involved in \system such as tuning sliders.}
\label{fig:avg_time_boxplot}
\end{figure}
\smallskip\noindent{\textbf{Participants' querying time.}} We measure the cumulative backend processing time per participant session. This
metric reflects backend request processing time rather than end-to-end
participant interaction time. We report this metric to demonstrate the level of
engagement and interaction participants had with the system. As shown in
Fig.~\ref{fig:avg_time_boxplot}, \system has a higher average backend
processing time than \SBERT, with a median of 42.05 seconds for \system compared
to 17.61 seconds for \SBERT. This suggests that participants interacted more
frequently and deeply with \system, likely due to its interactive features such
as the importance sliders and the ability to retrieve snippets based on user
input. In contrast, \SBERT's lower processing time may indicate less frequent
interactions, as it does not offer the same level of interactivity and
customization as \system. Furthermore, on average participants retrieved bundles
21.2 times with \system, while only 10.1 times with \SBERT, which also supports
the notion of higher engagement with \system.

\balance
\subsection{Qualitative Results after Using Both Tools}
\label{qualitative_results}
We analyzed the qualitative data by asking participants to provide feedback on
their experience with both tools. We asked them to share their preferences,
grievances, and suggestions for improvement. We also asked them to provide
feedback on the extracted snippets and how they supported their decision-making
process. 

\smallskip\noindent{\textbf{Participants preferred \system over \SBERT.}}
Compared to \SBERT, our participants preferred \system. When comparing \SBERT
and \system, many participants (at least 11/20 for each category;
Fig.~\ref{fig:appendix_user_results} (b)) preferred \system over \SBERT due to
its greater customization. P07 mentioned, \emph{``I absolutely loved the
sliders. Being able to rank what is most important to me instead of the system
trying to guess is great.''} While performing their tasks in \SBERT, several
participants (5/20) mentioned that it lacked customizability. P16 said,
\emph{``There was no way to adjust which factors were more or less important to
me. It was tedious to scroll through long Wikipedia articles looking for
relevant info to include, and sometimes articles did not address specific things
that I wanted to include. I sometimes filtered/searched for certain words or
terms, but there were no results, or none relevant.''} However, some
participants (4/20) preferred the simplicity that \SBERT provides. P20
mentioned, \emph{``It [\SBERT] didn't require much input to generate summaries
that captured what I was most interested in, which made it easier to review the
information and make a decision.''}

\smallskip\noindent{\textbf{Topic importance sliders helped to tune snippets.}} While
working on their tasks using \system, the participants had access to a set of
topic importance sliders to adjust extracted information. The majority of
participants (18/20) used the topic importance sliders to fine-tune their
snippets and ensure they reflected the information specific to participants'
needs. P08 explained how they used the sliders to focus on the economy and
climate for each state, \emph{``I raised the slider up so that the output
summaries would prioritize information that pertained to the economy and climate
of each state since those were the most important factors for me when it came to
choosing a state to work in.''} In contrast, P09 used the sliders to reduce the
importance of certain topics, \emph{``Topics that I wasn't interested in, I just
moved to lower importance. [...] That seems to have put more emphasis on certain
topics I cared about [in the summary].''} The participants (2/20) who did not
use the topic importance sliders stated that their choice was predetermined and
that changes to the snippets would not have swayed their decision. 

\smallskip\noindent{\textbf{Extracted snippet supported informed decision-making.}}
Participants across both conditions (14/19 for \SBERT and 14/20 for \system)
reported that the extracted snippets helped them make informed decisions by
enabling them to identify relevant information, gain insight into interesting
information, and validate their own preferences. After using \system, P04
mentioned, \emph{``The summaries show me the details and aspects that I'm
interested in while omitting many irrelevant details that I have no interest in
knowing. I was able to quickly see which states I find more appealing through
reading only the details that I would find relevant.''} Similarly, after using
\SBERT, P07 mentioned, \emph{``It was easy to see just the things that were most
important to me without the entire Wikipedia page in front of me. Having such
detail is nice, but it can be overwhelming. I was able to read the summaries and
say, Oh, that doesn't sound good. I'm going to avoid that, or Awesome! I am
going to continue learning more about this state.''} Participants were also
happy with the quality of the summary and how it organized their preferred data
quickly and concisely. P07 pointed out how such efficiencies helped with their
tasks, \emph{``Being able to switch between the states quickly meant I could see
almost instantly how those states compared to each other.''}

\smallskip\noindent{\textbf{Participants' feedback for improvement.}} The participants
suggested several improvements. One of the major grievances of participants was
the presence of irrelevant information in the extracted snippet. However, P08
found use in such irrelevant snippet while making decisions: \emph{``I read
information that showed that North Carolina had a high poverty rate, which I did
not remember inputting into the input summaries. However, this data helped me to
exclude North Carolina from my list of states to live in.''} In addition, some
noted the documents' text-heavy nature. They suggested that, while the snippets
were useful for identifying relevant information, creating the input snippets
was sometimes tedious and often led to confusion, as going through the Wikipedia
article to find candidate sentences for the input was time-consuming.
Participants also recommended following the document organization and adding an
option to filter sections while reading the articles. Additionally, after using
\system, 3/20 participants suggested that working with the tool has a steep
learning curve.  
P16 mentioned, \emph{``It was a bit confusing at first, even after the tutorial,
to know exactly how to use it [\system] and what to do for the best results. I
did not like having to scroll through really long Wikipedia articles for each
state in order to find which parts to highlight and use for the summary.''}

\end{document}